\pdfoutput=1                    
\documentclass{article}

\usepackage[final,nonatbib]{neurips_2020}

\usepackage[utf8]{inputenc} 
\usepackage[T1]{fontenc}    
\usepackage{hyperref}       
\usepackage{url}            
\usepackage{booktabs}       
\usepackage{amsfonts}       
\usepackage{nicefrac}       
\usepackage{microtype}      
\usepackage{enumitem}

\usepackage{amsmath,amsfonts,bm}

\def\eqref#1{equation~\ref{#1}}

\def\1{\bm{1}}

\def\vc{{\bm{c}}}

\def\vr{{\bm{r}}}
\def\vs{{\bm{s}}}

\def\vu{{\bm{u}}}

\def\vx{{\bm{x}}}

\def\mA{{\bm{A}}}

\def\mM{{\bm{M}}}

\def\mW{{\bm{W}}}

\DeclareMathAlphabet{\mathsfit}{\encodingdefault}{\sfdefault}{m}{sl}
\SetMathAlphabet{\mathsfit}{bold}{\encodingdefault}{\sfdefault}{bx}{n}

\DeclareMathOperator*{\argmin}{arg\,min}

\usepackage{acro}

\DeclareAcronym{nn}{
short = NN,
long = neural network,
}

\DeclareAcronym{cnn}{
short = CNN,
long = convolutional neural network,
}

\DeclareAcronym{cnns}{
short = CNNs,
long = convolutional neural networks,
}

\DeclareAcronym{pde}{
short = PDE,
long = partial differential equation,
}

\DeclareAcronym{2afc}{
short = 2AFC,
long = two-alternative forced choice,
}

\DeclareAcronym{hvs}{
short = HVS,
long = human visual system,
}

\DeclareAcronym{sph}{
short = SPH,
short-indefinite = an,
long = smoothed particle hydrodynamics,
}

\DeclareAcronym{flip}{
short = FLIP,
long = fluid-implicit-particle,
}

\DeclareAcronym{simple}{
short = SIMPLE,
long = semi-implicit method for pressure linked equations,
}

\DeclareAcronym{eno}{
short = ENO,
long = essentially non-oscillatory,
}

\DeclareAcronym{rmse}{
short = RMSE,
long = root-mean-square error,
}

\DeclareAcronym{psnr}{
short = PSNR,
long = peak signal-to-noise ratio,
}

\DeclareAcronym{ssim}{
short = SSIM,
long = structure similarity metric,
}

\DeclareAcronym{tgv}{
short = TGV,
long = Taylor-Green vortex,
}

\DeclareAcronym{ncm}{
short = NCM,
long = near-convergence consistency metric,
}

\newcommand{\myeqref}[1]{Eq.~\ref{#1}}

\newcommand{\myfigref}[1]{Fig.~\ref{#1}}
\newcommand{\mytabref}[1]{Table~\ref{#1}}

\newcommand{\myappref}[1]{App.~\ref{#1}}

\usepackage[normalem]{ulem}     
\usepackage{color}

\definecolor{R}{rgb}{0.9, 0, 0}
\definecolor{G}{rgb}{0, 0.6, 0}
\definecolor{B}{rgb}{0, 0, 1.0}

\definecolor{kCol}{rgb}{0.1, 0, 0.9}
\definecolor{nCol}{rgb}{0.8, 0.3, 0}
\definecolor{yCol}{rgb}{0.9, 0, 0.9}
\definecolor{todoCol}{rgb}{0.9, 0.0, 0}

\newcommand{\new}[1]{#1}

\usepackage{url}

\usepackage{breakurl}
\usepackage{amsmath}
\usepackage{subcaption}
\usepackage{graphicx}
\usepackage{placeins}
\usepackage{wrapfig}
\usepackage{mathrsfs}
\usepackage{overpic}

\newcommand{\mytitle}{Solver-in-the-Loop: Learning from Differentiable Physics to Interact with Iterative PDE-Solvers}

\newcommand{\myspacepara}{\vspace{-0.5em}}
\newcommand{\myspacesec}{\vspace{-0.8em}}
\newcommand{\myspacessec}{\vspace{-0.5em}}

\newcommand{\pde}{\mathcal{P}}         
\newcommand{\pdec}{\pde_{s}}
\newcommand{\manifsrc}{\mathscr{S}}    
\newcommand{\pder}{\pde_{R}}
\newcommand{\manifref}{\mathscr{R}}

\renewcommand{\vc}[1]{\vs_{#1}}            
\newcommand{\vcN}{\vs}                     
\newcommand{\vct}[1]{\tilde{\vs}_{#1}}     
\newcommand{\vctN}{\tilde{\vs}}            
\renewcommand{\vr}[1]{\mathbf{r}_{#1}}            
\newcommand{\vrN}{\mathbf{r}}                     

\newcommand{\project}{\mathcal{T}}           
\newcommand{\loss}{\mathcal{L}}              

\newcommand{\dt}{\Delta t}                   
\newcommand{\corrPre}{\mathcal{C}_{\text{pre}}}            
\newcommand{\corr}{\mathcal{C}}                         
\newcommand{\nnfunc}{F} 

\newcommand{\sol}[1]{SOL$_{#1}$}
\newcommand{\non}[1]{NON$_{#1}$}
\newcommand{\pre}[1]{PRE$_{\text{#1}}$}

\title{\mytitle}

\author{%
	Kiwon Um$^{1,2}$
	\And
	Robert Brand$^{1}$
	\And
	Yun (Raymond) Fei$^{3}$
	\And
	Philipp Holl$^{1}$
	\And
	Nils Thuerey$^{1}$
	\AND
	\textnormal{$^{1}$Technical University of Munich, $^{2}$LTCI, Telecom Paris, IP Paris, $^{3}$Columbia University}\\\\
	\texttt{kiwon.um@telecom-paris.fr, robert.brand@tum.de}\\\texttt{yf2320@columbia.edu}, \texttt{\{philipp.holl, nils.thuerey\}@tum.de}
}

\hypersetup{pdfinfo={
Title=\mytitle,
Author={Kiwon Um, Robert Brand, Yun (Raymond) Fei, Philipp Holl, Nils Thuerey}
}}

\begin{document}

\maketitle


\vspace{-3pt}
\vbox{%
	\centering {\small
  $^{1}$Technical University of Munich;
  $^{2}$LTCI, Telecom Paris, IP Paris;
  $^{3}$Columbia University } \\
  \vspace{4pt}
	\tt\href{https://github.com/tum-pbs/Solver-in-the-Loop}{github.com/tum-pbs/Solver-in-the-Loop}
}
\vspace{12pt}

\begin{abstract}
  Finding accurate solutions to partial differential equations (PDEs) is a
  crucial task in all scientific and engineering disciplines. It has recently
  been shown that machine learning methods can improve the solution accuracy by
  correcting for effects not captured by the discretized PDE. We target the
  problem of reducing numerical errors of iterative PDE solvers and compare
  different learning approaches for finding complex correction functions. We
  find that previously used learning approaches are significantly outperformed
  by methods that integrate the solver into the training loop and thereby allow
  the model to interact with the PDE during training. This provides the model
  with realistic input distributions that take previous corrections into
  account, yielding improvements in accuracy with stable rollouts of several
  hundred recurrent evaluation steps and surpassing even tailored supervised
  variants. We highlight the performance of the differentiable physics networks
  for a wide variety of PDEs, from non-linear advection-diffusion systems to
  three-dimensional Navier-Stokes flows.
\end{abstract}

\begin{figure}[b!]
  \includegraphics[width=0.95\columnwidth]{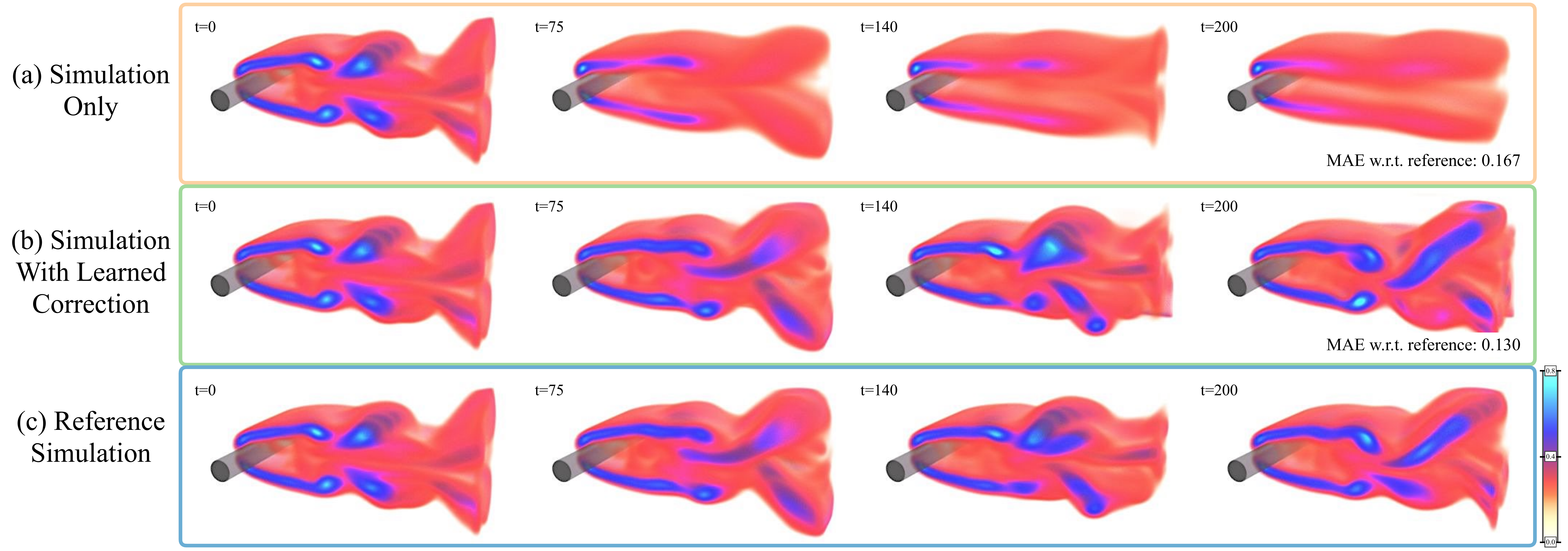}
  \caption{A 3D fluid problem (shown in terms of vorticity) for which the
    regular simulation introduces numerical errors that deteriorate the resolved
    dynamics (a). Combining the same solver with a learned corrector trained via
    differentiable physics (b) significantly reduces errors w.r.t. the reference
    (c).}
  \label{fig:teaser}
\end{figure}

\myspacesec{}
\section{Introduction}
\myspacesec{}

Numerical methods are prevalent in science to improve the understanding of our
world, with applications ranging from climate modeling
\cite{taylor2012overview,stocker2013climate} over simulating the efficiency of
airplane wings \cite{rhie1983numerical} to analyzing blood flow in a human body
\cite{johnston2004non}. These applications are extremely costly to compute due
to the fine spatial and temporal resolutions required in real-world scenarios.
In this context, deep learning methods are receiving strongly growing attention
\cite{morton2018deep,barsinai2019data,greydanus2019hamiltonian} and show promise
to account for those components of the solutions that are difficult to resolve
or are not well captured by our physical models. Physical models typically come
in the form of PDEs and are discretized in order to be processed by computers.
This step inevitably introduces numerical errors. Despite a vast amount of work
\cite{ghosal1996analysis,arnold2012ode} and experimental evaluations
\cite{brachet1983small,pang1997error}, analytic descriptions of these errors
remain elusive for most real-world applications of simulations.


In our work, we specifically target the numerical errors that arise in the
discretization of PDEs. We show that, despite the lack of closed-form
descriptions, discretization errors can be seen as functions with regular and
repeating structures and, thus, can be learned by neural networks. Once trained,
such a network can be evaluated locally to improve the solution of a PDE-solver,
i.e., to reduce its numerical error.

The core of most numerical methods contains some form of iterative process --
either in the form of repeated updates over time for explicit solvers or even
within a single update step for implicit solvers. Hence, we focus on iterative
PDE solving algorithms \cite{golub2012matrix}. \new{We show that neural networks can
achieve excellent performance if they take the reaction of the solver into
account.} This interaction is not possible with supervised learning on
pre-computed data alone. Even small inference errors of a supervised model can
quickly accumulate over time \cite{tompson2017,kim2019}, leading to a data
distribution that differs from the distribution of the pre-computed data. For
supervised learning methods, this causes deteriorated inference at best and
solver explosions at worst.


We demonstrate that neural networks can be successfully trained if they can
\emph{interact} with the respective PDE solver during training. To achieve this,
we leverage differentiable simulations
\cite{amosKolter2017,toussaint2018differentiable}. Differentiable simulations
allow a trained model to autonomously explore and experience the physical
environment and receive directed feedback regarding its interactions throughout
the solver iterations. Hence, our work fits into the broader context of machine
learning as differentiable programming, and we specifically target recurrent
interactions of highly non-linear PDEs with deep neural networks. This
combination bears particular promise: it improves generalizing capabilities of
the trained models by letting the PDE-solver handle large-scale changes to the
data distribution such that the learned model can focus on localized structures
not captured by the discretization. While physical models generalize very well,
learned models often specialize in data distributions seen at training time.
However, we will show that, by combining PDE-based solvers with a learned model,
we can arrive at hybrid methods that yield improved accuracy while handling
solution manifolds with significant amounts of varying physical behavior.

We show the advantages of training via differentiable physics for explicit and
implicit solvers applied to a broad class of canonical PDEs. For explicit and
semi-implicit solvers, we consider advection-diffusion systems as well as
different types of Navier-Stokes variants. We showcase models trained with up to
128 steps of a differentiable simulator and apply our model to complex
three-dimensional (3D) flows, as shown in \myfigref{fig:teaser}. Additionally,
we present a detailed empirical study of different approaches for training
neural networks in conjunction with iterative PDE-solvers for recurrent rollouts
of several hundred time steps. On the side of implicit solvers, we consider the
Poisson problem \cite{mathews1970mathematical}, which is an essential component
of many PDE models. Here, our method outperforms existing techniques on
predicting initial guesses for a conjugate gradient (CG) solver by receiving
feedback from the solver at training time.
\new{The source code for this project is available at \url{https://github.com/tum-pbs/Solver-in-the-Loop}.}


\myspacepara{} \myspacepara{}
\paragraph{Previous Work}
Combining machine learning techniques with PDE models has a long history in
machine learning
\cite{crutchfield1987equations,kevrekidis2003equation,brunton2016discovering}.
%
%
%
More recently, deep-learning-based methods were successfully applied to infer
stencils of advection-diffusion problems \cite{barsinai2019data}, to discover
PDE formulations \cite{long2017pde,raissi2018hiddenphys,sirignano2018dgm}, and
to analyze families of Poisson equations \cite{magill2018}. While identifying
governing equations represents an interesting and challenging task, we instead
focus on a general method to improve the solutions of chosen spaces of
solutions.

Other studies have investigated the similarities of dynamical systems and deep
learning methods \cite{weinan2017proposal} and employed conservation laws to
learn systems described by Hamiltonian mechanics
\cite{greydanus2019hamiltonian,cranmer2020lagrangian}.
Existing studies have also identified discontinuities in finite-difference
solutions with deep learning \cite{ray2018} and focused on improving the
iterative behavior of linear solvers \cite{hsieh2019learning}. So-called Koopman
operators likewise represent an interesting opportunity for deep learning
algorithms \cite{morton2018deep,li2019learning}. While these methods replace the
PDE-based time integration with a learned version, our models rely on and
interact with a PDE-solver that provides a coarse approximation to the problem.
Hence, our models always alternate between inference via an artificial neural
network (ANN) and a solver step. This distinguishes our work from studies of
recurrent ANN architectures
\cite{connor1994recurrent,sutskever2014sequence,vaswani2017attention} as the
PDE-solver can introduce significant non-linearities in-between evaluations of
the ANN.

%

We focus on chaotic systems for which fluid flow represents an exciting
and 
challenging problem domain that is highly relevant for industrial applications.
Deep learning methods have received significant amounts of attention in this
area \cite{kutz2017}.
For example, both steady \cite{guo2016} and unsteady \cite{morton2018deep}, as
well as multi-phase flows \cite{gibou2018} have been investigated with deep
learning based approaches. Turbulence closure modeling has been an area of
particular focus \cite{tracey2015machine,maulik2017,beck2018}. Additionally,
\ac{cnns} were studied for stochastic sub-grid modeling \cite{um2018},
airfoil flow problems \cite{thuerey2020deep,zhang2018},
and as part of generative networks to leverage the fast inference of
pre-trained models \cite{chu2017,xie2018,kim2019}. Other studies have targeted the
unsupervised learning of divergence-free corrections \cite{tompson2017} or
incorporated PDE-based loss functions to represent individual flow solutions via
ANNs \cite{raissi2018hiddenfluid,sirignano2018dgm}. In addition to temporal
predictions of turbulent flows
\cite{mohan2019compressed}, 
similar algorithms were more recently also employed for classification problems
\cite{he2020advectivenet}. 
However, to the best of our knowledge, the existing methods do not let ANNs
interact with solver in a recurrent manner. As we will demonstrate below, this
combination yields significant improvements in terms of inference accuracy.

While we focus on Eulerian, i.e., grid-based discretizations, the Lagrangian
viewpoint is a popular alternative. While a variety of studies has investigated
graph-based simulators, e.g., for rigid-body physics in the context of human
reasoning \cite{battaglia2013simulation,watters2017visual,bapst2019structured}
or weather predictions \cite{seo2020graphphysics}, particles are also a popular
basis for fluid flow problems
\cite{li2018learning,ummenhofer2020lagrangian,sanchez2020gns}. Despite our
Eulerian focus, Lagrangian methods could likewise benefit from incorporating
differentiable solvers into the training process.

Our work shares the motivation of previous work to use differentiable components
at training time
\cite{amosKolter2017,de2018end,toussaint2018differentiable,chen2018neural} and
frameworks for differentiable programming
\cite{schoenholz2019jax,hu2019difftaichi,innes2019differentiable,holl2020}.
Differentiable physics solvers were proposed for inverse problems in the context
of liquids \cite{schenck2018spnets}, cloth \cite{liang2019differentiable}, soft
robots \cite{hu2019difftaichi}, and molecular dynamics
\cite{wang2020differentiable}.
While these studies typically focus on optimization
problems or replace solvers with learned components, we focus on the interaction between
the two. Hence, in contrast to previous work, we always rely on a PDE-solver to
yield a coarse approximate solution and improve its performance via a trained
ANN.

\myspacesec{}
\section{Learning to Reduce Numerical Errors}\label{sec:flow-corr}
\myspacesec{}

Numerical methods yield approximations of a smooth function $\vu$ in a discrete
setting and invariably introduce errors. These errors can be measured in terms
of the deviation from the exact analytical solution. For discrete simulations of
PDEs, they are typically expressed as a function of the truncation, $O(\dt^k)$.
Higher-order methods, with large $k$, 
are preferable but difficult to arrive at in practice. For practical schemes, no
closed-form expression exists for truncation errors, and the errors often grow
exponentially as solutions are integrated over time. We investigate methods that
solve a discretized PDE $\pde$ by performing discrete time steps $\dt$. Each
subsequent step can depend on any number of previous steps,
$\vu(\vx,t+\dt) = \pde(\vu(\vx,t),\vu(\vx,t-\dt),...)$, where
$\vx \in \Omega \subseteq \mathbb{R}^d$ for the domain $\Omega$ in $d$
dimensions, and $t \in \mathbb{R}^{+}$.

\myspacepara{}
\paragraph{Problem Statement:}
We consider two different discrete versions of the same PDE $\pde$, with $\pder$
denoting a more accurate discretization with solutions $\vrN \in \manifref$ from
the \emph{reference manifold}, and an approximate version $\pdec$ with solutions
$\vcN \in \manifsrc$ from the \emph{source manifold}. We consider $\vrN$ and
$\vcN$ to be states at a certain instance in time, i.e., they represent phase
space points, and evolutions over time are given by a trajectory in each
solution manifold. As we focus on the discrete setting, a solution over time
consists of a \emph{reference sequence}
\new{$\{\vr{t}, \vr{t+\dt}, \cdots, \vr{t+k \dt}\}$} in the solution manifold
$\manifref$, and correspondingly, a more coarsely approximated \emph{source
  sequence} \new{$\{\vc{t},\vc{t+\dt},\cdots,\vc{t+k \dt}\}$} exists in the
solution manifold $\manifsrc$. We also employ a mapping operator $\project$ that
transforms a phase space point from one solution manifold to a suitable point in
the other manifold, e.g., for the initial conditions of the sequences above, we
typically choose $\vc{t} = \project \vr{t}$. We discuss the choice of $\project$
in more detail in the appendix, but in the simplest case, it can be obtained via
filtering and re-sampling operations.

By evaluating $\pder$ for $\manifref$, we can compute the points of the phase
space sequences, e.g., $\vr{t+\dt} = \pder(\vr{t})$ for an update scheme that
only depends on time $t$. Without loss of generality, we assume a fixed $\dt$
and denote a state $\vr{t+k \dt}$ after $k$ steps of size $\dt$ with $\vr{t+k}$.
Due to the inherently different numerical approximations,
$\pdec(\project \vr{t}) \ne \project \vr{t+1}$ for the vast majority of states.
In chaotic systems, such differences typically grow exponentially over time
until they saturate at the level of mean difference between solutions in the two
manifolds. We use an \new{$L^2$-norm} in the following to quantify the
deviations, i.e.,
\new{$\loss (\vc{t},\project \vr{t})=\Vert\vc{t}-\project \vr{t}\Vert_2$}. Our
learning goal is to arrive at a correction operator $\corr ( \vc{} )$ such that
a solution to which the correction is applied has a lower error than an
unmodified solution:
$\loss ( \pdec( \corr (\project \vr{t_0}) ) , \project \vr{t_1}) < \loss (
\pdec( \project \vr{t_0} ), \project \vr{t_1})$. The correction function
$\corr (\vcN | \theta)$ is represented as a deep neural network with weights
$\theta$ and receives the state $\vcN$ to infer an additive correction field
with the same dimension. To distinguish the original phase states $\vcN$ from
corrected ones, we denote the latter with $\vctN$, and we use an exponential
notation to indicate a recursive application of a function, i.e.,
\begin{equation}
  \vc{t+n} = \pdec(\pdec(\cdots \pdec( \project \vr{t}  )\cdots)) = \pdec^n ( \project \vr{t} ) \ .
\end{equation}



\begin{wrapfigure}[14]{r}{0.6\linewidth}
  \centering
  \begin{overpic}[width=\linewidth]{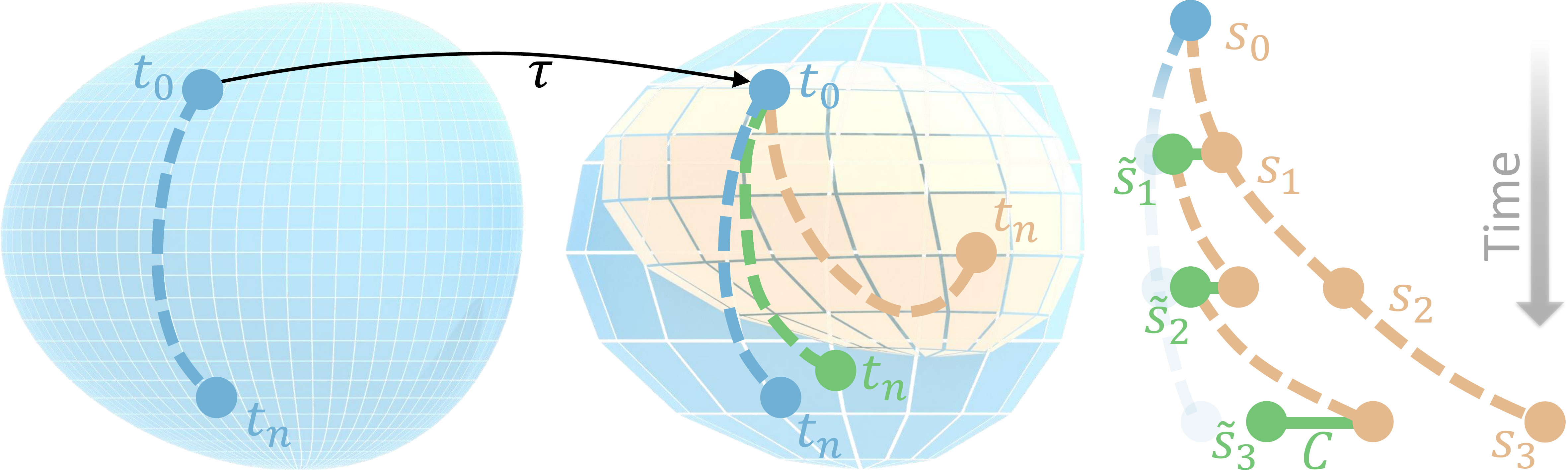}
    \put(0, 2){\footnotesize $\manifref$}  
    \put(36, 2){\footnotesize $\manifsrc$} 
  \end{overpic}
  \caption{Transformed solutions of the reference sequence computed on
    $\manifref$ (blue) differ from solutions computed on the source manifold
    $\manifsrc$ (orange). A correction function $\corr$ (green) updates the
    state after each iteration to more closely match the projected reference
    trajectory on $\manifsrc$.}
  \label{fig:methods}
\end{wrapfigure}

Within this setting, any type of learning method naturally needs to compare
states from the source domain with the reference domain in order to bridge the
gap between the two solution manifolds. How the evolution in the source manifold
at training time is computed, i.e., if and how the corrector interacts with the
PDE, has a profound impact on the learning process and the achievable final
accuracy. We distinguish three cases: no interaction, a pre-computed form of
interaction, and a tight coupling via a differentiable solver in the training
loop.


\vspace{-0.5em}
\begin{itemize}[leftmargin=1.5em]
  \itemsep0em

\item \textbf{Non-interacting (NON)}: The learning task purely uses the
  unaltered PDE trajectories, i.e., $\vc{t+n} = \pdec^n ( \project \vr{t} )$
  with $n$ evaluations of $\pdec$. These trajectories are fully contained in the
  source manifold $\manifsrc$. Learning from these states means that a model
  will not see any states that deviate from the original solutions. As a
  consequence, models trained in this way can exhibit undesirably strong error
  accumulations over time. This corresponds to learning from the difference
  between the orange and blue trajectories in \myfigref{fig:methods}, and most
  commonly applied supervised approaches use this variant.

\item \textbf{Pre-computed interaction (PRE)}: To let an algorithm learn from
  states that are closer to those targeted by the correction, i.e., the
  reference states, a pre-computed or analytic correction is applied. Hence, the
  training process can make use of phase space states that deviate from those in
  $\manifsrc$, as shown in green in \myfigref{fig:methods}, to improve inference
  accuracy and stability. This approach can be formulated as
  $\vc{t+n} = (\pdec \corrPre)^n ( \project \vr{t} )$ with a pre-computed
  correction function $\corrPre$. In this setting, the states $\vcN$ are
  corrected without employing a neural network, but they should ideally resemble
  the states achievable via the learned correction later on. As the modified
  states $\vcN$ are not influenced by the learning process, the training data
  can be pre-computed. A correction model $\corr (\vcN | \theta)$ is trained via
  $\vctN$ that replaces $\corrPre$ at inference time.

\item \textbf{Solver-in-the-loop (SOL)}: By integrating the learned function
  into a differentiable physics pipeline, the corrections can interact with the
  physical system, alter the states, and receive gradients about the future
  performance of these modifications. The learned function $\corr$ now depends
  on states that are modified and evolved through $\pde$ for one or more
  iterations. A trajectory for $n$ evaluations of $\pdec$ is given by
  $\vct{t+n} = ( \pdec \corr )^n ( \project \vr{t} )$, with
  $\corr (\vctN | \theta)$. The key difference with this approach is that
  $\corr$ is trained via $\vctN$, i.e., states that were affected by previous
  evaluations of $\corr$, and it affects $\vctN$ in the following iterations. As
  for (PRE), this learning setup results in a trajectory like the green one
  shown in \myfigref{fig:methods}, however, in contrast to before, the learned
  correction itself influences the evolution of the trajectory, preventing a gap
  for the data distribution of the inputs.

\end{itemize}
\vspace{-0.5em}

In addition to these three types of interaction, a second central parameter is
the look-ahead trajectory per iteration and mini-batch of the optimizer during
learning. A subscript $n$ denotes the number of steps over which the future
evolution is recursively evaluated, e.g., \sol{n}. The objective function, and
hence the quality of the correction, is evaluated with the training goal to
minimize $\sum_{i=t}^{t+n} \loss (\vc{i},\vr{i})$. Below, we will analyze a
variety of learning methodologies that are categorized via learning methodology
(NON, PRE or SOL) and look-ahead horizon $n$.



\myspacesec{}
\section{Experiments}
\myspacesec{}

We now provide a summary and discussion of our experiments with the different
types of PDE interactions for a selection of physical models. Full details of
boundary conditions, parameters, and discretizations of all five PDE scenarios
are given in App. B. 

\myspacessec{}
\subsection{Model Equations and Data Generation}
\myspacessec{}

We investigate a diverse set of constrained advection-diffusion models of which
the general form is
\begin{equation}
  \label{eq:model-adv-diff}
  \partial{\vu} / \partial{t}
  = - \vu \cdot\nabla\vu +
  \nu \nabla\cdot \nabla \vu + \mathbf{g}
  \quad \text{subject to} \quad \mM \vu = 0,
\end{equation}
where $\vu$ is the velocity, $\nu$ denotes the diffusion coefficient (i.e.,
viscosity), and $\mathbf{g}$ denotes external forces. The constraint matrix
$\mM$ contains an additional set of equality constraints imposed on $\vu$. In
total, we target four scenarios: pure non-linear advection-diffusion (Burger's
equation), two-dimensional Navier-Stokes flow, Navier-Stokes coupled with a
second advection-diffusion equation for a buoyancy-driven flow, and a 3D
Navier-Stokes case. Also, we discuss CG solvers in the context of differentiable
operators below.

For each of the five scenarios, we implement the non-interacting evaluation
(NON) by pre-computing a large-scale data set that captures a representative and
non-trivial space of solutions in $\manifsrc$. The reference solutions from
$\manifref$ are typically computed with the same numerical method using a finer
discretization (4x in our setting, with effective resolutions of $128^2$ and
higher). The PDEs are parametrized such that the change of discretization leads
to substantial differences when integrated over time. For several of the 2D
scenarios, we additionally train models with data sets of trajectories that have
been corrected with other pre-computated correction functions. For these PRE
variants, we use a time-regularized, constrained least-squares corrector
\cite{henderson1975best} to obtain corrected phase state points. For the SOL
variants, we employ a differentiable PDE-solver that runs mini-batches of
simulations and provides gradients for all operations of the solving process
within the deep learning framework. This allows gradients to freely propagate
through the PDE-solver and coupled neural networks via automatic
differentiation. For $n>1$, i.e., PDE-based look-ahead at training time, the
gradients are back-propagated through the solver $n-1$ times, and the difference
w.r.t. a pre-computed reference solution is evaluated for all intermediate
results.


\myspacessec{}
\subsection{Training Procedure}
\myspacessec{}

The neural network component $\nnfunc (\vc{} \, | \, \theta)$ of the correction
function 
is realized with a fully convolutional architecture. As our focus lies on the
methodology for incorporating PDE models into the training, the architectures
are intentionally kept simple. However, they were chosen to yield high accuracy
across all variants.
Our networks typically consist of 10 convolutional layers with 16 features each,
interspresed with ReLU activation functions using kernel sizes of $3^d$ and
$5^d$. The networks parameters $\theta$ are optimized with a fixed number of
steps with an ADAM optimizer \cite{kingma2014} and a learning rate of $10^{-4}$.
For validation, we use data sets generated from the same parameter distribution
as the training sets. All results presented in the following use test data sets
whose parameter distributions differ from the ones of the training data set.

We quantify the performance of the trained models by computing the mean absolute
error between a computed solution and the corresponding projected reference
for $n$ consecutive steps of a simulation. We report absolute error values for
different models in comparison to an unmodified source trajectory from
$\manifsrc$. Additionally, relative improvements are given w.r.t. the difference
between unmodified source and reference solutions.
An improvement by 100\% would mean that the projected reference is reproduced
perfectly, while negative values indicate that the modified solution deviates
more from the reference than the original source trajectory.


\newcommand{\heightSumFig}{0.29\linewidth}
\newcommand{\heightSumLeg}{0.096\linewidth}
\begin{figure}[tb]
  \centering \hfill
  \begin{overpic}[height=\heightSumFig]{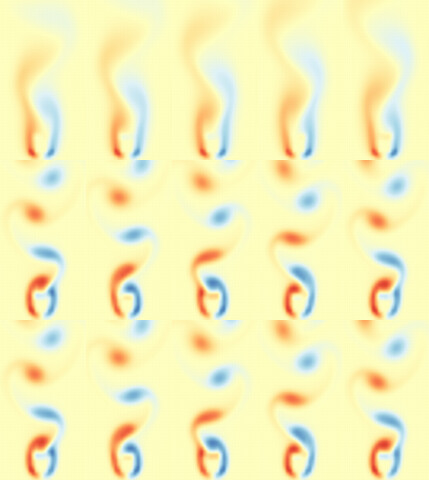}
    \put(80,92){\footnotesize (a)}
    \put(-5,66){\scriptsize\rotatebox{90}{\makebox[0.097\linewidth]{Source}}}
    \put(-5,33){\scriptsize\rotatebox{90}{\makebox[0.097\linewidth]{Corrected}}}
    \put(-5,0) {\scriptsize\rotatebox{90}{\makebox[0.097\linewidth]{Reference}}}
  \end{overpic}
  \begin{overpic}[height=\heightSumLeg]{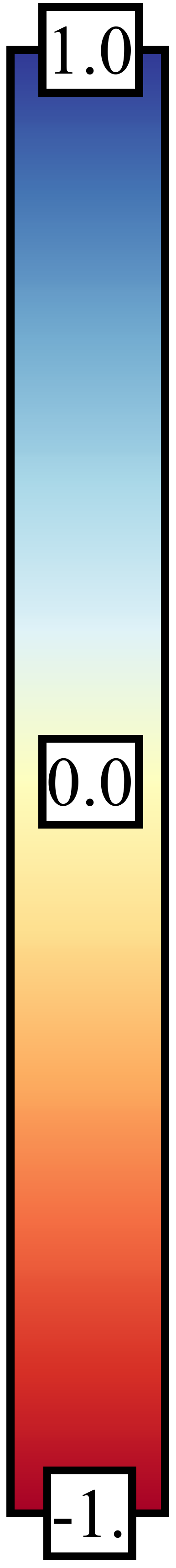}\end{overpic} 
  \hspace{1pt} 
  \begin{overpic}[height=\heightSumFig]{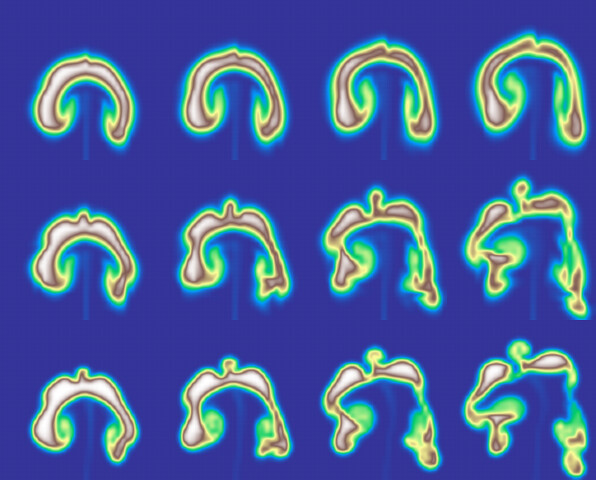}
    \put(92,75){\footnotesize {\color{white}(b)}}
  \end{overpic}
  \begin{overpic}[height=\heightSumLeg]{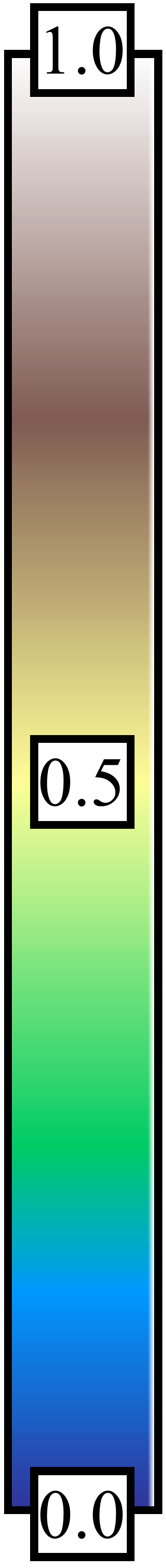}\end{overpic} 
  \hspace{1pt} 
  \begin{overpic}[height=\heightSumFig]{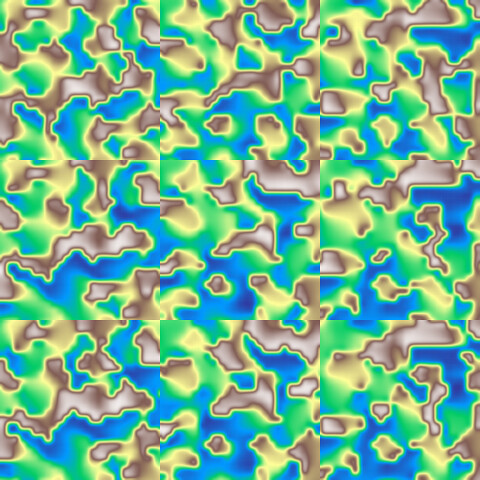}
    \put(90,93){\footnotesize {\color{black}(c)}} 
  \end{overpic}
  \begin{overpic}[height=\heightSumLeg]{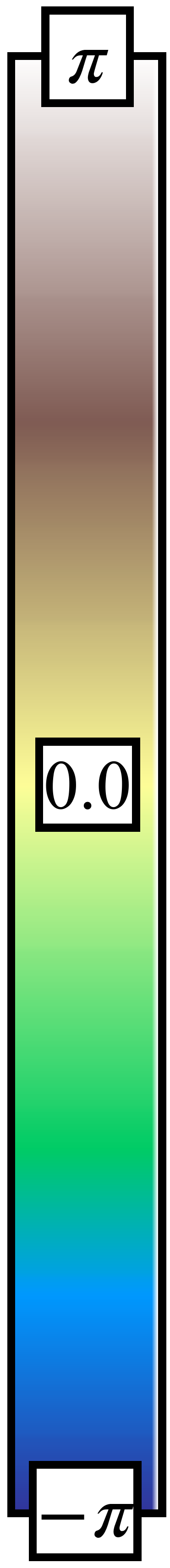}\end{overpic} 
  \caption{Our PDE scenarios cover a wide range of behavior including (a) vortex
    shedding, (b) complex buoyancy effects, and (c) advection-diffusion systems.
    Shown are different time steps (l.t.r.) in terms of vorticity for (a),
    transported density for (b), and angle of velocity direction for (c).}
  \label{fig:visualResultSummary}
\end{figure}

\myspacesec{}
\section{Results}
\myspacesec{}


%
Our experiments show that learned correction functions can achieve substantial
gains in accuracy over a regular simulation. When training the correction
functions with differentiable physics, this additionally yields further
improvements of more than 70\% over supervised and pre-computed approaches from
previous work. A visual overview of the different tests is given in
\myfigref{fig:visualResultSummary}, and a summary of the full evaluation from
the appendix is provided in \myfigref{fig:mainResultSummary} and
\mytabref{tab:main}. In the appendix, we also provide error measurements w.r.t.
physical quantities such as kinetic energy and frequency content.
The source code of our experiments and analysis will be published upon
acceptance. 

\myspacepara{}
\paragraph{Unsteady Wake Flow}
%
The PDE scenario for unsteady wake flows represents a standard benchmark case
for fluids \cite{rajani2009numerical,morton2018deep} and involves a continuous
inflow with a fixed, circular obstacle, which induces downstream vortex shedding
with distinct frequencies depending on the Reynolds number. For coarse
discretizations, the approximation errors distort the flow leading to
deteriorated motions or suppressed vortex shedding altogether. An example flow
configuration is shown in \myfigref{fig:visualResultSummary}a. In this scenario,
the simplest method (NON) yields stable training and a model that already
reduces the mean absolute error (MAE) from 0.146 for a regular simulation
without correction (SRC) to an MAE of 0.049 when applying the learned
correction. The pre-computed correction (PRE) improves on this behavior via its
time regularization with an error of 0.031. A \sol{32} model trained with a
differentiable physics solver for 32 time steps in each iteration of ADAM yields
a significantly lower error of 0.013. This means, the numerical errors of the
source simulation w.r.t. the reference were reduced by more than a factor of 10.
%
Despite the same architecture and weight count for all three models, the overall
performance varies strongly, with the \sol{32} version outperforming the simpler
variants by 73\% and more.
An example of the further evaluations provided in the appendix is given in
\myfigref{fig:sum-karman-freq}.


\myspacepara{}
\paragraph{Buoyancy-driven Flow}
%
We evaluate buoyancy-driven flows as a scenario with increased complexity. In
addition to an incompressible fluid, a second, non-uniform marker quantity is
advected with the flow that exerts a buoyancy force. This coupled system of
equations leads to interesting and complex swirling behavior over time. We
additionally use this setup to highlight that the reference solutions can be
obtained with different discretization schemes. We use a higher-order advection
scheme in addition to a 4$\times$ finer spatial discretization to compute the
reference data.

Interestingly, the correction functions benefit from particularly long rollouts
at training time in this scenario. Models with simple pre-computed or unaltered
trajectories yield mean errors of 1.37 and 1.07 compared to an error of 1.59 for
the source simulation, respectively. Instead, a model trained with
differentiable physics with 128 steps (\sol{128}) successfully reduces the error
to 0.62, an improvement of more than 59\% compared to the unmodified simulation.

\myspacepara{}
\paragraph{Forced Advection-Diffusion}
%
A third scenario employs Burger's equation as a physical model. We mimic the
setup from previous work \cite{barsinai2019data} to inject energy into the
system via a forcing term with a spectrum of 
sine waves. This forcing prevents the system from dissipating to relatively
static and slowly moving configurations. While the PRE and NON versions yield
clear improvements, the SOL versions do not significantly outperform the simpler
baselines. This illustrates a limitation of long rollouts via differentiable
physics: Learned correction functions need to be able to anticipate future
behavior to make high-quality corrections. The randomized forcing in this
example severely limits the number of future steps that can accurately be
predicted given one state. This behavior contrasts with other physical systems
without external disturbances, where a single state uniquely determines its
evolution.
\new{We show in the appendix that the SOL models with an increased number of
  interaction steps pay off when the external disturbances are absent.}

\myspacepara{}
\paragraph{Conjugate Gradient Solver}
%
We turn to iterative solvers for linear systems of equations to illustrate
another aspect of learning from differentiable physics: its importance for the
propagation of boundary condition effects. As our learning objective, we target
the inference of initial guesses for CG solvers \cite{hestenes1952cg}. Following
previous work \cite{tompson2017}, we target Poisson problems of the form
$\nabla \cdot \nabla p = \nabla \cdot \vu$, which arise for projections of a
velocity $\vu$ to a divergence-free state. Instead of fully relying on an ANN to
produce the pressure field $p$, we instead target the learning objective to
produce an initial guess, which is improved by a regular CG solver until a given
accuracy threshold is reached.

This goal can be reached by directly minimizing the right-hand side term
$\nabla \cdot \vu$, similar to physics-based loss terms proposed in a variety of
studies \cite{raissi2018hiddenfluid,sirignano2018dgm}. Alternatively, we can
employ a differentiable CG solver and formulate the learning goal as minimizing
the same residual after $n$ steps of the CG solver (similar to the \sol{n}
models above). While the physics-based loss version reduces the initial
divergence more successfully, it fares badly when interacting with the CG
solver: compared to the SOL version, it requires 63\%
more steps to reach a desired accuracy. Inspecting the inferred solutions
reveals that the former model leads to comparatively large errors near
boundaries, which are small for each grid cell but significantly influence the
solution on a large scale. The SOL version immediately receives feedback about
this behavior via the differentiable solver iterations. I.e., the differentiable
solver provides a look-ahead of how different parts of the solution affect
future states. In this way, it can anticipate problems such as those in the
vicinity of boundary conditions.


\myspacepara{}
\paragraph{Three-dimensional Fluid Flow}
%
Lastly, we investigate a 3D case of incompressible flow. The overall setup is
similar to the unsteady wake flow in two dimensions outlined above, but the
third dimension extends the axes of rotation in the fluid from one to three,
yielding a very significant increase in complexity. As a result, the flow behind
the cylindrical obstacle quickly becomes chaotic and forms partially turbulent
eddies, as shown in \myfigref{fig:teaser}. This scenario requires significantly
larger models to learn a correction function, and the NON version does not
manage to stabilize the flow consistently. Instead, the \sol{16} version
achieves stable rollouts for several hundred time steps and successfully
corrects the numerical inaccuracies of the coarse discretization, improving the
numerical accuracy of the source (SRC) simulation by more than 22\% across a
wide range of configurations.




\begin{figure}[t!]
  \centering \subcaptionbox{\scriptsize Unsteady wake\label{fig:sum-karman}}
  {\includegraphics[height=0.23\columnwidth,page=4]{figs/summary-karman-plot}}
  \hfill \subcaptionbox{\scriptsize Buoyancy-driven\label{fig:sum-buoy}}
  {\includegraphics[height=0.23\columnwidth,page=3]{figs/out-vgf-jmsmaller-steps-smoke-uns-jmsmaller-msteps}}
  \hfill \subcaptionbox{\scriptsize Advection-diffusion\label{fig:sum-burgers}}
  {\includegraphics[height=0.23\columnwidth,page=3]{figs/summary-burgers-plot}}
  \hfill \subcaptionbox{\scriptsize CG solver\label{fig:sum-cg}}
  {\includegraphics[height=0.23\columnwidth,page=1]{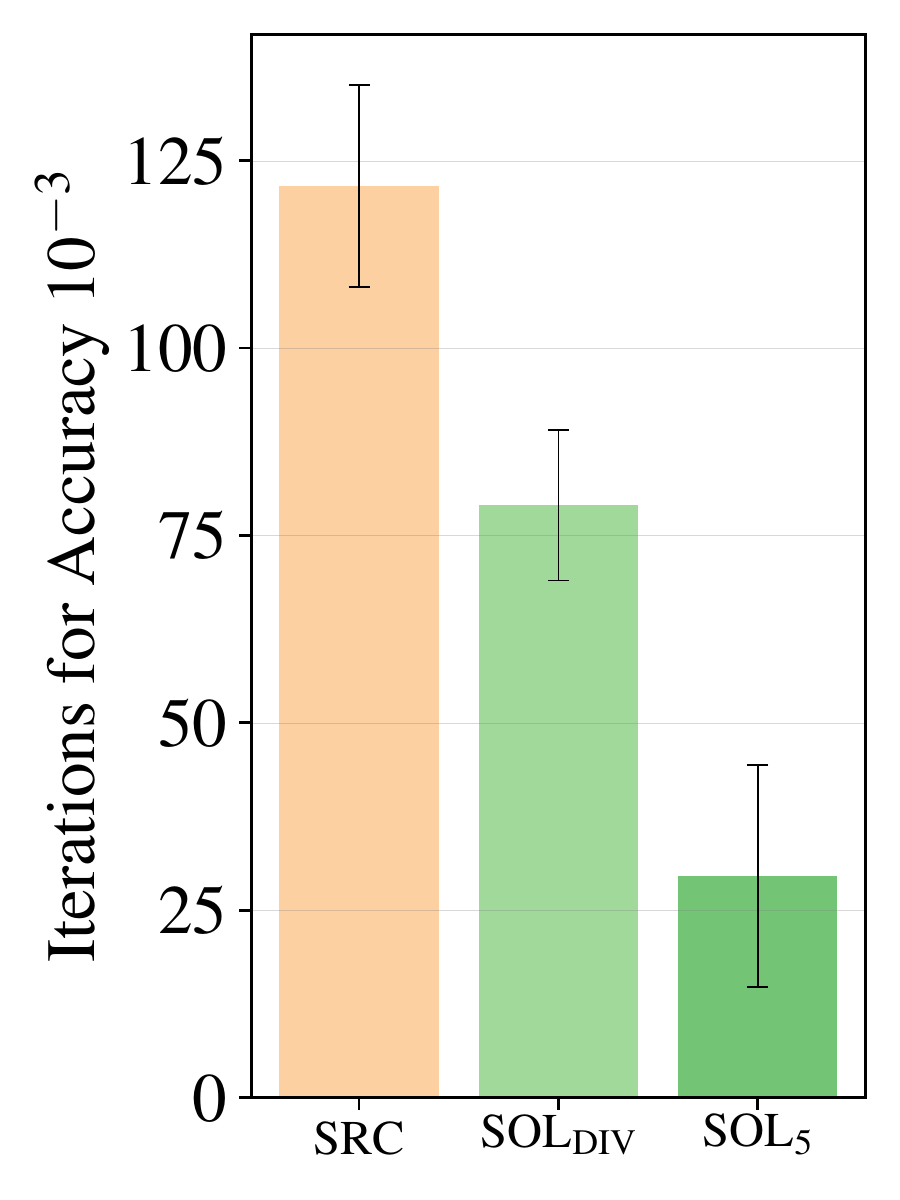}} \hfill
  \subcaptionbox{\scriptsize 3D wake\label{fig:sum-karman3d}}
  {\includegraphics[height=0.23\columnwidth,page=1]{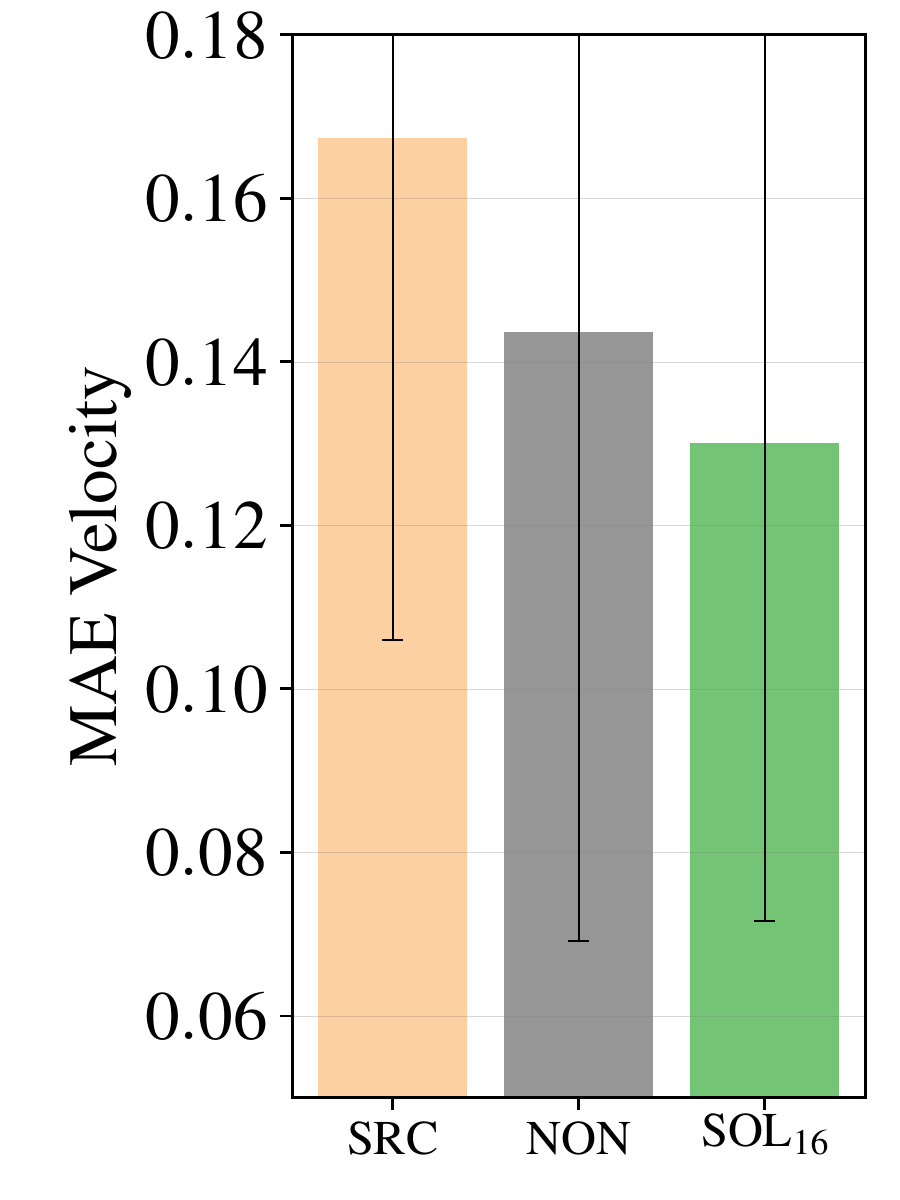}}
  \\
  \subcaptionbox{\scriptsize Unsteady wake,
    look-ahead\label{fig:sum-karman-rel}}
  {\includegraphics[height=0.20\columnwidth,page=6]{figs/summary-karman-plot}}
  \hfill \subcaptionbox{\scriptsize Buoyancy-driven,
    look-ahead\label{fig:sum-buoy-rel}}
  {\includegraphics[height=0.20\columnwidth,page=6]{figs/out-vgf-jmsmaller-steps-smoke-uns-jmsmaller-msteps}}
  \hfill \subcaptionbox{\scriptsize Unsteady wake, frequency
    error\label{fig:sum-karman-freq}}{\includegraphics[height=0.20\columnwidth,page=1]{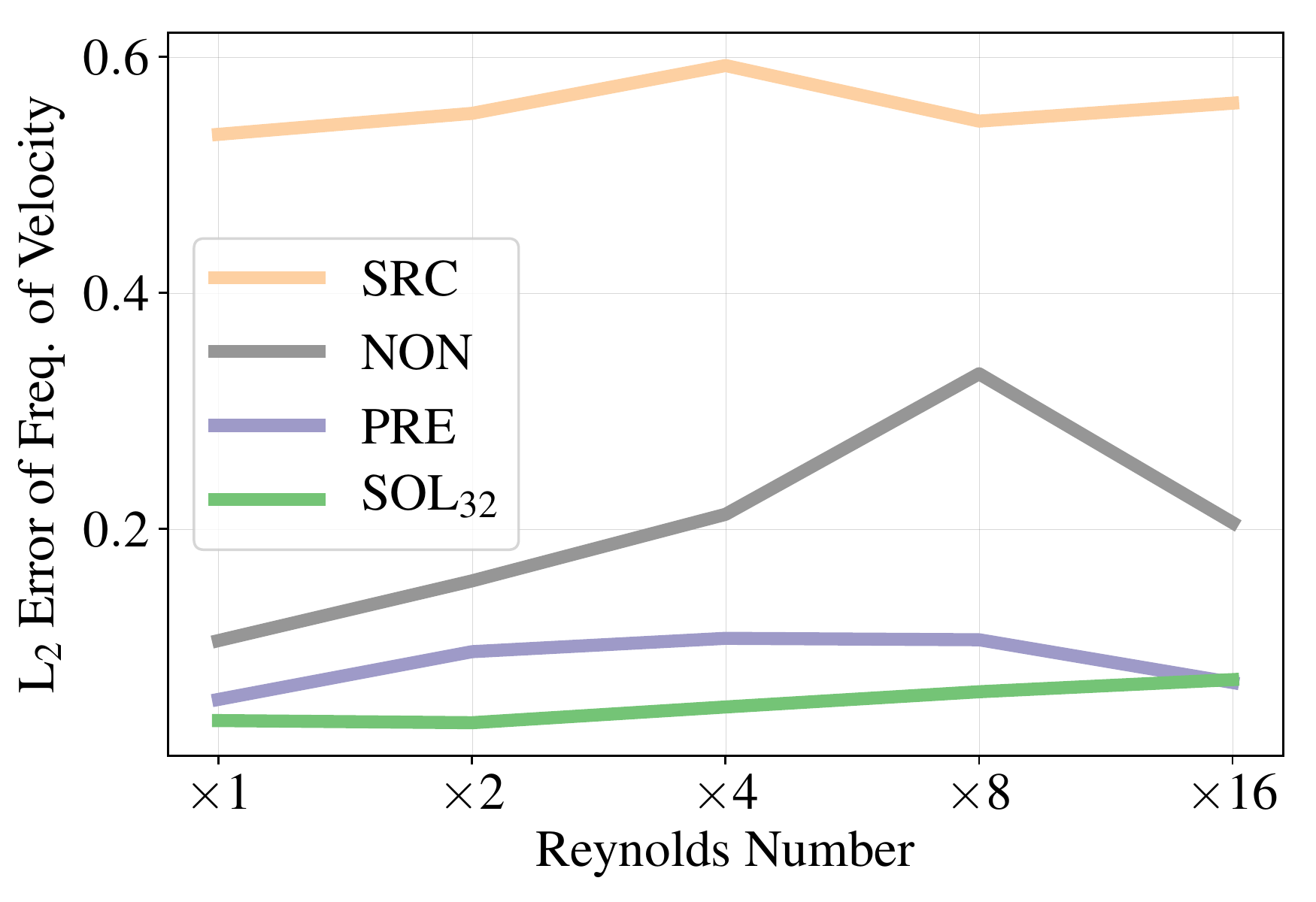}}
  \caption{(a)-(e) Numerical approximation error w.r.t. reference solution for
    unaltered simulations (SRC) and with learned corrections. The models trained
    with differentiable physics and look-ahead achieve significant gains over
    the other models. (f,g) Relative improvement over varying look-ahead
    horizons. (h) A frequency-based evaluation for the unsteady wake flow
    scenario.}
  \label{fig:mainResultSummary}
\end{figure}

\myspacesec{}
\section{Ablations and Discussion}
\myspacesec{}

We performed an analysis of the proposed training via differentiable physics to
highlight which hyperparameters most strongly influence results. Specifically,
we evaluate varying look-ahead horizons, different model architectures, training
via perturbations, and pre-computed variants.

\myspacepara{}
\paragraph{Future Look-Ahead}
For systems with deterministic behavior,
long rollouts via differentiable physics at training time yield significant
improvements, as shown in \myfigref{fig:sum-karman-rel} and
\ref{fig:sum-buoy-rel}. While training with a few (1 to 4) steps yields
improvements of up to 40\% for the buoyancy-driven flow scenario, this number
can be raised significantly by increasing the look-ahead at training time. A
performance of more than 54\% can be achieved by 64 recurrent solver iterations,
while raising the look-ahead to 128 yields average improvements of 60\%.
%
Our tests consistently show that, without changing the number of weights or the
architecture of a network, the gradients provided by the longer rollout times
allow the network to anticipate the behavior of the physical system better and
react to it. Throughout our tests, similar performances could not be obtained by
other means.

\myspacepara{}
\paragraph{Generalization}
The buoyancy scenario also highlights the very good generalizing capabilities of
the resulting models. All test simulations were generated with an
out-of-distribution parametrization of the initial conditions, leading to
substantially different structures, and velocity ranges over time.

\myspacepara{}
\paragraph{Training with Noise}
An interesting variant to stabilize physical predictions in the context of Graph
Network-based Simulators 
was proposed by Sanchez et al. \cite{sanchez2020gns}. They report that
perturbations of input features with 
noise lead to more stable long-term rollouts. We mimic this setup in our
Eulerian setting by perturbing the inputs to the neural networks with
$\mathcal{N}(0,\sigma)$ for varying strengths $\sigma$. While a sweet spot with
improvements of 34.5\% seems to exist around $\sigma=10^{-4}$, the increase in
performance is small compared to a model with less perturbations (30.6\%), as
training with an increased look-ahead for the SOL models gives improvements up
to 60.0\%.


\myspacepara{}
\paragraph{Training Stability}
The physical models we employ introduce a large amount of complexity into the
training loop.
Especially during the early stages of training, an inferred correction can
overly distort the physical state. Performing time integration via the
PDE 
then typically leads to exponential increases of existing oscillations and a
diverging calculation. Hence, we found it important to pre-train networks with
small look-aheads (we usually use \sol{2} models),
and then continue training with longer recurrent iterations for the look-ahead.
While this scheme can be applied hierarchically, we saw no specific gains from,
e.g., starting a \sol{32} training with a \sol{2} model versus a \sol{16} model.

\myspacepara{}
\paragraph{Runtime Performance}
The training via differentiable physics incurs an increased computational cost
at training time, as the PDE model has to be evaluated for $n$ steps for each
learning iteration, and the calculation of the gradients is typically of similar
complexity as the evaluation of the PDE itself.
However, this incurs only moderate costs in our tests. For example, for the
buoyancy-driven flow, the training time increases from 0.21 seconds per
iteration on average for \sol{2} to 0.42s for \sol{4}, and 1.25s for \sol{16}.
The look-ahead additionally provides $n$ times more gradients at training time,
and the inference time of the resulting models is not affected. Hence, the
training cost can quickly pay off in practical scenarios by yielding more
accurate results without any increase in cost at inference time.


Computing solutions with the resulting hybrid method which alternates PDE
evaluations and ANN inference also provides benefits in terms of evaluation
performance:
%
A pre-trained, fully convolutional CNN has an $\mathcal O(n)$ cost for $n$
degrees of freedom, in contrast to many PDE-solvers with a super-linear
complexity.
\new{For example, a simulation as shown in \myfigref{fig:teaser} involving the
  trained model took 13.3s on average for 100 time steps, whereas a CPU-based
  reference simulation required 913.2s. A speed-up of more than 68$\times$.}

\begin{table}[t!]
  \caption{A summary of the quantitative evaluation for the five PDE scenarios.
    \sol{s} denotes a variant with shorter look-ahead compared to SOL. ($^*$ For
    the CG solver scenario, iterations to reach an accuracy of 0.001 are given.
    Here, SOL$_s$ denotes the physics-based loss version.)}
  \scriptsize
  \label{tab:main}
  \begin{center}
    \addtolength{\tabcolsep}{-1pt}
    \begin{tabular}{cccccccccc}
      \toprule
      \textbf{Exp.} & \multicolumn{5}{c}{\textbf{Mean absolute error of velocity}} & \multicolumn{4}{c}{\textbf{Rel. improvement}} \\

      \cmidrule(lr){2-6}\cmidrule(l){7-10}
                    & SRC             & PRE             & NON             & \sol{s}         & SOL             & PRE  & NON  & \sol{s} & SOL     \\
      \midrule
      Wake Flow    & 0.146$\pm$0.004 & 0.031$\pm$0.010 & 0.049$\pm$0.012 & 0.041$\pm$0.009 & 0.013$\pm$0.003 & 79\% & 67\% & 72\%    & 91\%    \\
      Buoyancy     & 1.590$\pm$1.033 & 1.373$\pm$0.985 & 1.080$\pm$0.658 & 0.944$\pm$0.614 & 0.620$\pm$0.390 & 19\% & 29\% & 41\%    & 60\%    \\
      Adv.-diff.   & 0.248$\pm$0.019 & 0.218$\pm$0.017 & 0.159$\pm$0.015 & 0.152$\pm$0.015 & 0.158$\pm$0.017 & 12\% & 36\% & 39\%    & 36\%    \\
      $^*$CG Solver& 121.6$\pm$13.44 & -               & -               & 79.03$\pm$10.02 & 29.59$\pm$14.83 & -    & -    & 35\%    & 76\%    \\
      3D Wake      & 0.167$\pm$0.061 & -               & 0.144$\pm$0.074 & -               & 0.130$\pm$0.058 & -    & 14\% & -       & 22\%    \\
      \bottomrule
    \end{tabular}
    \addtolength{\tabcolsep}{1pt}
  \end{center}
\end{table}


\myspacesec{}
\section{Conclusions}
\myspacesec{}


We have demonstrated how to achieve significant reductions of numerical errors
in PDE-solvers by training ANNs with long look-ahead rollouts and differentiable
physics solvers. The resulting models yield substantially lower errors than
models trained with pre-computed data. We have additionally provided a first
thorough evaluation of different methodologies for letting PDE-solvers interact
with recurrent ANN evaluations.

Identical networks yield significantly better results purely by having the
solver in the learning loop. This indicates that the numerical errors have
regular structures that can be learned and corrected via learned
representations. The resulting networks likewise improve generalization for
out-of-distribution samples and provide stable, long-term recurrent predictions.
Our results have the potential to enhance learning physical priors for a variety of deep learning tasks. 
%
\new{
Beyond engineering applications and medical simulations, a particularly interesting application of our approach is weather prediction \cite{rasp2020weatherbench}, where a simple differentiable solver could be augmented with a learned correction function to recover the costly predictions of operational forecasting systems.} 

\new{
Overall, we hope that the demonstrated gains in accuracy will help to establish trained neural networks as components in the
numerical toolbox of computational science.
}



\section*{Broader Impact}

PDE-based models are very commonly used and can be applied to a wide range of
applications, including weather and climate, epidemics, civil engineering,
manufacturing processes, and medical applications. Our work has the potential to
improve how these PDEs are solved. As PDE-solvers have a long history, there is
a wide range of established tools, some of which still use COBOL and FORTRAN.
Hence, it will not be easy to integrate deep learning methods into the existing
solving pipelines, but in the long run, our method could yield solvers that
compute more accurate solutions with a given amount of computational resources.

Due to the wide range of applications of PDEs, our methods could also be used in
the development of military equipment (machines and weapons) or other harmful
systems.
However, our method shares this danger with all numerical methods. For the
discipline of computational science as a whole, we see more positive aspects
when computer simulations become more powerful. Nonetheless, we will encourage
users of our method likewise to consider ethical implications when employing
PDE-solvers with learning via differentiable physics.

\begin{ack}
This work is supported by the ERC Starting Grant {\em realFlow} (StG-2015-637014).
\end{ack}




\bibliography{references}
\bibliographystyle{abbrv}

\clearpage
\appendix

\newcommand{\mytoptitlebar}{
  \hrule height 2pt
  \vskip 0.25in
  \vskip -\parskip%
}
\newcommand{\mybottomtitlebar}{
  \vskip 0.29in
  \vskip -\parskip
  \hrule height 1pt
  \vskip 0.09in%
}
\mytoptitlebar
{\centering
{\Large Appendix for \emph{\mytitle} \par}}
\mybottomtitlebar

Below, we give additional details regarding the steps and numerical methods employed in 
each of the interaction variants discussed in the main text. 
We present details of the simulation setups for the five scenarios and give more detailed 
results for each case. Lastly, we discuss performance and list details of our neural network architectures.


As our experiments in the main text already demonstrate, 
deep learning algorithms that can closely interact with a differentiable 
PDE solver can yield substantially improved performance.
This illustrates how crucial it is for deep learning algorithms that co-exist 
or interact with numerical solvers in a recurrent manner
to anticipate shifts in the distributions of input 
features. We present additional results and show how 
interactions between PDE solvers and deep neural networks can be formulated. 
These interactions help to bridge the gap between distribution shifts that exist between 
different discretizations of a PDE. 
We will demonstrate that avoiding distribution shifts is essential for a model 
to infer a correction successfully. In our iterative setting, this, in turn, helps 
to keep the distributions aligned over the course of many iterations.


\section{Correction Functions for PDEs}

For completeness, we provide a brief summary of our notation.
We consider reference solutions $\vrN$ of
the PDE $\pde$ that are contained in the phase space manifold
$\manifref$ with reference trajectories over time denoted by
$\{\vr{t},\vr{t+1},\cdots,\vr{t+k}\}$ 
for $k$ steps of size $\dt$.
A more coarsely approximated solution of the same problem is denoted by $\vcN$
in the manifold $\manifsrc$ with trajectories
$\{\vc{t},\vc{t+1},\cdots,\vc{t+k}\}$. 
We typically initialize the source state from the reference version 
via a transfer operator $\project$ with $\vc{t} = \project \vr{t}$ as initial condition.
A transfer from source to reference states is denoted by $\project^T$.

The learning objective is to find the best possible correction function 
$\corr (\vcN \, | \, \theta)$ given the weights $\theta$ and a network architecture.
Without loss of generality, we assume that the correction 
function is applied additively, i.e., $\vctN = \vcN + \corr (\vcN \, | \, \theta)$, 
where the tilde in $\vctN$ indicates the corrected state. 
A new state is computed in combination with the PDE 
via $\vct{t+1} = \pdec(\vc{t}) + \corr ( \pdec(\vc{t}) \, | \, \theta)$ for which we
use the short form $(\pdec  \corr ) (\vc{t})$ below. Multiple recurrent 
evaluations of a function are denoted by $\vct{t+k} = ( \pdec \corr )^k ( \vc{t} )$ for $k$ steps
starting from an unaltered source state $\vc{t}$.

For training neural networks, we use an \new{$L^2$-based}
loss, i.e., \new{$\loss (\vct{t},\project \vr{t})=\Vert\vct{t}-\project \vr{t}\Vert^2$},
which is typically evaluated 
for $n$ steps via  $\sum_{i=t}^{t+n} \loss (\vct{i},\vr{i})$ in order to find a 
solution to the minimization problem: $\argmin_{\theta} \sum_{i=t}^{t+n} \loss (\vct{i},\vr{i})$.

We consider constrained advection-diffusion PDEs:
$\partial{\vu} / \partial{t} = - \vu \cdot\nabla\vu + \nu \nabla\cdot \nabla \vu + \mathbf{g}$ subject to $\mM \vu = 0$.
%
Here, $\vu$, $\nu$, and $\mathbf{g}$ denote velocity, diffusivity, and external forces, respectively.
The constraint matrix $\mM$ contains an additional set of equality constraints imposed on $\vu$.


\subsection{Learning Without Interaction}

In the main text, we use learning via non-interacting trajectories 
as a baseline learning setup. In this case, a model is 
trained to minimize differences between states $\vcN$ and $\vrN$ in a fully supervised 
manner. These versions are denoted by NON.

Despite its simplicity, different variants of this learning setup can be considered.
In the simplest case, we initialize the source simulation from the corresponding reference
version, evaluate the PDE once, and then train a model via a large number of such cases.
In our notation, this means learning from states computed as 
$\vc{t+1} = \pdec( \project \vr{t} )$. This effectively takes into account only
a single evaluation of the source PDE, and a model can only learn from numerical 
differences that build up within this single step. Hence, a variant of this 
approach is to allow reference and target version to evolve over the course of 
multiple steps such that the errors in the source states $\vcN$ show up more clearly 
with respect to $\vrN$. Similar to the look-ahead discussed in the main text, 
we can use $\vc{t+n} = \pdec^n ( \project \vr{t} )$ as a training data set. 
We denote such versions that have no interaction but consider multiple steps of unaltered 
coarse evolution as \non{n} below. Note that the previously discussed NON version 
could be denoted by \non{1}, but we keep the label NON for consistency with 
the main text in the following.

For all choices of $n$, we obtain the following minimization problem for learning via 
non-interacting solvers:
\begin{equation} \label{eq:non-min}
\new{\argmin_{\theta} \sum_{i=0}^{n} \Vert \vc{t+i} + \corr ( \vc{t+i} \, | \, \theta) - \project \vr{t+i} \Vert^2}.
\end{equation}

Another non-interacting variant could be trained by reversing the 
setup above and initializing reference trajectories from source states,
i.e., $\vr{t+n} = \pder^n ( \project^T \vc{t} )$. 
Like before, a model could be trained in a supervised fashion from a data set of $\vcN$
and $\vrN$ states computed in this way. However, as the interesting structures that 
make up the reference solutions typically take very long time spans to form (if they are 
achievable at all), this variant is clearly sub-optimal. Hence, due to the poor performance 
of the \non{n} versions, we have not included this reversed NON variant in our experiments.

The NON models presented in the main text so far already 
allow for a first quantification of the problems caused by the distribution shifts 
of the input features: across the two-dimensional 
fluid flow cases, the unaltered source simulations deviate by more 
than 50\% in terms of MAE from the corrected simulations. 
This means that, after applying the corrections, the model receives 
inputs that strongly differ from those seen at training time.
In terms of content of the input feature vectors,
the MAE measurements show a change of over 50\%. 
Nonetheless, we expect the model to reconstruct 
the reference states despite receiving inputs that are significantly different
from the inputs seen at the time of training.
Not surprisingly, the models only have limited success achieving this goal.

%

%


\subsection{Pre-computed Interactions}

As an improvement over the non-interacting 
versions above, we consider a class of models learning 
from data generated via pre-computed interactions, denoted by PRE.
The pre-computations have the goal of reducing the gap between 
source and reference trajectories. The pre-computation
changes the source trajectories and thus provides the learning optimization 
with modified inputs that are closer to the reference at inference time. This scenario
is common practice, e.g., for weather predictions, where simulations need 
to be aligned with real-world measurements, i.e., {\em data assimilation} algorithms
\cite{jones1997latent,stephan2008assimilation,xi2011automatic}. 
As the data set has to be prepared only once, 
computationally expensive pre-computation is often still feasible
as this overhead will not influence the performance at 
inference time. However, in the context of machine learning, pre-computed corrections 
can only provide limited improvements as the correction during the pre-computation 
phase can only partially mimic the behavior of the actual, learned version.


For PRE models, two correction functions are used: 
one for preparing the training data set denoted by $\corrPre$ and the learned 
correction $\corr$. The training data set is computed as 
$\vct{t+n} = (\pdec \corrPre)^n ( \project \vr{t} )$, where $n$ denotes 
the number of steps for independent simulation trajectories in the source and reference 
manifolds. 
Note that, in this context, due to the corrections being applied at 
the time of data generation, there is hope for longer unrolling periods (i.e., larger $n$)
to have a positive effect on the learning outcome (in contrast to the \non{n} versions above).
%
At inference time, $\corrPre$ is no longer used, and trajectories are instead 
computed as $\vct{t+n} = ( \pdec \corr )^n ( \vc{t} )$, in line with the NON variants.
Hence, in total, four versions of a trajectory from a single initial phase space point $\vr{t}$
exist: a source trajectory, a source trajectory corrected by pre-computation via $\corrPre$, 
a source trajectory corrected by the learned correction function $\corr$, and the reference trajectory.

We first describe how to include a pre-computation correction 
for spatial corrections while taking into account simulation constraints
before including the temporal dimension. 
For both, we adopt a constrained version of
{\em best linear unbiased estimates} \cite{henderson1975best}, which are widely used for data assimilation.

\subsubsection{Pre-computed Spatial Regularization}\label{sec:space-reg}

For a constraint-aware interpolation that can serve as a correction operator, 
consider two vector spaces $\mathbf{R} \in \mathbb{R}^\chi$ and
$\mathbf{S} \in \mathbb{R}^\xi$ 
with different dimensionalities $\xi,\chi \in \mathbb{N}$ with $\xi < \chi$.
Both vector spaces satisfy the constraint $\mM$,
i.e., $\mM\mathbf{r}=0$ for $\forall \mathbf{r} \in \mathbf{R}$, 
and $\mM\mathbf{s}=0$ for $\forall \mathbf{s} \in \mathbf{S}$. 
Given a finer vector field $\mathbf{c}_R$,
e.g., containing the reference solutions,
we aim to find the closest vector
field $\mathbf{c}_S$ ($\in \mathbf{S}$) to $\mathbf{c}_R$ ($\in \mathbf{R}$). 
Consider an interpolation operator $\mW$ that introduces new data points
within a vector field $\mathbf{c}_S$ ($\in \mathbf{S}$), i.e.,
$\mW\mathbf{c}_S \in \mathbb{R}^\chi$. We, then, strive to minimize the
distance between $\mW\mathbf{c}_S$ and $\mathbf{c}_R$ such that
$\mathbf{c}_S$ can best represent the information of $\mathbf{c}_R$ without
violating the constraints.
Thus, we aim for computing $\mathbf{c}_S$ with
\begin{equation}
  \label{eq:objective}
  \argmin_{\mathbf{c}_S} || \mW\mathbf{c}_S - \mathbf{c}_R ||^2 \quad
  \text{subject to}\quad \mM \mathbf{c}_S = 0.
\end{equation}
This represents a constrained optimization problem with equality constraints,
which we can solve via Lagrange multipliers $\mathbf{\lambda}$ as follows:
\begin{equation}
  \label{eq:lm-objective}
  \Phi = ||\mW\mathbf{c}_S - \mathbf{c}_R ||^2 +
  ( \mM \mathbf{c}_S)^{\top}\mathbf{\lambda}.
\end{equation}
This results in a system of equations:
\begin{equation}
  \label{eq:ls}
  \begin{bmatrix}
    \mW^{\top}\mW                                         &  -\mM  \\
    -\mM^{\top}                                           &  0
  \end{bmatrix}
  \begin{bmatrix}
    \mathbf{c}_S                                                    \\
    \mathbf{\lambda}
  \end{bmatrix}
  =
  \begin{bmatrix}
    \mW^{\top}\mathbf{c}_R                                         \\
    0
  \end{bmatrix}.
\end{equation}
Using the Schur complement, we can simplify this system to speed up
calculations:
\begin{eqnarray}
  \mM^{\top}(\mW^{\top}\mW)^{-1}\mM\mathbf{\lambda} & = & 
  \mM^{\top}(\mW^{\top}\mW)^{-1}\mW^{\top}\mathbf{c}_R , \\ 
  \label{eq:schur-sol}
  \mathbf{c}_S                                      & = & 
  (\mW^{\top}\mW)^{-1}(\mW^{\top}\mathbf{c}_R - \mM\mathbf{\lambda}).
\end{eqnarray}
%

In our setting, given source states $\vcN$ and reference states $\vrN$, 
we can thus compute a correction vector field via 
$\mathbf{c}_{t} = (\mW^{\top}\mW)^{-1}(\mW^{\top} (\vr{t} - \mW\vc{t}) - \mM\mathbf{\lambda})$,
e.g., using $\mM = (\nabla \cdot)$ for Navier-Stokes scenarios.
In order to train a model $\corr (\vcN \, | \, \theta)$ to infer the corrections, 
we can directly use the pre-computed correction vectors:
\begin{equation} \label{eq:argminPreLm}
\new{\argmin_{\theta} \sum_{i=0}^{n} \Vert \mathbf{c}_{t+i} - \corr ( \vct{t+i} | \theta)  \Vert^2}.
\end{equation}
We will denote versions using this pre-computation scheme for $\corrPre$ with
spatial regularization as \pre{SR}.

\subsubsection{Pre-computed Spatiotemporal Regularization}\label{sec:temp-reg}

The vector fields we target are obtained from a numerical simulation, 
where the underlying PDE is solved for a finite number of steps from an
initial condition. 
%
In the context of deep learning, an important
aspect to consider is the sensitivity \cite{murphy2004quantification} of the
targeted function (i.e., the correction) with respect to the data at hand, i.e.,
in our case, the state of a source simulation. 
The pre-computation process described in the previous section is typically done on a per-time-step basis,
and hence correction vector fields can vary significantly even for smooth changes of the
source simulation. That means the correction function can have a very nonlinear and
difficult to learn relationship with the observable data in a simulation.

In order to address this difficulty, we include 
a temporal regularization by limiting the changes over time for each sample point in space.
Consequently, we regularize our correction vector fields such that
they change smoothly in time by penalizing temporal change of the correction
vector field within the Lagrange multiplier framework. We minimize $d\mathbf{c}_S/dt$
together with the constrained transfer from fine to coarse discretizations:
\begin{equation}
  \label{eq:new-objective}
  \argmin_{\mathbf{c}_S}
   \left( || \mW\mathbf{c}_S - \mathbf{c}_R ||^2 + \beta ||
    \frac{d\mathbf{c}_S}{dt} ||^2 \right)\quad
  \text{subject to}\quad \mM \mathbf{c}_S = 0.
\end{equation}
Here, $\beta$ is the temporal regularization coefficient. 
A finite difference approximation of the temporal derivative of the correction field, 
i.e., $d\mathbf{c}_S/dt$, yields the following system of equations:
\begin{equation}
  \label{eq:ls-new}
  \begin{bmatrix}
    \mW^{\top}\mW + \beta\frac{2}{\Delta t}\mathbf{I} & -\mM \\
    -\mM^{\top}                                       & 0
  \end{bmatrix}
  \begin{bmatrix}
    \mathbf{c}_S                                             \\
    \mathbf{\lambda}
  \end{bmatrix}
  =
  \begin{bmatrix}
    \mW^{\top}\mathbf{c}_R + \beta\frac{2}{\Delta t}\mathbf{c}_S^{t-1} \\
    0
  \end{bmatrix},
\end{equation}
where $\Delta{t}$ is the time step size, $\mathbf{I}$ is the identity matrix, and
$\mathbf{c}_S^{t-1}$ denotes the correction vector field evaluated
at the previous time step.
Following \myeqref{eq:argminPreLm}, this data is pre-computed and used for training 
a neural network in a supervised manner.
Models trained with data from this spatiotemporal pre-computation as $\corrPre$ are denoted
by PRE, and we have used a coefficient of $\beta=1.0$
for all PRE models of our submission.

\subsection{Solver-in-the-Loop Interactions via Differentiable Physics}

The main goal of training via differentiable physics 
is to bridge the gap that arises from changes in the input data distribution
and directly train with the environment that the learned model is supposed to work 
with at inference time. Hence, the learning process 
aims to solve the minimization problem
\begin{equation} \label{eq:diffPhysTrain}
\new{\argmin_{\theta} \sum_{i=0}^{n-1} \Vert \pdec(\vct{t+i}) + \corr ( \pdec( \vct{t+i} ) | \theta) - \project \vr{t+i+1} \Vert^2},
\end{equation}
where the phase space trajectories are computed via 
$\vct{t+k} = ( \pdec \corr )^k ( \project \vr{t} )$. This formulation 
illustrates that
a cyclic dependency between the corrected 
states $\vctN$ and the learned correction function $\corr$ exists for the ``solver-in-the-loop'' interactions
of this section. As both the deep neural network for $\corr$ and likewise the PDE $\pdec$
are potentially highly non-linear operators, the corresponding coupled minimization problem for 
calculating the weights of $\corr$ is challenging. However, our results clearly show 
that stable optimizations can be achieved in practice and that they lead to very significant 
improvements of the learned representation.

The recurrent training requires differentiable physics solvers that allow for a
back-propagation of gradients through the discretized physical simulation. 
In this work, we employ a differentiable PDE solver from the open source $\Phi_\textrm{Flow}$
library \cite{holl2020}. This solver builds on the automatic differentiation of
the underlying machine learning framework to compute analytic derivatives and augments them
with custom derivatives where necessary. For example, the pressure correction step 
of a Navier-Stokes solver is provided with a custom gradient for performance reasons.
This setup allows for a straightforward integration of solver functionality into 
machine learning models and enables end-to-end training in recurrent settings.
Although all of our examples use the $\Phi_\textrm{Flow}$ solver, we do not leverage 
any special functionality apart from gradients being provided for all steps 
of the PDE solve. Hence, our results should carry over to other types of differentiable 
physics solvers.

It is worth noting that, in the setup discussed so far, 
the reference solver does not need to be differentiable; i.e., the phase 
space points in $\manifref$ could be provided by a black-box approximation
without gradients as long as a differentiable solver for 
the source manifold $\manifsrc$ exists. 
We demonstrate the split setup
using an external solver for the buoyancy-driven flows below.

Our implementation directly follows \myeqref{eq:diffPhysTrain}. For each mini-batch,
we start with a collection of reference states $\vrN$ for which recurrent trajectories 
of $(\pdec \corr)^n$ are unrolled for $n$ steps. The loss with respect to corresponding 
reference states is computed over all intermediate states of the trajectory.
Back-propagation, then, unrolls
the differences through the sequence of solver steps to update the weights of the 
neural network that provides the correction function.

Under the assumption that the training process converges, this entirely 
removes the problem of distribution shift.
Once the learned correction $\corr$ converges to a steady-state, 
it is trained with exactly the phase state inputs that are produced at inference time.
The MAE of the test data samples again provides a measure of the discrepancies.
Compared to the differences of around 50\% for non-interacting variants (measured 
between source states and corrected states), the deviations grow to 75\% and above for SOL versions.
Nonetheless, even this larger difference in terms of input distributions is 
unproblematic here as the network receives the modified states at training time.
However, we noticed that, during our training runs, the 
final states typically do not fully converge, but still show smaller oscillations in terms of 
performance. While this could be prevented via learning-rate decay, we believe 
the slightly changing states provide robustness similar to dropout or
manual injections of noise \cite{sanchez2020gns}.

\new{While the error accumulates and typically
grows over the course of a full trajectory, our key hypothesis here is that a learned approach can nonetheless identify
and correct a large part of the error function based on information from a single phase-space input. 
For the PDEs we consider, a single state uniquely describes its future evolution.
We have experimented with additionally providing varying numbers of previous states 
$\mathbf{s}_{t-k}, ... ,\mathbf{s}_{t-1}$ as input to our model. Our 
tests have not shown improvements from these additional states
and indicate that the components of the error function that are learned with our 
approach can be reliably inferred from a single state $\mathbf{s}_t$. 
}

\section{Experiments}
\label{app:exp}

To acquire our data sets, we generate a set of simulation sequences with varying initial conditions.
These sequences are used for obtaining pairs of source and reference velocity fields for training. 
%
The following PDEs typically work with a continuous 
velocity field $\vu$ with $d$ dimensions and components, i.e.,
$\vu(\vx,t): \mathbb{R}^d \rightarrow \mathbb{R}^d $.
For discretized versions below, $d_{i,j}$ will denote the dimensionality 
of a field such as the velocity with $i \in \{s,r\}$ denoting source/inference manifold 
and reference manifold, respectively.
This yields 
$\vc{} \in \mathbb{R}^{d \times d_{s,x} \times d_{s,y} \times d_{s,z} }$ and
$\vr{} \in \mathbb{R}^{d \times d_{r,x} \times d_{r,y} \times d_{r,z} }$
with domain size $d_{x},d_{y},d_{z}$ for source and reference.
Typically, $d_{r,i} > d_{s,i}$ and $d_{z}=1$ for $d=2$. 
For all PDEs, we use non-dimensional parametrizations as outlined below,
and the components of the velocity vector are denoted by $x,y,z$ subscripts, i.e.,
$\vu = (u_x,u_y,u_z)^T$ for $d=3$.

The mapping function $\project$
denotes a projection to the source manifold by $\project \vr{t}$,
and we assume that the transpose transforms to the reference manifold, i.e., $\project^T \vc{t}$.
The mapping function is typically neither bijective nor unique, i.e.,
$\project^T \project \vr{t} \ne \vr{t}$, however, within this work, we are primarily
concerned with retrieving projected references of the form $\project \vr{t}$.
The potential null-space of $\project^T$ is an interesting topic for super-resolution
approaches \cite{fukami2019super}. 
We found that a bi- or tri-linear
spatial downsampling from reference to source space is efficient to compute
and yields sufficient accuracy for the transfer in our experiments.
In order to make comparisons with the source simulations easier, we visualize 
the projected reference solution, i.e., $\project \vr{t}$, in the following.


\subsection{Unsteady Wake Flow in Two Dimensions}\label{app:expKarman2d}

For the unsteady wake flow setup, we use the incompressible Navier-Stokes equations 
for Newtonian fluids:
\begin{eqnarray}
  \label{eq:model-ns}
  \begin{aligned}
    \frac{\partial u_x}{\partial{t}} + \vu \cdot \nabla u_x =
    - \frac{1}{\rho}\nabla{p} + \nu \nabla\cdot \nabla u_x  \\
    \frac{\partial u_y}{\partial{t}} + \vu \cdot \nabla u_y =
    - \frac{1}{\rho}\nabla{p} + \nu \nabla\cdot \nabla u_y  \\
    \text{subject to} \quad \nabla \cdot \vu = 0,
  \end{aligned}
\end{eqnarray}
where $\rho$, $p$, $\nu$, and $g$ denote density, pressure, viscosity, and
external forces, respectively. The constraint, $\nabla \cdot \vu = 0$, is
particularly important and introduces additional complexity
as it restricts motions to the space of divergence-free (i.e., volume preserving) motions.
The flow is integrated over time with operator splitting, and pressure is solved 
implicitly with a Chorin projection \cite{chorin1967numerical}.
The domain $\Omega$ has an extent of $1 \times 2$ with open boundary conditions
and a velocity inflow $\vu_{\text{in}}=(0,1)^T$ at the bottom face of the domain.
A circular obstacle with diameter of $0.1$ is located at position $(1/2,1/2)^T$.
For reference simulations, the domain is discretized with 
$d_{r,x}=128$ and $d_{r,y}=256$ cells using a staggered layout for the velocity components. 
The source domain instead contains 
$d_{s,x}=32$ and $d_{r,y}=64$ cells. Data sets from both contain sequences of 500 time steps
each. For the training data, the viscosity coefficient $\nu$ is chosen
to yield Reynolds numbers 
Re$_{\text{train}} \in \{97.7, 195.3, 390.6, 781.3, 1562.5, 3125.0\}$;
i.e., there is a factor of more than 30 between smallest and largest Reynolds numbers
in the training data.
The test data set instead contains the Reynolds numbers 
Re$_{\text{test}} \in \{146.5 , 293.0 , 585.9 , 1171.9, 2343.8\}$,
which are denoted as $\times1$, $\times2$, $\times4$, $\times8$, and $\times16$ below, respectively.
%



\paragraph{Training Procedure}


The neural network of $\corr$ is fully convolutional. It consists of 
five ResBlocks \cite{he2015} with 5$\times$5 kernels. 
The convolutional layers have two times 32 features per block 
(details of the architecture are given in \myappref{app:models}). 
Overall, the model has around 260k trainable parameters.
\new{In addition to the velocity, the model receives a constant field containing 
the Reynolds number in order to distinguish the different physical regimes.}

\begin{figure}[tb]
  \centering
  \begin{overpic}[width=0.96\linewidth]{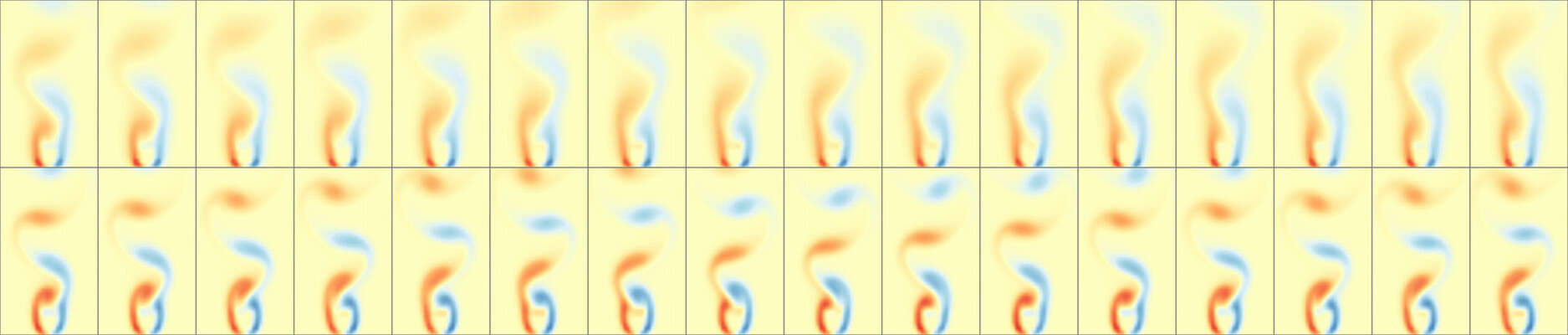}
    \put(-1.6,0)    {\scriptsize\rotatebox{90}{\makebox[10.66\unitlength]{Reference}}}
    \put(-1.6,10.66){\scriptsize\rotatebox{90}{\makebox[10.66\unitlength]{SRC}}}
    \put(100.5,0){\includegraphics[height=10.66\unitlength]{figs/summary-legend-karman}}
  \end{overpic}
  \caption{An example sequence of the unsteady wake flow from the training
      data set for time steps $t \in \{50, 60, \cdots, 200\}$.
  }
  \label{fig:appx:karman-images-train}
\end{figure}

With the Reynolds number range above, we generate 500 time steps as 
training data, which contain temporal dynamics with ca. eight vortex shedding cycles for each case,
\new{i.e., they cover a similar number of eddy turnover times.}
This leads to roughly 98 million cells of data in the reference trajectories, 
which are down-sampled to 6.1 million cells with lower resolution of the source data.
Example flow fields are shown in \myfigref{fig:appx:karman-images-train}.

All SOL models are trained with the differentiable physics solver for 99.8k iterations 
with a batch size of 3 and a learning rate of $10^{-4}$.
The NON model uses the same training modalities replacing the differentiable PDE solver
with the supervised loss of \myeqref{eq:non-min}.
On the other hand, all PRE models are trained in a supervised manner for 36k iterations 
with a batch size of 32 and initial learning rate of $10^{-3}$ that is lowered to 
$5\times10^{-7}$ over the course of the training. 
Here, we augment the training data via randomized horizontal flipping and 
use 5\% of the training data as validation samples.
To show the stability of training, we train three models for each case 
below with different random seeds.


\paragraph{Results}

We present results for the unsteady wake flow scenario
using models trained via different interaction methodologies 
and evaluate each model on the test set of Reynolds numbers Re$_{\text{test}}$. 
Each simulation is computed for 500 time steps using the source solver in combination with 
a correction from a trained neural network. Mean errors are computed 
in comparison to reference phase space states, i.e., $\project \vrN$.
We compute the errors over the three trained models for each variant.

In this scenario, the NON model already leads
to a significant reduction of the overall velocity error. While the regular source 
simulation (SRC) shows a MAE of 0.146 with respect to the projected reference 
states $\project \vrN$, the NON model reduces this error to 0.049. These errors (and the 
following measurements) are mean values for all five test Reynolds numbers, which 
were not seen at training time. The results are visualized in 
\myfigref{fig:appx:karman-plot}, and corresponding numeric values are given in \mytabref{tab:karman}.

The pre-computed variants improve on this behavior,
roughly halving the remaining error.
The pre-computed variant without temporal regularization (\pre{SR}) gives a worse performance
than the one with spatiotemporal regularization (PRE) but, nonetheless, fares better than the 
NON version.

\begin{figure}[tb]
  \centering
  \begin{overpic}[width=0.96\linewidth]{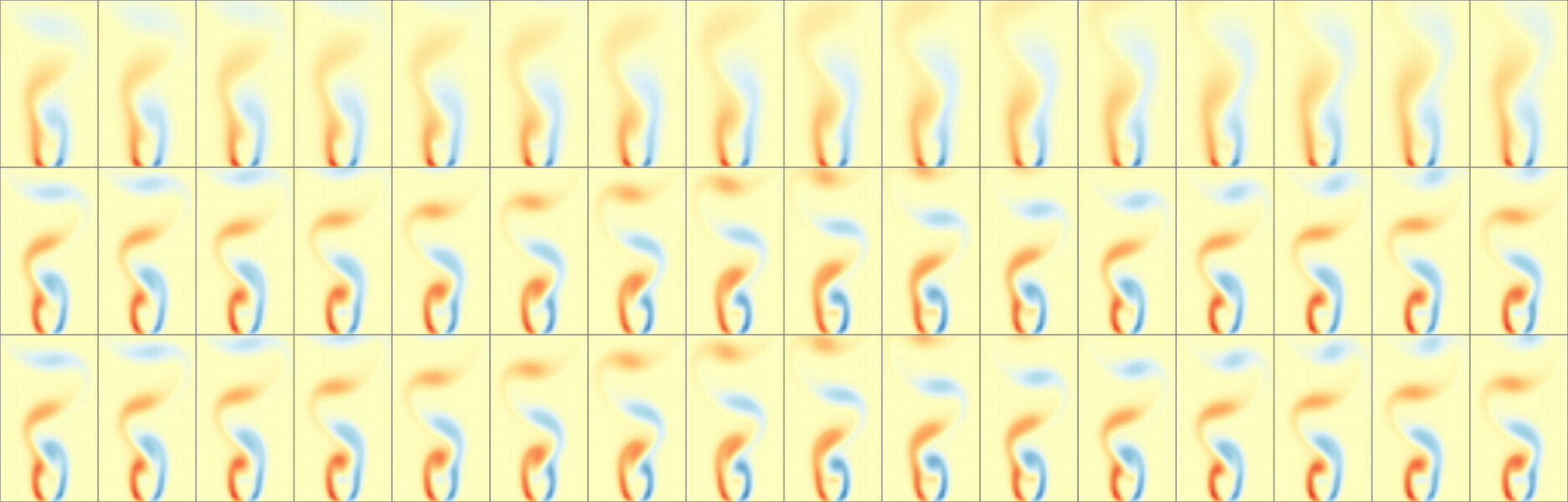}
    \put(0,30)      {(a)}
    \put(-1.6,0)    {\scriptsize\rotatebox{90}{\makebox[10.66\unitlength]{Reference}}}
    \put(-1.6,10.66){\scriptsize\rotatebox{90}{\makebox[10.66\unitlength]{\sol{32}}}}
    \put(-1.6,21.33){\scriptsize\rotatebox{90}{\makebox[10.66\unitlength]{SRC}}}
    \put(100.5,0){\includegraphics[height=10.66\unitlength]{figs/summary-legend-karman}}
  \end{overpic}
  \\\vspace{0.5em}
  \begin{overpic}[width=0.96\linewidth]{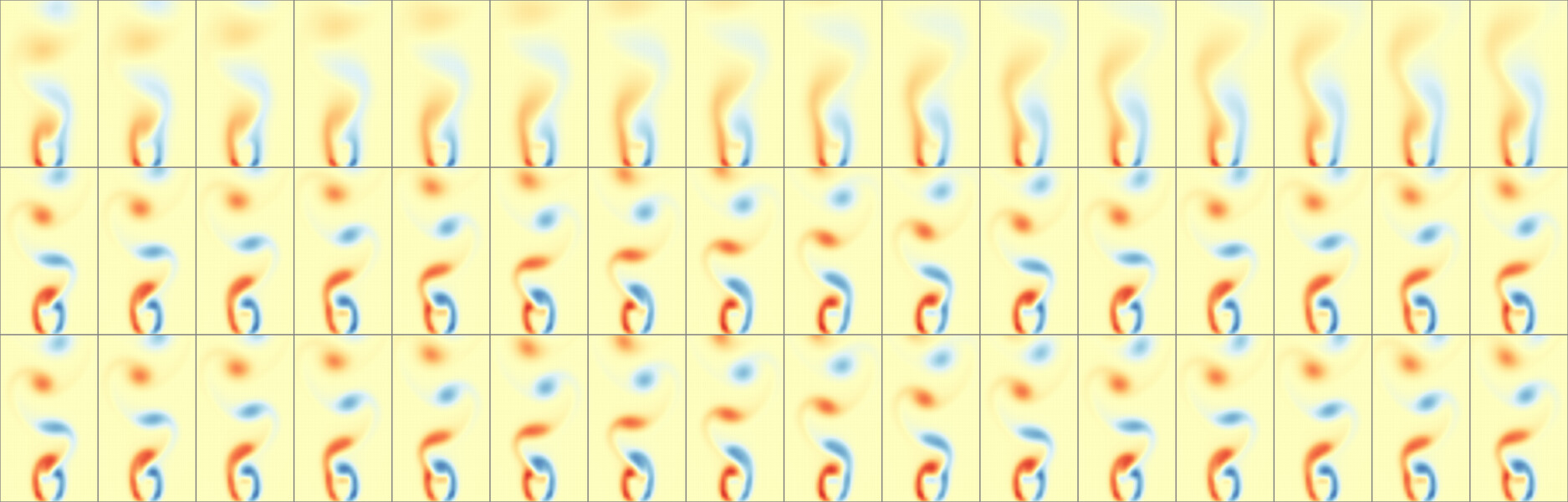}
    \put(0,30)      {(b)}
    \put(-1.6,0)    {\scriptsize\rotatebox{90}{\makebox[10.66\unitlength]{Reference}}}
    \put(-1.6,10.66){\scriptsize\rotatebox{90}{\makebox[10.66\unitlength]{\sol{32}}}}
    \put(-1.6,21.33){\scriptsize\rotatebox{90}{\makebox[10.66\unitlength]{SRC}}}
    \put(100.5,0){\includegraphics[height=10.66\unitlength]{figs/summary-legend-karman}}
  \end{overpic}
  \caption{Time steps of test cases for the unsteady wake flow for $t \in \{50, 60, \cdots, 200\}$: (a) Re = $\times1$ and (b) Re = $\times16$. 
    }
  \label{fig:appx:karman-images-test}
\end{figure}

\myfigref{fig:appx:karman-plot} additionally shows results for different 
SOL versions trained with the solver-in-the-loop interaction. While the \sol{4} version 
fares better than NON, it is only roughly on par with \pre{SR}. Increasing the number 
of look-ahead steps, however, increases the performance substantially
with the \sol{32} model exhibiting a final MAE of only 0.013.
Several visual examples of simulated flows from the five 
test cases used in these evaluations are shown in \myfigref{fig:appx:karman-images-test}.
It is visible that the SOL version matches the behavior of the reference 
solution much more closely.

We additionally break down the errors with respect to the different Reynolds 
numbers of the five cases in \myfigref{fig:appx:karman-plot-re}. Despite a factor of
16 between the Reynolds numbers, there is no significant decrease in performance across 
the different cases. 
Only the NON version exhibits slightly larger errors for higher Reynolds numbers.
On the other hand, the performance is largely uniform for the SOL versions.

Due to the distinct vortex shedding characteristics of the flow, 
it is interesting to evaluate the flow field in terms of its frequency spectrum.
As an example, \myfigref{fig:appx:karman-ucen-step} shows the $u_x$ velocity component 
over the course of 500 simulation steps at the center of domain, i.e., behind the
obstacle, for one of our test data sets.
We show the corresponding evaluation in \myfigref{fig:appx:karman-plot-freq}.
Interestingly, especially the PRE versions fare better in terms of frequency 
errors. Here the relatively expensive
pre-computation step shows a performance gain. Nonetheless, the models trained via differentiable 
physics likewise learn to control the frequency behavior when training 
with a sufficient number of look-ahead steps 
as the \sol{32} model yields
a substantially lower frequency error than the PRE model.


We additionally show results for a smaller model for 
a simpler sequential convolutional network with
57k trainable parameters in \myfigref{fig:appx:karman-plot-small}.
The overall relative ordering of the interaction methods remains the same.
The non-interacting method
performs worse than pre-computation, which in turn is outperformed 
by the differentiable physics interaction. However, the overall 
performance is reduced, e.g., the NON model only reduces the error by ca. 30\%.
The \sol{16} version still outperforms the other versions.
Overall, not surprisingly, the reduced weight count significantly
reduces the representational capabilities of the neural networks and leads
to a deteriorated performance. Nonetheless, training via interactions with differentiable 
physics is beneficial for inference performance. 

To conclude, approximate solutions of the unsteady wake flow case can be corrected substantially by learned
models, and especially training with differentiable physics in the loop yields significantly 
reduced errors in long simulated sequences. The \sol{32} version with a larger model reduces the MAE 
with respect to the reference solution to less than 9\% (on average) of the error induced by the source 
simulation.

\newcommand{\myFigHKarman}{0.30\columnwidth}
\begin{figure}[htb]
  \centering
  \subcaptionbox{ Velocity error for different models                          }{\includegraphics[height=\myFigHKarman,page=3]{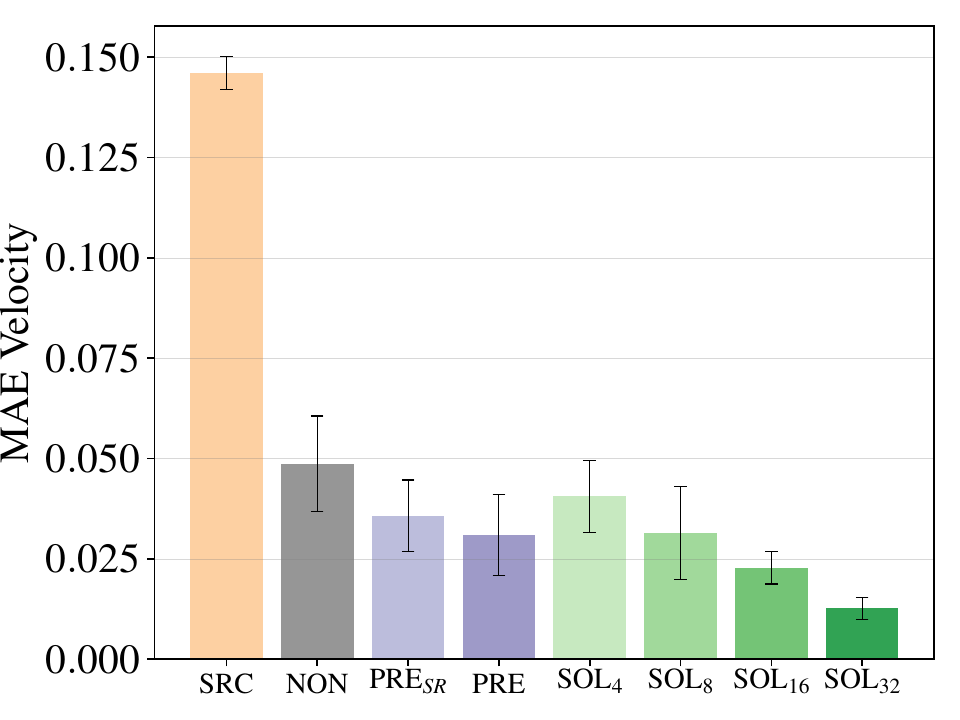}}
  \subcaptionbox{ Velocity improvement (relative to SRC) for different models  }{\includegraphics[height=\myFigHKarman,page=7]{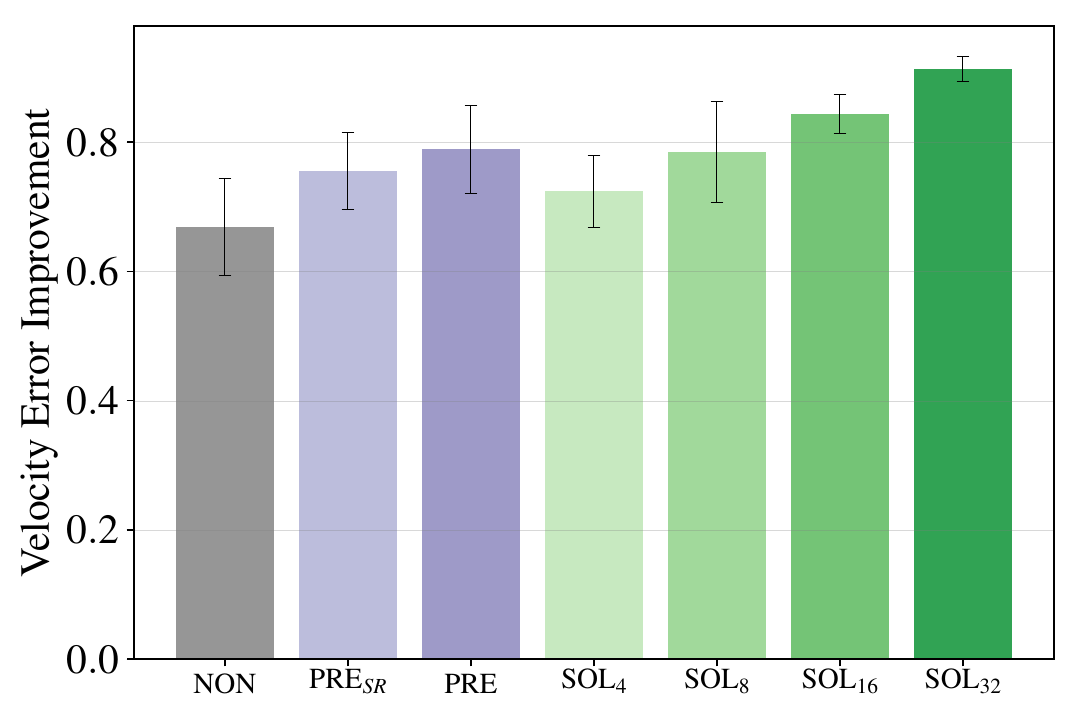}}
  \caption{
    Different models applied to five test cases over 500 time steps for the unsteady wake flow scenario.
    The \sol{32} reduces the error introduced by SRC by a factor of 11.2 on average.
  }
  \label{fig:appx:karman-plot}
\end{figure}

\begin{figure}[htb]
  \centering
  \subcaptionbox{ Velocity error per Reynolds number                         }{\includegraphics[height=\myFigHKarman,page=2]{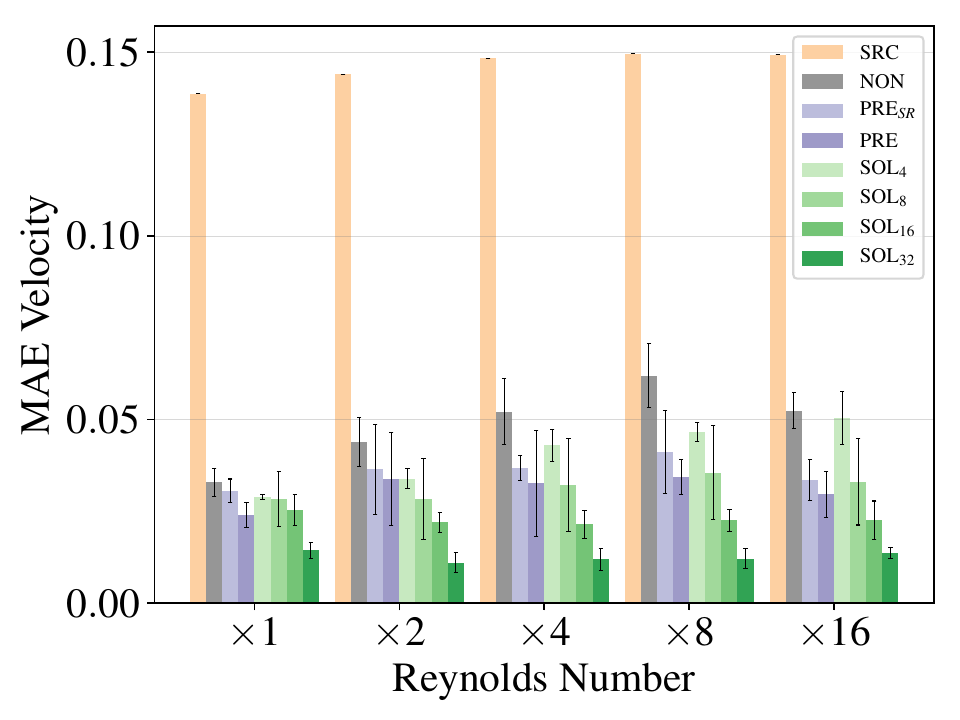}}
  \subcaptionbox{ Velocity improvement (relative to SRC) per Reynolds number }{\includegraphics[height=\myFigHKarman,page=5]{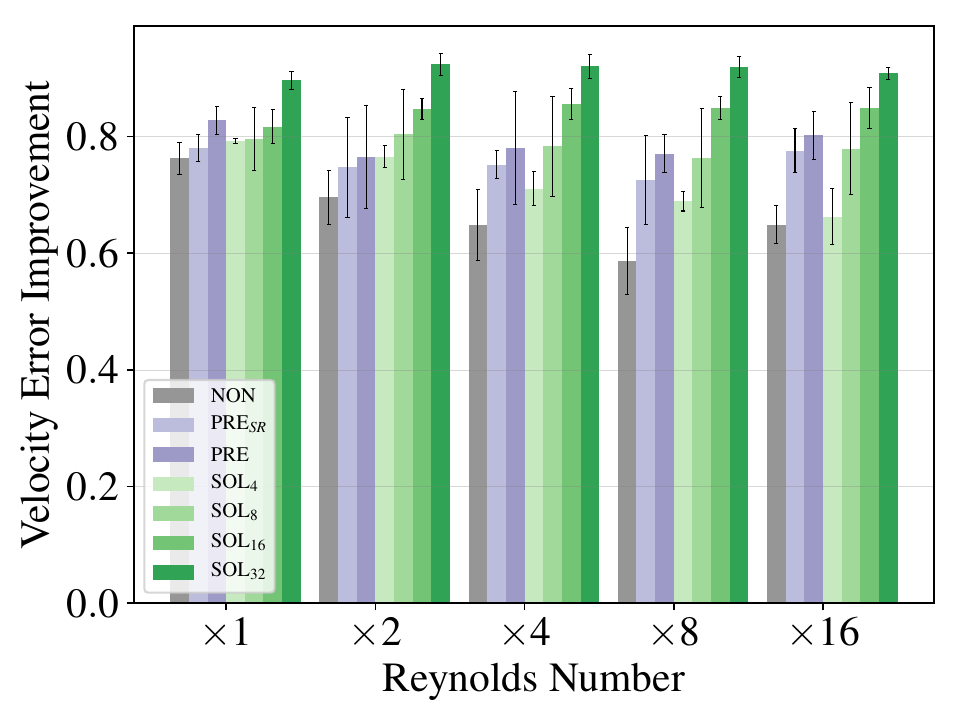}}
  \caption{
    Separate evaluations for five different test cases of the unsteady wake flow scenario.
  }
  \label{fig:appx:karman-plot-re}
\end{figure}

\newcommand{\tweakSpaceKtwo}{\vspace{-0.1cm}}
\begin{figure}[htb]
  \centering
  \includegraphics[height=\myFigHKarman,page=1]{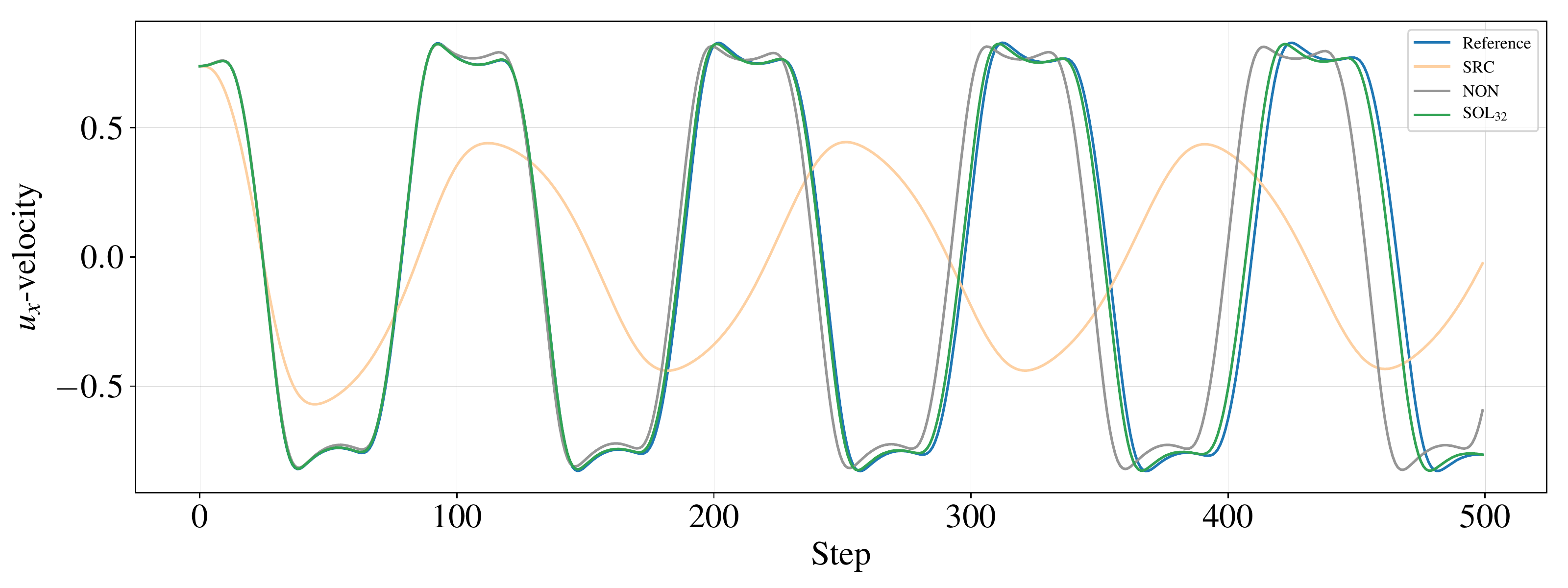}
  \tweakSpaceKtwo\caption{
    $u_x$-velocity at the center of domain for one test data set (Re = $\times4$).
  }
  \label{fig:appx:karman-ucen-step}
\end{figure}

\begin{figure}[htb]
  \centering
  \subcaptionbox{ Frequency error for different models }{\includegraphics[height=\myFigHKarman,page=9]{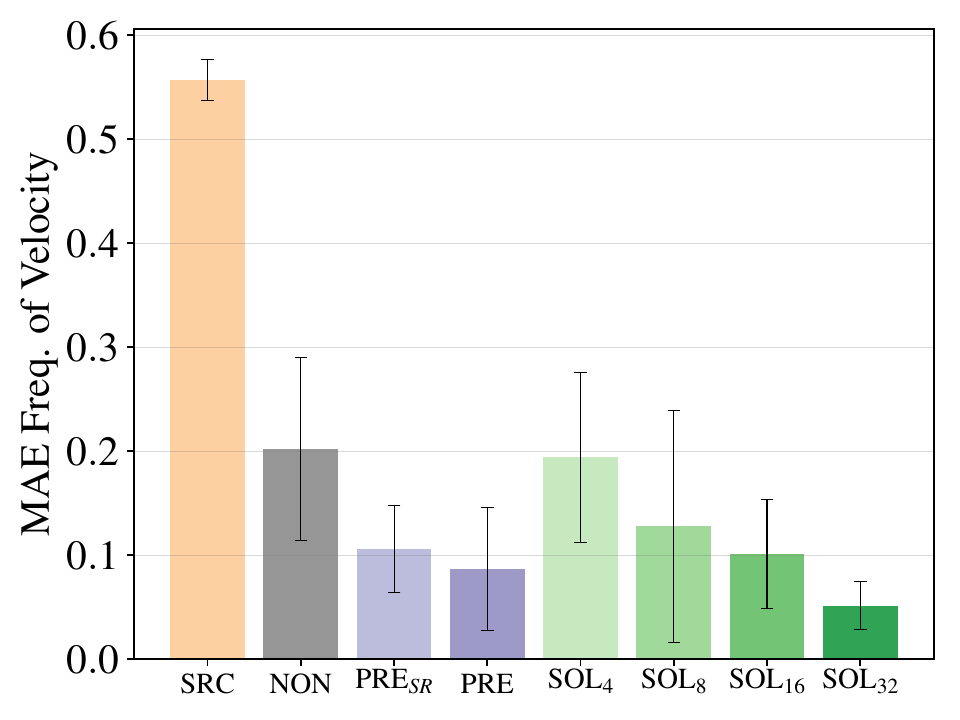}}
  \subcaptionbox{ Frequency error per Reynolds number  }{\includegraphics[height=\myFigHKarman,page=8]{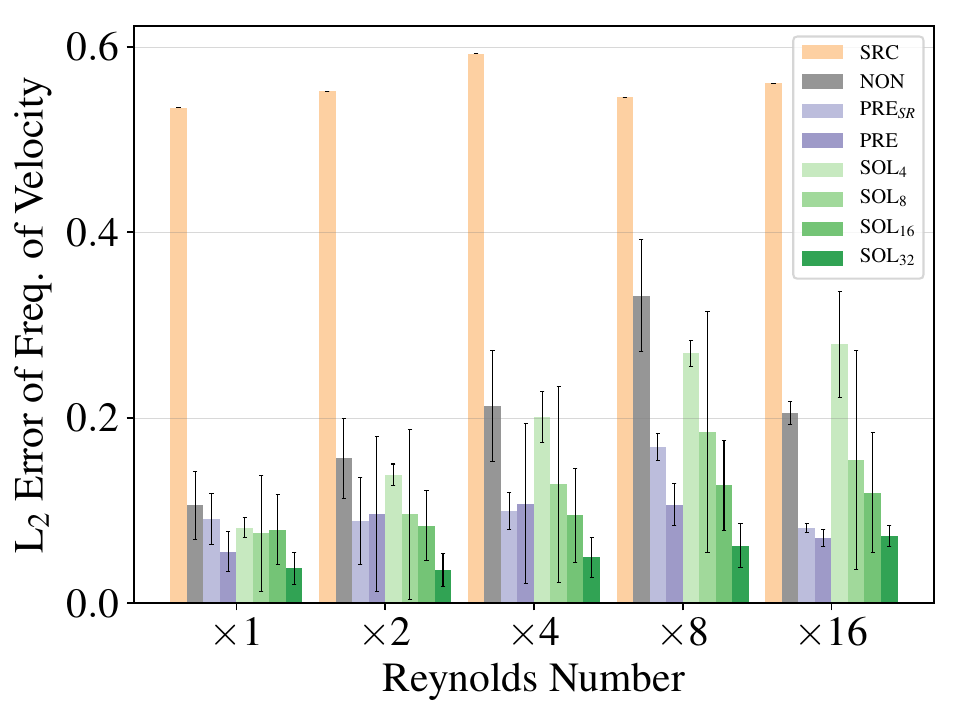}}
  \tweakSpaceKtwo\caption{
    Frequency-domain evaluation for the unsteady wake flow scenario. 
    Shown for the five test cases over 500 time steps.
  }
  \label{fig:appx:karman-plot-freq}
\end{figure}

\begin{figure}[htb]
  \centering
  \subcaptionbox{ Velocity error for different smaller models      }{\includegraphics[height=\myFigHKarman,page=3]{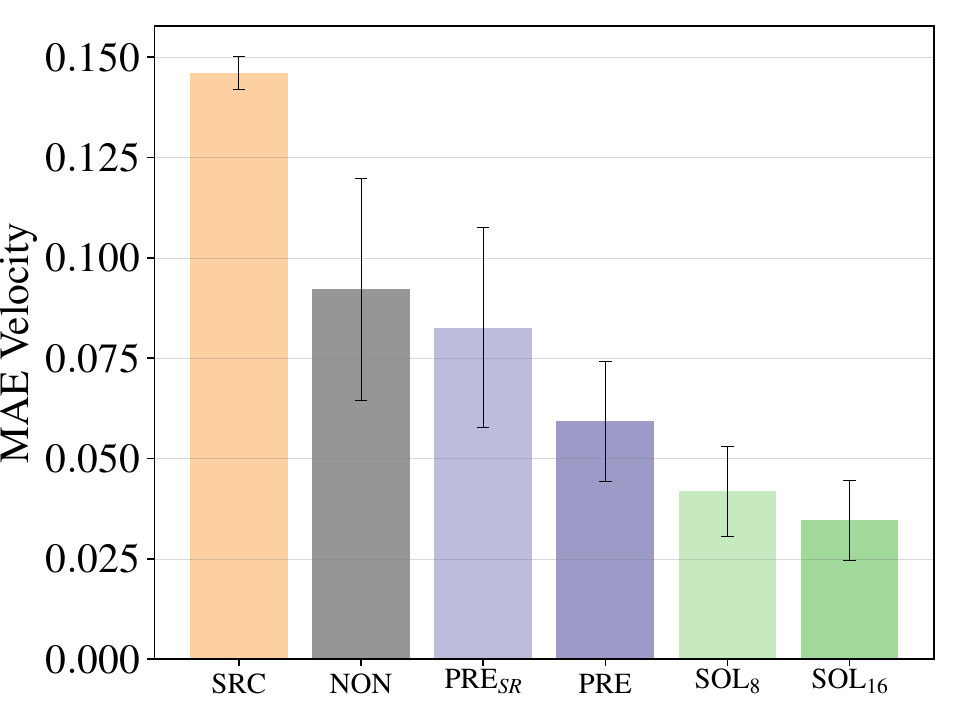}}
  \subcaptionbox{ Smaller models per Reynolds number               }{\includegraphics[height=\myFigHKarman,page=2]{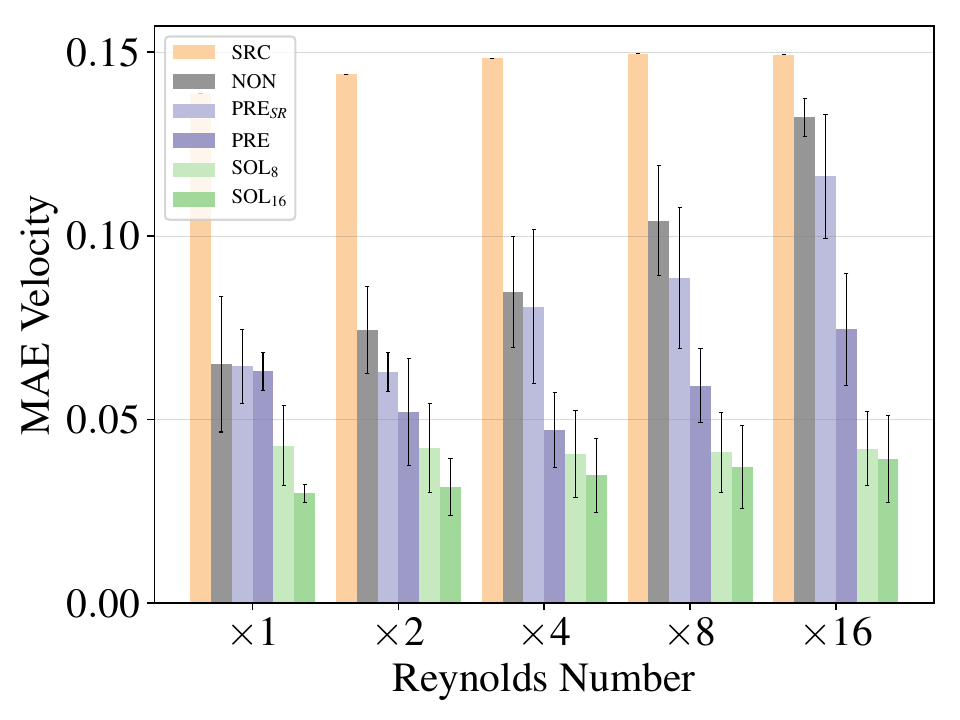}}
  \tweakSpaceKtwo\caption{
    Different models with a smaller network size (57k trainable weights) applied to five test cases over 500 time steps for the unsteady wake flow scenario.
  }
  \label{fig:appx:karman-plot-small}
\end{figure}

\begin{table}[tb]
  \caption{Quantitative evaluation of different models for the unsteady wake flow scenario.}
  \label{tab:karman}
  \begin{center}
    \begin{tabular}{ccccccccc}
      \toprule
      \textbf{Model} & \multicolumn{8}{c}{\textbf{MAE Velocity}, Mean (std. dev.)}                      \\
      \cmidrule{2-9}
                     & SRC     & NON     & \pre{SR} & PRE     & \sol{4} & \sol{8} & \sol{16} & \sol{32} \\
      \midrule                                                            
      Regular        & 0.146   & 0.049   & 0.036    & 0.031   & 0.041   & 0.031   & 0.023    & 0.013    \\
                     & (0.004) & (0.012) & (0.009)  & (0.010) & (0.009) & (0.012) & (0.004)  & (0.003)  \\
      Smaller        & 0.146   & 0.092   & 0.083    & 0.059   & -       & 0.042   & 0.035    & -        \\
                     & (0.004) & (0.028) & (0.025)  & (0.015) & -       & (0.011) & (0.010)  & -        \\
      \midrule
                     & \multicolumn{8}{c}{\textbf{\new{$L^2$ Error} of $u_x$ Velocity in Frequency Domain}} \\
      \cmidrule{2-9}
                     & SRC     & NON     & \pre{SR} & PRE     & \sol{4} & \sol{8} & \sol{16} & \sol{32} \\
      \midrule                                                           
      Regular        & 0.557   & 0.202   & 0.106    & 0.087   & 0.194   & 0.128   & 0.101    & 0.051    \\
      Smaller        & 0.557   & 0.275   & 0.244    & 0.158   & -       & 0.093   & 0.155    & -        \\
      \bottomrule
    \end{tabular}
  \end{center}
\end{table}


\clearpage

\subsection{Buoyancy-driven Fluid Flow}\label{app:expBuoy}

This scenario encompasses a volume of hot smoke rising in a 
closed container. The motion of the smoke volume is driven by buoyancy forces
computed via a marker field that is passively advected in the flow, and 
which marks a region of fluid with lower density.
Assuming a small relative change of density between the marker and the 
bulk, we compute the resulting forces with a Boussinesq model. 
Hence, this scenario is likewise based on the Navier-Stokes equations, 
but due to the additional coupled system, it leads to significantly 
more chaotic and complex behavior than the unsteady wake flow.
In order to target solutions with complex motions, we do not explicitly solve for
viscosity effects, but rely on the numerical viscosity inherent in the discretization.
%
%
This yields the following PDE:
\begin{eqnarray}
  \label{eq:boussinesq}
  \frac{\partial u_x}{\partial{t}} + \vu \cdot \nabla u_x =
  - \frac{1}{\rho}\nabla{p},
  \quad 
  \frac{\partial u_y}{\partial{t}} + \vu \cdot \nabla u_y =
  - \frac{1}{\rho}\nabla{p}  + \eta d
  \nonumber
  \\
  \text{subject to} \quad \nabla \cdot \vu = 0,
  \quad
  \frac{\partial d}{\partial{t}} + \vu \cdot \nabla d = 0 \ , 
\end{eqnarray}
where $\eta$ denotes the buoyancy factor for the Boussinesq model.

We also use this scenario to demonstrate that the reference data can be computed by 
a discretization or algorithm that differs from the one used to compute the source trajectories.
More specifically, we use second-order pressure projection 
scheme for the reference trajectory solutions \cite{zehnder2018},
which was shown to lead to an improved conservation of energy \cite{hairer2006geometric}.
In addition, we use a less dissipative advection scheme for the source 
and reference solvers \cite{selle2008}.

\begin{figure}[tb]
  \centering
  \begin{overpic}[width=0.96\linewidth]{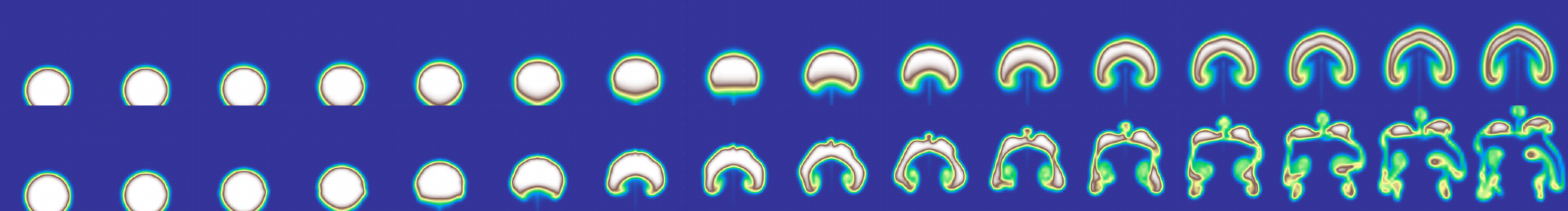}
    \put(-1.6,0)   {\scriptsize\rotatebox{90}{\makebox[6.71\unitlength]{Ref.}}}
    \put(-1.6,6.71){\scriptsize\rotatebox{90}{\makebox[6.71\unitlength]{SRC}}}
    \put(100.5,0){\includegraphics[height=6.71\unitlength]{figs/summary-legend-buoy}}
  \end{overpic}
  \caption{An example sequence of the buoyancy scenario from the training data
    set for time steps $t \in \{0, 25, \cdots, 375\}$. 
  }
  \label{fig:appx:smoke-images-train}
\end{figure}

The domain has an extend of $1 \times 2$ units, where the marker density is injected 
in the lower quadrant.
The reference simulations use a staggered discretization with 
$d_{r,x}=128$ and $d_{r,y}=256$, while the source simulations 
use a domain with $d_{s,x}=32$ and $d_{r,y}=64$.
We randomize the initial size of the marker volumes with circular shapes with a radius 
$r \sim \mathcal{U}(0.1,0.25)$, where $\mathcal{U}$ denotes a uniform distribution.
The training data set consists of 
48 different initial conditions simulated for 1000 steps each.
Several examples are shown in \myfigref{fig:appx:smoke-images-train}.
%
For the test scenes, we change the initial marker distribution $d$
to obtain five simulations containing two circles with
$r \sim \mathcal{U}(0.05,0.1)$ and another five simulations with 
$r \sim \mathcal{U}(0.2,0.3)$. Thus, we obtain ten test scenes, half of which 
have a reduced marker quantity compared to the training data and five with 
an increased quantity. As the $d$ determines the forces induced by the Boussinesq
model, this leads to simulations that are slower and faster, respectively, 
than those in the training set.

\paragraph{Training Procedure}

The neural network architecture for $\corr$ follows the one described above, 
but instead uses four ResBlocks with 16 features each and contains ca. 36k trainable weights.
As both velocity $\vu$ and marker $d$ determine the dynamics of the flow, the network 
receives both fields as input, but still only infers a correction for the velocity;
i.e., $d$ is modified only via advection through $\vu$, not directly by $\corr$.
%
All SOL and NON models are trained for 294k iterations 
with a batch size of 4 and a learning rate of $10^{-4}$.
We evaluate the models on validation set with 5 simulations and 300 time steps 
drawn from the same initial marker distribution as the training data,
and keep the model with the lowest validation loss.

To speed up the pre-computations, we only compute $\corrPre$ 
for cells $i,j$ in the domain with $d_{i,j}>10^{-4}$ (we validate this choice below).
The PRE variants of $\corr$ are then trained on the resulting, regularized 
data for 300k iterations with a batch size of 32 using horizontal flipping as data augmentation.

\paragraph{Results}
We evaluate different models which are applied to 300 time steps of ten 
test conditions. Errors with respect to the reference solutions are computed and averaged 
across the resulting 3k phase field states.
Numeric error values for the following tests can be found in \mytabref{tab:smoke}.

We evaluate the different baseline versions (NON and PRE) in comparison to the source 
simulation (which underlies all other variants) and compare them to SOL versions 
with increasing look-ahead.
The resulting errors and relative improvements 
are shown in \myfigref{fig:appx:smoke-plot} and given numerically in \mytabref{tab:smoke}.
It is apparent that the SOL versions yield very significant improvements over the other learned variants.
Besides the velocity errors, we also provide an evaluation of the passively 
advected marker density $d$. This quantity is crucial for the dynamics of the flow,
but cannot be influenced directly by the neural networks. Hence, it provides 
an additional view on how well the inferred corrections manage to reduce the numerical 
errors of the source simulation. %
The corresponding evaluation highlights 
that both velocity and density improvements increase consistently with SOL variants that 
were trained with larger look-aheads.
We also evaluate the different
models in terms of kinetic energy of the flows. As the kinetic energy is agnostic 
to the direction of the flow, the residual errors of the different variants do not show up as clearly 
as in the other evaluations.
However, while the density and kinetic energy improvements are 
smaller than those for the velocity fields, the \sol{128} model nonetheless
clearly outperforms the other variants.

\begin{figure}[tb]
  \centering
  \begin{overpic}[width=0.96\linewidth]{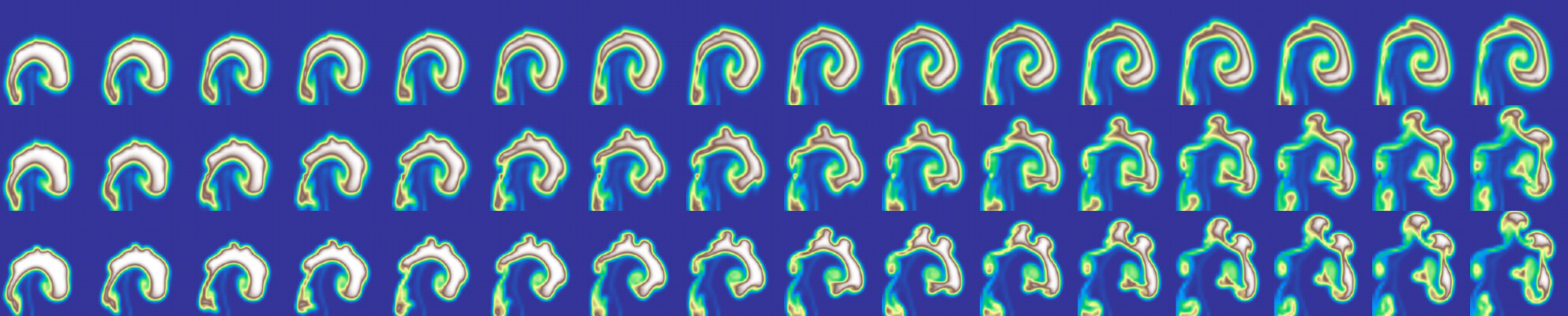}
    \put(0,18)      {{\color{white}(a)}}
    \put(-1.6,0)    {\scriptsize\rotatebox{90}{\makebox[6.71\unitlength]{Ref.}}}
    \put(-1.6,6.71){\scriptsize\rotatebox{90}{\makebox[6.71\unitlength]{\sol{128}}}}
    \put(-1.6,13.43){\scriptsize\rotatebox{90}{\makebox[6.71\unitlength]{SRC}}}
    \put(100.5,0){\includegraphics[height=6.71\unitlength]{figs/summary-legend-buoy}}
  \end{overpic}
  \\\vspace{0.5em}
  \begin{overpic}[width=0.96\linewidth]{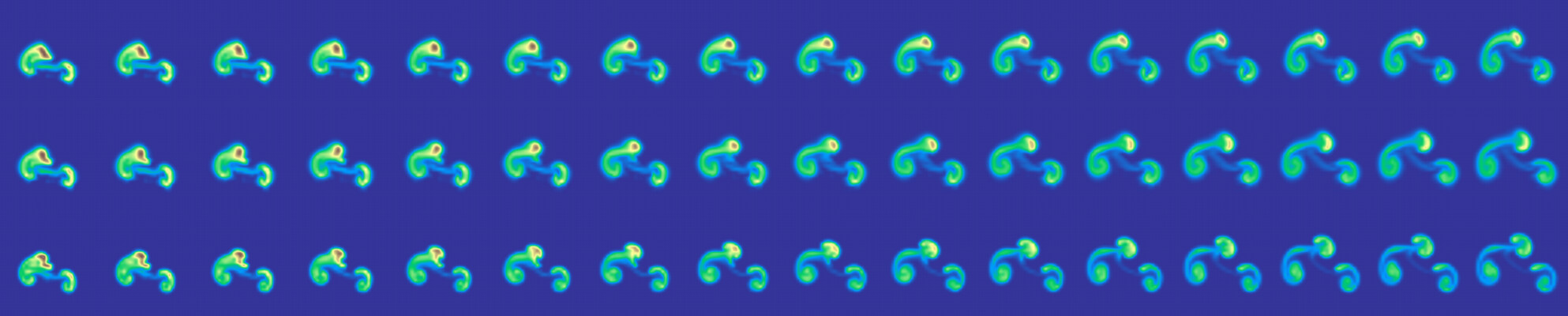}
    \put(0,18)      {{\color{white}(b)}}
    \put(-1.6,0)    {\scriptsize\rotatebox{90}{\makebox[6.71\unitlength]{Ref.}}}
    \put(-1.6,6.71){\scriptsize\rotatebox{90}{\makebox[6.71\unitlength]{\sol{128}}}}
    \put(-1.6,13.43){\scriptsize\rotatebox{90}{\makebox[6.71\unitlength]{SRC}}}
    \put(100.5,0){\includegraphics[height=6.71\unitlength]{figs/summary-legend-buoy}}
  \end{overpic}
  \\\vspace{0.5em}
  \begin{overpic}[width=0.96\linewidth]{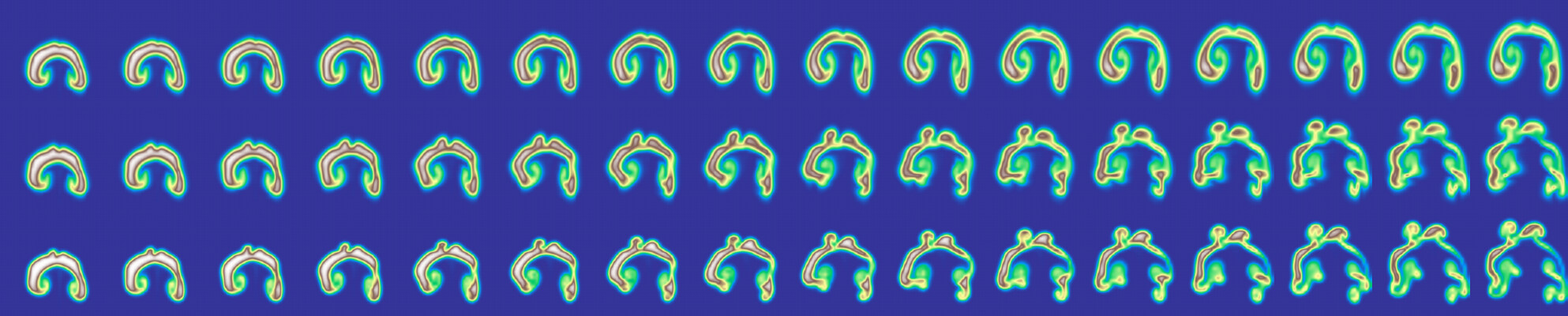}
    \put(0,18)      {{\color{white}(c)}}
    \put(-1.6,0)    {\scriptsize\rotatebox{90}{\makebox[6.71\unitlength]{Ref.}}}
    \put(-1.6,6.71){\scriptsize\rotatebox{90}{\makebox[6.71\unitlength]{\sol{128}}}}
    \put(-1.6,13.43){\scriptsize\rotatebox{90}{\makebox[6.71\unitlength]{SRC}}}
    \put(100.5,0){\includegraphics[height=6.71\unitlength]{figs/summary-legend-buoy}}
  \end{overpic}
  \caption{Several time steps $t \in \{50, 60, \cdots, 200\}$ of three buoyancy-driven fluid flow test cases
    (a)-(c).}
  \label{fig:appx:smoke-images-test}
\end{figure}

Visualized evolutions of several test simulations 
are shown in \myfigref{fig:appx:smoke-images-test}. Here, the bi-modal nature 
of the test data with smaller (b) and larger (a,c) initial marker density configurations
is shown. The different initial conditions lead to smaller and larger average velocities
and, hence, highlight that the trained model generalizes very well.


\paragraph{Ablations}
An evaluation of different neural network architectures for the buoyancy-driven 
flows with \sol{2} interaction illustrates how improvements stagnate beyond a certain network size and depth.
For example, a model with more than 100k weights and almost three times the size of the regular 
model only yields an improvement of 3.6\%. 
Another increase by a factor of four only gives 0.3\% improvement.
The corresponding graphs can be found in \myfigref{fig:appx:smoke-sizes}.
Decreasing the network size, on the other hand, yields a performance that is
8.7\% lower or even more for the smallest model. This motivates our choice 
to focus on the architecture with 36k trainable parameters, which was 
used for all other test with the buoyancy-driven flows.

As discussed in the main text, we also evaluated a method proposed by 
Sanchez et al. \cite{sanchez2020gns} to perturb inputs to network with noise 
in order to stabilize predictions. This approach shares our goal to 
reduce the shift of distributions for the input data such that the trained 
networks can produce more reliable estimates 
as they encounter new inputs
at inference time. However, in contrast to the Lagrangian graph-based physics
predictions, the added noise did not lead to large gains in our context.
We test a variety of trained \sol{2} networks for which noise was injected
into the input features, i.e., cell-wise samples of velocity and
marker density, from a component-wise normal distribution $\mathcal{N}(0,\sigma)$
with standard deviation $\sigma$.

Details of the results are visualized in \myfigref{fig:appx:smoke-noise}.
As can be seen in the results,
there is only a slight positive effect across a wide range of different noise strengths.
The networks with $\sigma \sim 10^{-4}$ show the best results. 
However, the improvements of up to 34.6\% via noise perturbations
are surpassed by the \sol{n} models, where the best one yields an improvement of 59.8\%.
%
We think that the gains of our interacting model compared to injecting noise come from
the systematic improvements of the SOL training, which potentially 
provides more reliable inputs at training time than stochastic perturbations.
The fully convolutional nature of the networks additionally provides 
regularization at training time.

We have also evaluated how sub-optimal choices for solver interactions 
affect the inference performance.
We train several NON models that are allowed 
to evolve for $n$ time steps without interaction,
while computing a regular \new{$L^2$ loss} via \myeqref{eq:non-min}. These versions 
are denoted with \non{dn} for $n$ steps of diverging evolution. 
In addition, we evaluate a model \pre{\text{SR}} using a pre-computed interaction without 
temporal regularization (i.e., only spatial) and one version (\pre{\text{F}}) that uses the full spatiotemporal 
regularization without a density threshold; i.e., it requires several times more pre-computation 
by solving the Lagrange-multiplier minimization for the full spatial domains.
Especially, the \non{dn} variants perform badly and exhibit large errors, with \non{d8} significantly distorting 
the flow behavior, instead of improving it. 
The corresponding evaluations 
are visualized in \myfigref{fig:appx:smoke-plot-diverging}.
It is likewise apparent that the additional PRE variants deteriorate the ability of the ANNs to 
correct the numerical errors of the source simulations.

To summarize, despite the complexity of the buoyancy-driven flows and 
the difficult reference trajectories produced by a higher-order PDE solver,
the numerical errors of the source simulation can be reduced very successfully 
by training with the solver in the training loop.

\newcommand{\myFigHSmoke}{0.25\columnwidth}

\newcommand{\myFigHSmokeS}{0.175\columnwidth}
\begin{figure}[htb]
  \centering
  \subcaptionbox{Velocity error\label{fig:appx:smoke-plot:models-vel}}{\includegraphics[height=\myFigHSmokeS,page=2]{figs/out-vgf-jmsmaller-steps-smoke-uns-jmsmaller-msteps}}
  \subcaptionbox{Marker advection error\label{fig:appx:smoke-plot:models-dens}}{\includegraphics[height=\myFigHSmokeS,page=1]{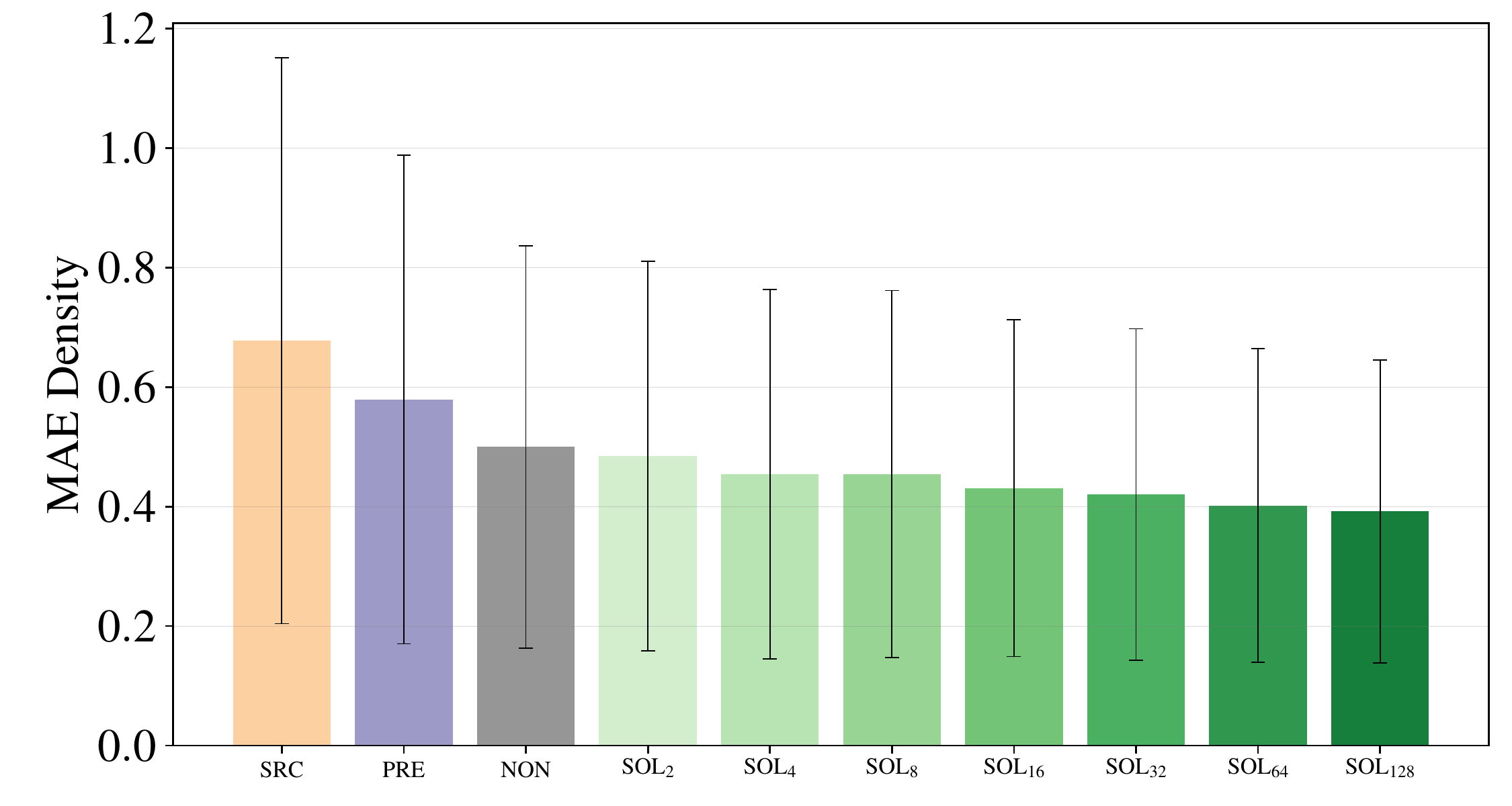}}
  \subcaptionbox{Kinetic energy error\label{fig:appx:smoke-plot:models-ke}}{\includegraphics[height=\myFigHSmokeS,page=4]{figs/out-vgf-jmsmaller-steps-smoke-uns-jmsmaller-msteps}}
  \\
  \subcaptionbox{Velocity improvement\label{fig:appx:smoke-plot:models-vel-impr}}{\includegraphics[height=\myFigHSmokeS,page=6]{figs/out-vgf-jmsmaller-steps-smoke-uns-jmsmaller-msteps}}
  \subcaptionbox{Impr. of marker advection\label{fig:appx:smoke-plot:models-den-impr}}{\includegraphics[height=\myFigHSmokeS,page=5]{figs/out-vgf-jmsmaller-steps-smoke-uns-jmsmaller-msteps}}
  \subcaptionbox{Impr. of kin. energy\label{fig:appx:smoke-plot:models-ke-impr}}{\includegraphics[height=\myFigHSmokeS,page=7]{figs/out-vgf-jmsmaller-steps-smoke-uns-jmsmaller-msteps}}
  \caption{Velocity, marker advection, and kinetic energy errors for different
    models, especially for different SOL versions with increasing look-ahead. In
    the second row, we show improvements relative to the source version SRC.}
  \label{fig:appx:smoke-plot}
\end{figure}


\begin{figure}[htb]
  \centering
  \subcaptionbox{Trainable weights\label{fig:appx:smoke-plot:sizesw}}{\includegraphics[height=\myFigHSmoke,page=9]{figs/out-vgf-jmsmaller-steps-smoke-uns-varymodel}}
  \subcaptionbox{Velocity error\label{fig:appx:smoke-plot:sizesv}}{\includegraphics[height=\myFigHSmoke,page=2]{figs/out-vgf-jmsmaller-steps-smoke-uns-varymodel}}
  \subcaptionbox{Marker error\label{fig:appx:smoke-plot:sizesd}}{\includegraphics[height=\myFigHSmoke,page=1]{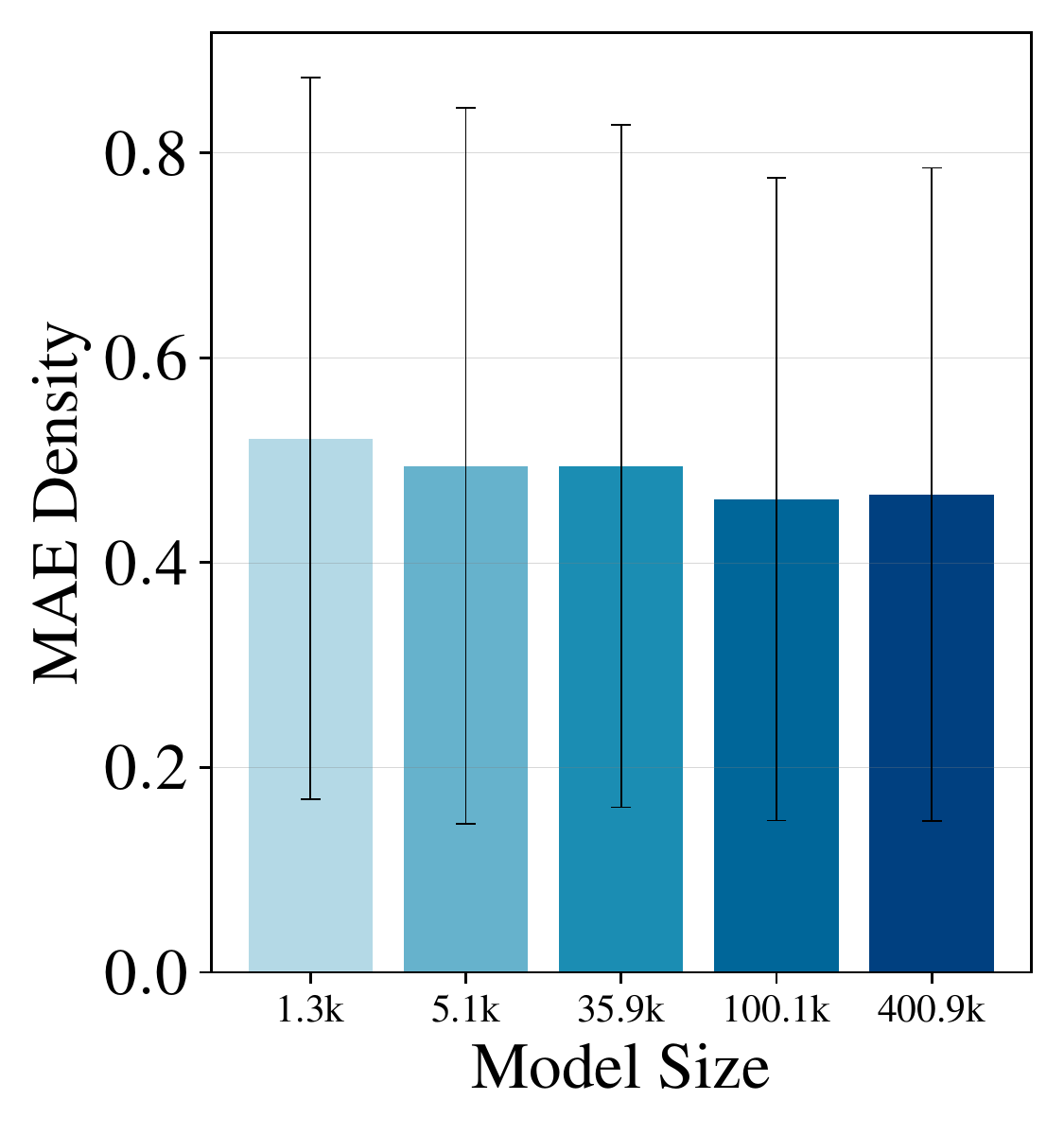}}
  \subcaptionbox{Velocity improvement\label{fig:appx:smoke-plot:sizesvi}}{\includegraphics[height=\myFigHSmoke,page=5]{figs/out-vgf-jmsmaller-steps-smoke-uns-varymodel}}
  \caption{\sol{2} training with different architectures that strongly vary the
    number of trainable parameters (a). While the smaller two models lead to a
    clear drop in accuracy, the larger two architectures yield small gains
    despite the increased weight count.}
  \label{fig:appx:smoke-sizes}
\end{figure}

\newcommand{\myFigHSmokeN}{0.22\columnwidth}
\begin{figure}[htb]
  \centering
  \subcaptionbox{Velocity improvement \label{fig:appx:smoke-plot:noisev}}{\includegraphics[height=\myFigHSmokeN,page=5]{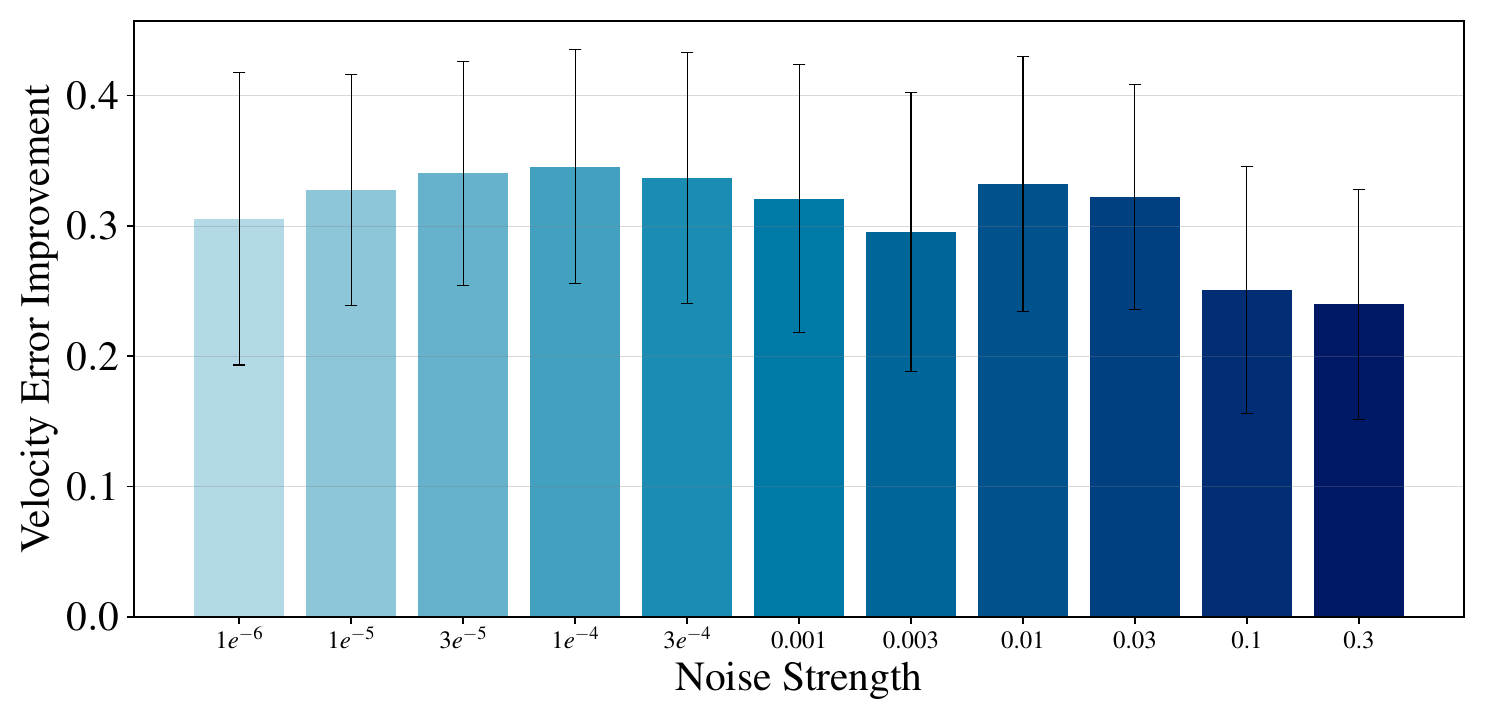}}
  \subcaptionbox{Improvement of marker advection  \label{fig:appx:smoke-plot:noised}}{\includegraphics[height=\myFigHSmokeN,page=4]{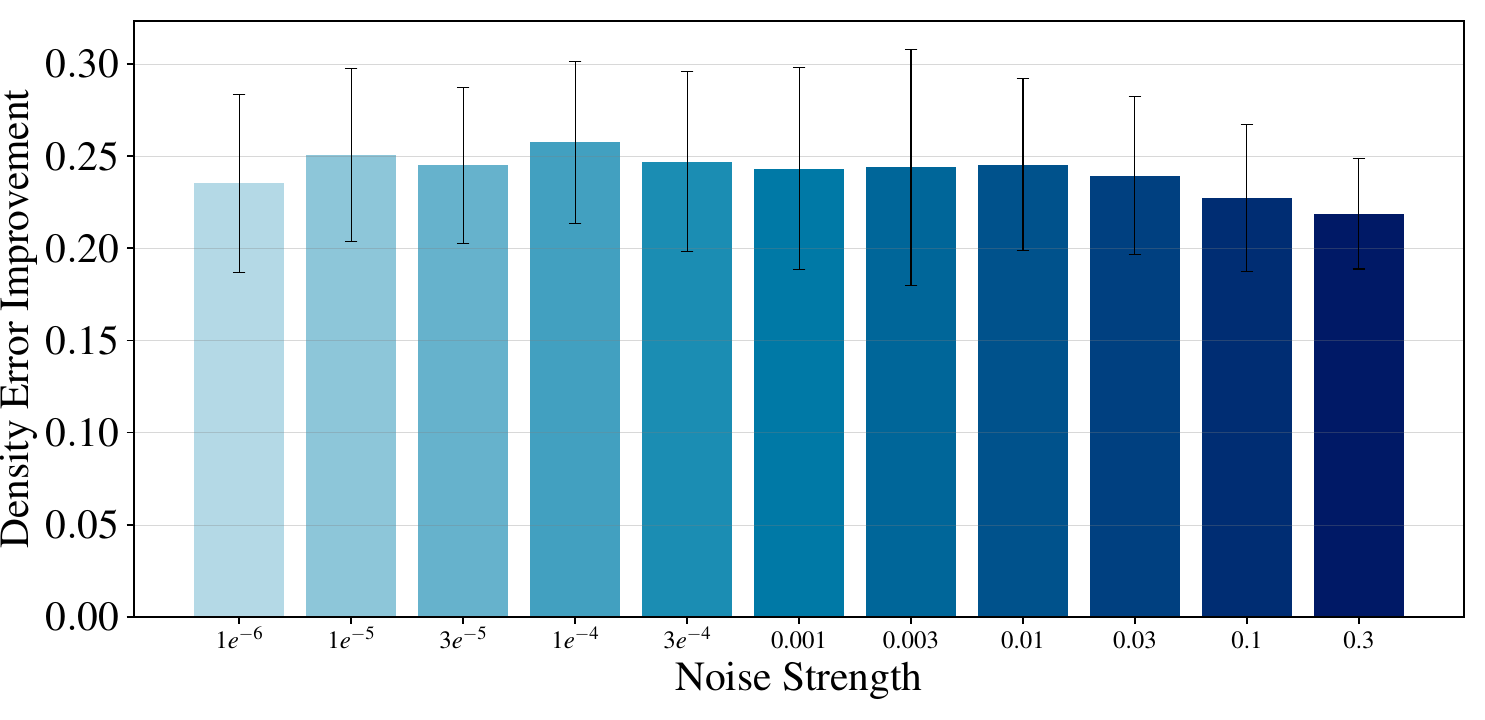}}
  \caption{Varying levels of noise injected into the input features for \sol{2}
    at training time. While values around $10^{-4}$ lead to slight positive
    effects, the improvements are negligible compared to those achievable by the
    SOL variants.}
  \label{fig:appx:smoke-noise}
\end{figure}

\newcommand{\myFigHSmokeB}{0.22\columnwidth}

\begin{figure}[htb]
  \centering
  \subcaptionbox{Velocity error \label{fig:appx:smoke-plot:vel}}{\includegraphics[height=\myFigHSmokeB,page=2]{figs/out-vgf-jmsmaller-diverging-beta0}}
  \subcaptionbox{Density error \label{fig:appx:smoke-plot:den}}{\includegraphics[height=\myFigHSmokeB,page=1]{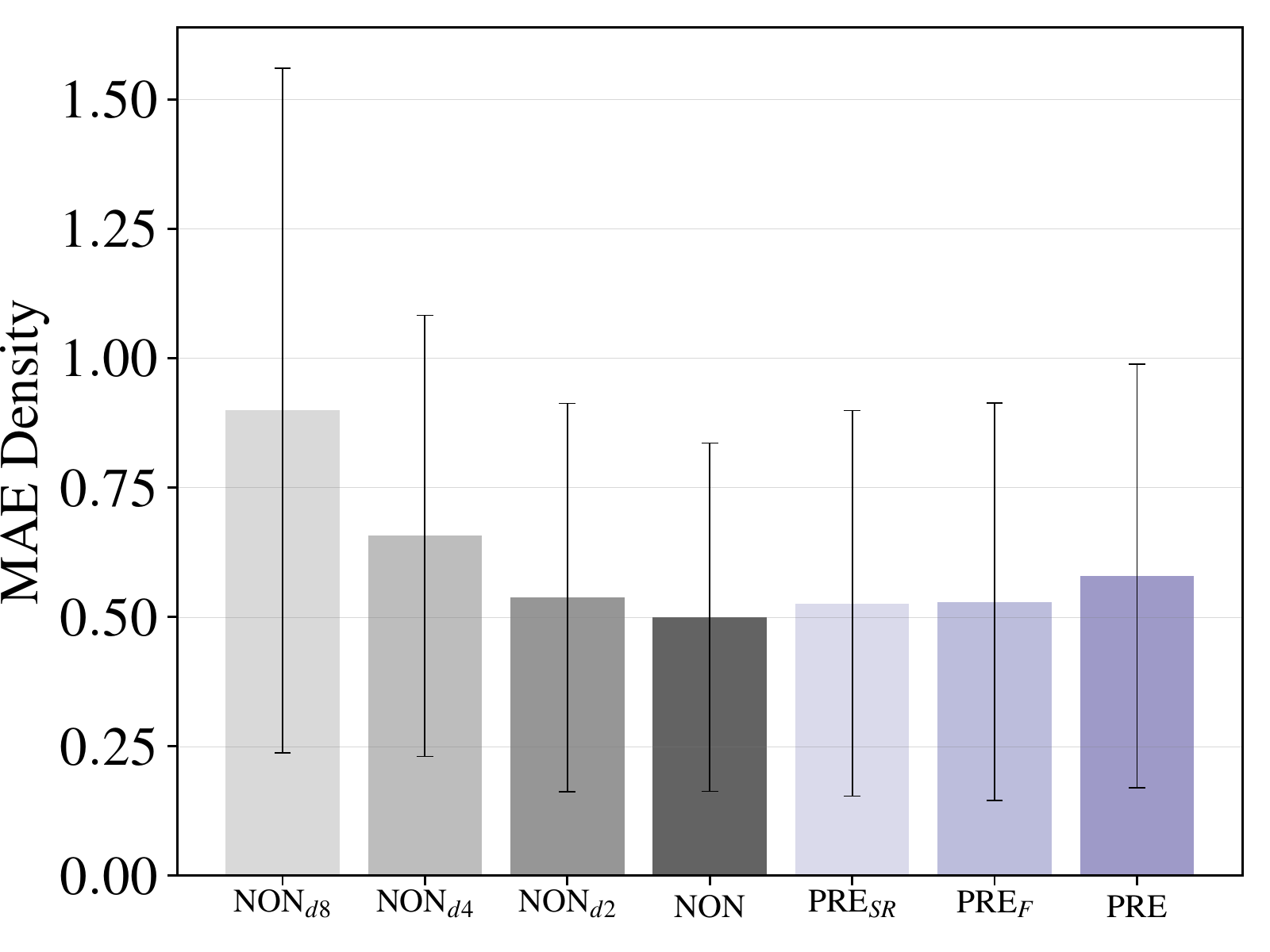}} 
  \subcaptionbox{Velocity improvement \label{fig:appx:smoke-plot:imp}}{\includegraphics[height=\myFigHSmokeB,page=5]{figs/out-vgf-jmsmaller-diverging-beta0}}
  \caption{A comparison of models trained with a variety of sub-optimal
    interaction schemes for the buoyancy scenario. \non{dn} allows
    non-interacting models to evolve and diverge over $n$ steps, while \pre{SR}
    employs only spatial regularization in the pre-computation. \pre{F}
    resembles PRE, but was trained without a density threshold. Especially, the
    changes relative to SRC in (c) highlight that the \non{dn} variants have a
    negative effect.}
  \label{fig:appx:smoke-plot-diverging}
\end{figure}

\begin{table}[tb]
  \caption{Quantitative evaluation of models for the buoyancy-driven flow
    scenario. M$_{XS,S,L,XL}$ denote different model sizes, while
    $\sigma_{1,2,3}$ denote models trained with noise of
    $\sigma=10^{-3,-4,-5}$.}
  \label{tab:smoke}
  \begin{center}
    \begin{tabular}{ccccccccc}
      \toprule
      \textbf{Quantity} & \multicolumn{8}{c}{\textbf{MAE Velocity}, Mean (std. dev.)}                      \\
      \cmidrule{2-9}
                     & SRC     & NON     & PRE      & \sol{2} & \sol{16} & \sol{32} & \sol{64} & \sol{128} \\
      \midrule                                                            
      Velocity       &  1.590  &  1.079  &  1.373  &  1.027  &  0.859  &  0.775  &  0.695  &  0.620  \\
                     & (1.032) & (0.658) & (0.985) & (0.656) & (0.539) & (0.482) & (0.420) & (0.389) \\
      Marker $d$     &  0.677  &  0.499  &  0.579  &  0.484  &  0.430  &  0.419  &  0.401  &  0.391  \\
                     & (0.473) & (0.336) & (0.409) & (0.325) & (0.281) & (0.277) & (0.262) & (0.253) \\
      \midrule
                     &M$_{XS}$ &M$_{S}$  &  M$_{L}$& M$_{XL}$&$\sigma_1$&$\sigma_2$&$\sigma_3$& \non{d4} \\
      \midrule                                                           
      Velocity       &  1.228  &  1.193  &  0.982  &  0.969  &  1.070  &  1.056  &  1.078  &  3.196  \\
                     & (0.746) & (0.826) & (0.646) & (0.626) & (0.683) & (0.700) & (0.706) & (1.404) \\
      Marker $d$     &  0.521  &  0.494  &  0.461  &  0.466  &  0.503  &  0.496  &  0.503  &  0.656  \\
                     & (0.352) & (0.349) & (0.313) & (0.318) & (0.341) & (0.339) & (0.345) & (0.426) \\
      \bottomrule
    \end{tabular}
  \end{center}
\end{table}


\clearpage

\subsection{Forced Advection-Diffusion}\label{app:expBurgers}

In the forced advection-diffusion scenario, we target a PDE environment 
with a constant, randomized forcing term. 
This forcing continuously injects energy into the dissipative system
and takes the form of a spectrum of parametrized bands of sine waves.
In this scenario, we target Burgers' equation. It represents
a well-studied advection-diffusion PDE:
\begin{eqnarray}
  \label{eq:burgers}
  \new{\frac{\partial u_x}{\partial{t}} + \vu \cdot \nabla u_x =
  \nu \nabla\cdot \nabla u_x + g_x(t)}, 
  \quad 
  \new{\frac{\partial u_y}{\partial{t}} + \vu \cdot \nabla u_y =
  \nu \nabla\cdot \nabla u_y + g_y(t)},
\end{eqnarray}
where $\nu$ and $\mathbf{g}$ denote diffusion constant and external forces, respectively.
Our setup resembles a 2D variant of the tests employed
by the work on learning data-driven discretizations \cite{barsinai2019data};
correspondingly, we extend the forcing terms described there to 2D.
We generate the forces from 20 overlapping sine functions each with a random direction, amplitude, and phase shift:
\begin{equation}
    g_x(t) = \sum_{i=1}^{20} \cos(\alpha_i) a_i \sin(\omega_i t - k x + \phi_i),
    \quad
    g_y(t) = \sum_{i=1}^{20} \sin(\alpha_i) a_i \sin(\omega_i t - k x + \phi_i).
\end{equation}
This PDE scenario does not involve any equality constraints, i.e., $\mM=0$.

Similar to the previous scenarios, we discretize the system 
on a staggered grid and compute the
advection operator with a semi-Lagrangian scheme \cite{stam1999}.
The domain has a square, normalized size of $1 \times 1$
with reference trajectories computed via a resolution of
$d_{r,x}\!=\!d_{r,y}\!=\!128$. The source domain correspondingly uses 
$d_{s,x}\!=\!d_{s,y}\!=\!32$.

\paragraph{Training Procedure and Results}

\begin{figure}[tb]
  \centering
  \begin{overpic}[width=0.96\linewidth]{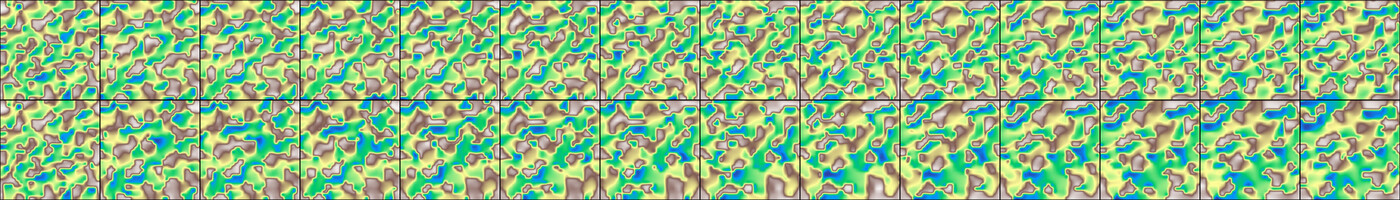}
    \put(-1.6,0)   {\scriptsize\rotatebox{90}{\makebox[7.14\unitlength]{Ref.}}}
    \put(-1.6,7.14){\scriptsize\rotatebox{90}{\makebox[7.14\unitlength]{SRC}}}
    \put(100.5,0){\includegraphics[height=7.14\unitlength]{figs/summary-legend-burgers}}
  \end{overpic}
  \caption{An example sequence from the training data set of the forced
    advection-diffusion test case.}
  \label{fig:appx:burgers-images-train}
\end{figure}

As training data, ten simulations of 200 steps each are used.
An example sequence of the data is shown in \myfigref{fig:appx:burgers-images-train}.
The SOL and NON models are trained for 38.4k steps with a batch size of \new{five}
with a learning rate of $10^{-4}$,
while the 
PRE model is trained for 25k steps with a batch size of 32
using an initial learning rate of $10^{-3}$ that was lowered to 
$5\times10^{-7}$ over the course of the training. 
The PRE model additionally 
uses 5\% of the training data set for validation.
The test data set contains five cases with different initial conditions 
and force fields over the course of 200 time steps.
All models use a neural network architecture 
with five ResBlocks with 32 features each.

\begin{figure}[tb]
  \centering
  \begin{overpic}[width=0.96\linewidth]{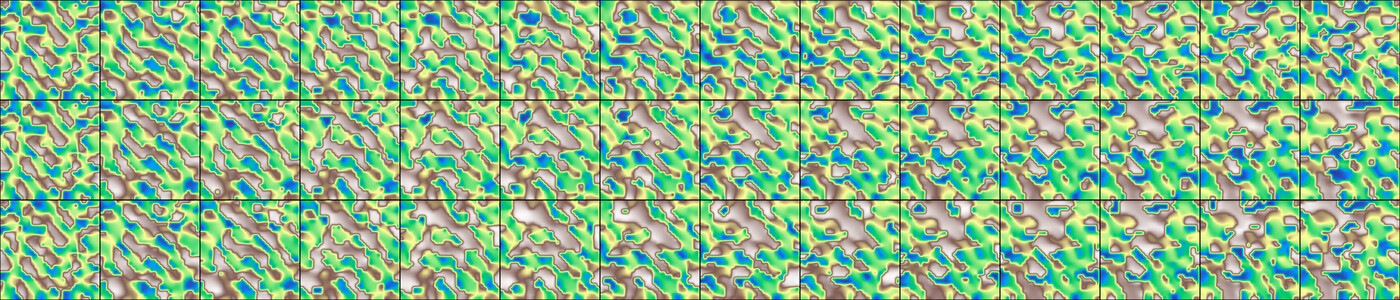}
    \put(100,19)    {(a)}
    \put(-1.6,0)    {\scriptsize\rotatebox{90}{\makebox[7.14\unitlength]{Ref.}}}
    \put(-1.6,7.14) {\scriptsize\rotatebox{90}{\makebox[7.14\unitlength]{\sol{2}}}}
    \put(-1.6,14.28){\scriptsize\rotatebox{90}{\makebox[7.14\unitlength]{SRC}}}
    \put(100.5,0){\includegraphics[height=7.14\unitlength]{figs/summary-legend-burgers}}
  \end{overpic}
  \\\vspace{0.5em}
  \begin{overpic}[width=0.96\linewidth]{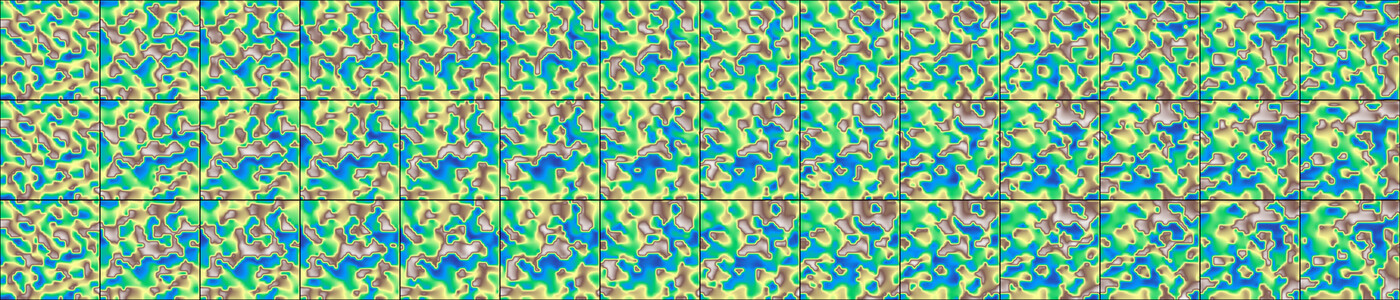}
    \put(100,19)    {(b)}
    \put(-1.6,0)    {\scriptsize\rotatebox{90}{\makebox[7.14\unitlength]{Ref.}}}
    \put(-1.6,7.14) {\scriptsize\rotatebox{90}{\makebox[7.14\unitlength]{\sol{2}}}}
    \put(-1.6,14.28){\scriptsize\rotatebox{90}{\makebox[7.14\unitlength]{SRC}}}
    \put(100.5,0){\includegraphics[height=7.14\unitlength]{figs/summary-legend-burgers}}
  \end{overpic}
  \\\vspace{0.5em}
  \begin{overpic}[width=0.96\linewidth]{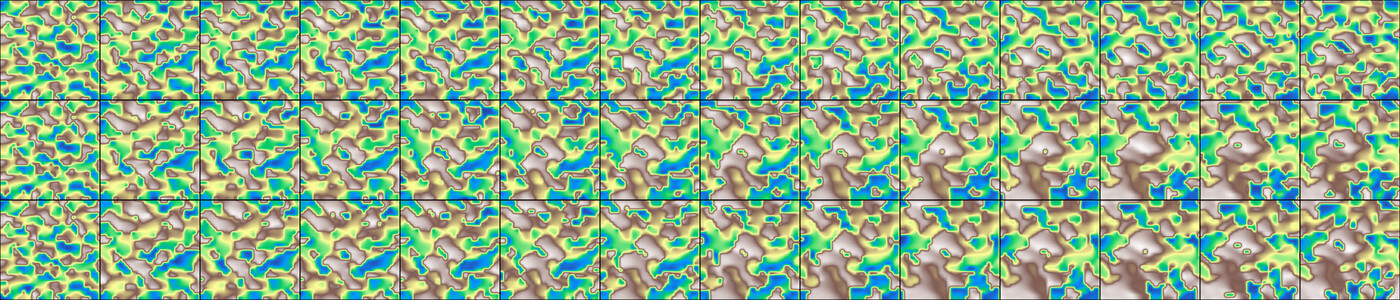}
    \put(100,19)    {(c)}
    \put(-1.6,0)    {\scriptsize\rotatebox{90}{\makebox[7.14\unitlength]{Ref.}}}
    \put(-1.6,7.14) {\scriptsize\rotatebox{90}{\makebox[7.14\unitlength]{\sol{2}}}}
    \put(-1.6,14.28){\scriptsize\rotatebox{90}{\makebox[7.14\unitlength]{SRC}}}
    \put(100.5,0){\includegraphics[height=7.14\unitlength]{figs/summary-legend-burgers}}
  \end{overpic}
  \caption{
    Time steps of three test cases (a)-(c) from the forced advection-diffusion scenario.
    }
  \label{fig:appx:burgers-images-test}
\end{figure}

As summarized in the main text, the learned correction functions can 
significantly decrease the numerical errors of the source simulation. 
Across the different test cases (partly shown in \myfigref{fig:appx:burgers-images-test}),
the best models achieve a 
reduction by over 67\%. The corresponding MAE measurements 
are given in \mytabref{tab:burgers}, 
and \myfigref{fig:appx:burgers-plot}
provides an overview of the performance per test case.
While the PRE model shows a lower performance, most likely due to an overly 
strong temporal regularization, the NON model is close to the best 
SOL model in this case with an MAE of 0.159 compared to 0.148 for \sol{2}.
Interestingly, this behavior matches the results of Bar-Sinai et al. \cite{barsinai2019data}.
They experimented with up to \new{eight} recurrent steps 
of a 1D Burgers' simulation, but did not report significant advantages from training 
with the 1D solver in the loop.

\new{
  In contrast, we found that more interactions show their advantage in a
  deterministic scenario, where we exclude the external forces from the Burgers'
  equation above, i.e., \myeqref{eq:burgers}. 
  As this versions exhibits less chaotic behavior,
  the SRC version generally shows smaller errors
  compared to the SRC version in the forced scenario. 
  The SOL versions now yield 
  further improvements when trained with more look-ahead:
  \sol{4} yields an improvement of 10\% over SRC, \sol{16} yields 12\%,
  while the \sol{32} version reduces the error by 17\%. 
  \mytabref{tab:burgers} shows the corresponding MAE measurements.
}

Our results highlights that deep learning via physical 
simulations works particularly well when the ANNs can actually learn 
to predict the behavior of the dynamics and, thus, compensate for the 
numerical errors that will occur. If, on the other hand, external and 
unpredictable influences such as the randomized forcing terms dominate the behavior,
the model has a reduced chance to predict the right correction function.


\begin{table}[htb]
  \caption{Quantitative evaluation of different models for the forced
    advection-diffusion scenario. \new{MAE values without forcing are given with
      a $\times 100$ factor.}}
  \label{tab:burgers}
  \begin{center}
    \begin{tabular}{ccccccc}
      \toprule
      \multicolumn{7}{c}{\textbf{MAE Velocity}, Mean (std. dev.)}                       \\
      \midrule
      With forcing    & SRC     & PRE     & NON     & SOL$_2$ & SOL$_4$    & SOL$_8$    \\
      \cmidrule{2-7}
                      & 0.248   & 0.218   & 0.159   & 0.148   & 0.152      & 0.158      \\
                      & (0.019) & (0.017) & (0.015) & (0.016) & (0.015)    & (0.017)    \\
      \midrule
      Without forcing & SRC     & NON     & SOL$_4$ & SOL$_8$ & SOL$_{16}$ & SOL$_{32}$ \\      
      \cmidrule{2-7}
      $(\times 100)$  & 0.306   & 0.272   & 0.276   & 0.277   & 0.268      & 0.253      \\
                      & (0.020) & (0.028) & (0.037) & (0.040) & (0.030)    & (0.020)    \\
      \bottomrule
    \end{tabular}
  \end{center}
\end{table}

\newcommand{\myFigHBurgers}{0.35\columnwidth}
\begin{figure}[htb]
  \centering
  \subcaptionbox{Velocity error per test case}{\includegraphics[height=\myFigHBurgers,page=1]{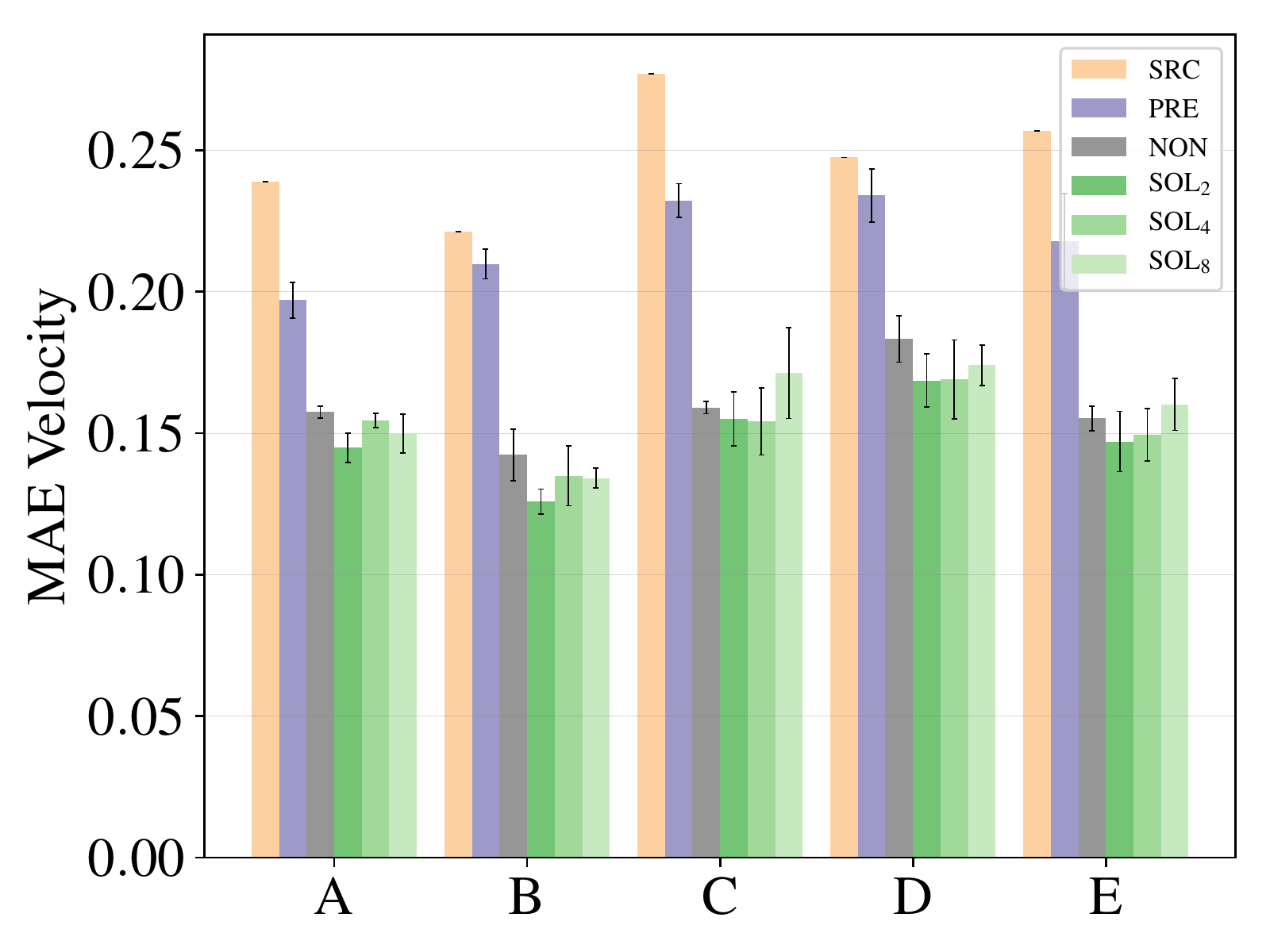}}
  \subcaptionbox{Velocity improvement (relative to SRC) per test case}{\includegraphics[height=\myFigHBurgers,page=4]{figs/summary-burgers-plot}}
  \caption{Separate evaluations for five different test cases of the forced advection-diffusion scenario.}
  \label{fig:appx:burgers-plot}
\end{figure}




\clearpage

\subsection{Inference of Initial Guesses for Conjugate Gradient Solvers}\label{app:expCgSolver}

\newcommand{\res}{\new{r}} 
\newcommand{\press}{\new{p}} 
\newcommand{\pressh}{\new{\hat{\press}}} 

In this section, we investigate the interaction of learning models with conjugate gradient (CG) solvers \cite{hestenes1952cg}. 
We target Poisson problems, 
which often arise many PDEs, e.g., in electrostatics or 
in fluid flow problems where the pressure $\press$ is computed
via $\nabla \cdot \nabla \press = \nabla \cdot \vu$.
Specifically, we explore the iteration behavior of the CG solver given an initial state predicted by a trained model.
To this end, we compare three main methods: A solver-in-the-loop (\sol{n}) approach, a non-interacting supervised approach (NON),
and a differentiable physics-based (\sol{\text{DIV}}), which is trained to directly minimize the PDE residual.
%
In general, the CG solver iterations converge toward a reference pressure field $\press$ such that $\mA \press = \nabla \cdot \vu$ with
$\mA=\nabla \cdot \nabla$. For an intermediate solution $\pressh$, the residual $\res = \nabla \cdot \vu - \mA \pressh$ measures how far away the approximated pressure $\pressh$ is from the true solution.
Thus, as the solver converges, $\res$ decreases and the difference $\pressh - \press$ converges to zero. In the following, we employ the neural network $\corr$ to infer 
a pressure field given a velocity sample $\vu$, i.e., $\pressh=\corr(\vu)$.
We focus on 2D cases, i.e., $\vu{} \in \mathbb{R}^{2 \times d_{x} \times d_{y}}$ and
$\press,\res \in \mathbb{R}^{d_{x} \times d_{y}}$.

\paragraph{Loss Functions}
The NON version employs a regular supervised loss, i.e., the difference of the predicted pressure $\hat{p}$ from the pre-computed reference pressure $p$ for $j$ different samples:
\begin{equation} \label{eq:cg_loss1}
    \new{\loss_{\text{NON}} = \Vert \corr(\vu) - \press \Vert^2}.
\end{equation}


We additionally compare to a variant that is often referred to as \emph{unsupervised} in previous work,
and which is in line with other physics-based or physics-informed loss constructions \cite{raissi2018hiddenfluid,sirignano2018dgm}.
Specifically, the \sol{\text{DIV}} version
replicates the setup described in
\cite{tompson2017} and uses the PDE $\nabla^2 p - \nabla \cdot \vu = 0$ as loss for the training of a neural network.
Given an input velocity $\vu^*$, the goal is to infer a 
pressure function $\hat{p}(\nabla \cdot \vu^*)$ such that the PDE residual is minimized:
\begin{equation}
    \begin{aligned}
    \new{\loss_{\text{SOL}_{\text{DIV}}} 
        = \Vert \nabla \cdot \vu^* - \nabla\cdot\nabla \corr(\vu)  \Vert^2}.
    \end{aligned}
    \label{eq:cg_loss2}
\end{equation}
This version represents a different form of
differentiable PDE solvers, namely including them in the loss formulation, and hence we denote it with \sol{\text{DIV}}. However, due to a lack of iterating calculations for this variant,
a more appropriate name would be \emph{``solver-in-the-loss''} rather than \emph{``solver-in-the-loop''}.


As a third variant, we employ a solver-in-the-loop interaction that 
employs a differentiable CG solver and uses a learning objective 
to minimize the PDE residual after $n$ iterations of the CG solver. In this scenario,
$\pdec$ represents a linear operator, i.e., one step of the CG method 
to approximate 
$\nabla^{-2} \, (\nabla \cdot \vu)$,
and the loss function is given by:
\begin{equation} 
  \new{\loss_{\text{SOL}_{n}} = \Vert \pdec^n( \corr(\vu) ) - \corr(\vu) \Vert^2}.
  \label{eq:cg_loss3}
\end{equation}
%
%
Thus, the \sol{n} and \sol{\text{DIV}} both minimize the same residual divergence $\res$; while the \sol{\text{DIV}} loss aims to do so directly, the \sol{n} version instead sees how the iterative solver performs.
At training time, the \sol{n} variant receives gradients through $n$ iterations of
the iterative solver via back-propagation.

\paragraph{Training Procedure} 
The trained models in this section all use the same convolutional U-net architecture \cite{unet2015} with 22 layers of strided convolutions and 5$\times$5 kernels, containing around 127k trainable parameters (see \myappref{app:models} for details).
The training data set was generated using the conjugate gradient solver from the $\Phi_\textrm{Flow}$ framework \cite{holl2020}. 
It is comprised of 3k fluid simulations on a domain with $d_{x} = d_{y} = 64$ and closed boundaries. Each simulation consists of a randomly generated density and velocity field, which are integrated over time for 16 steps. 
%
Each model was trained for 300k steps with a learning rate of $2 \times 10^{-4}$ and training batch size of 16. 
The reference solutions were pre-computed with a CG solver using an accuracy threshold of $10^{-6}$ for the residual norm.

\paragraph{Results}
We now compare the different loss functions by their performance in conjunction with the CG solver.
We compute averages for 100 test cases each time, i.e., samples that were not seen at training time.
As baseline, we denote a CG solving process that starts from a zero guess as SRC.

\begin{figure}[htb]
  \centering
  \begin{overpic}[width=0.45\linewidth]{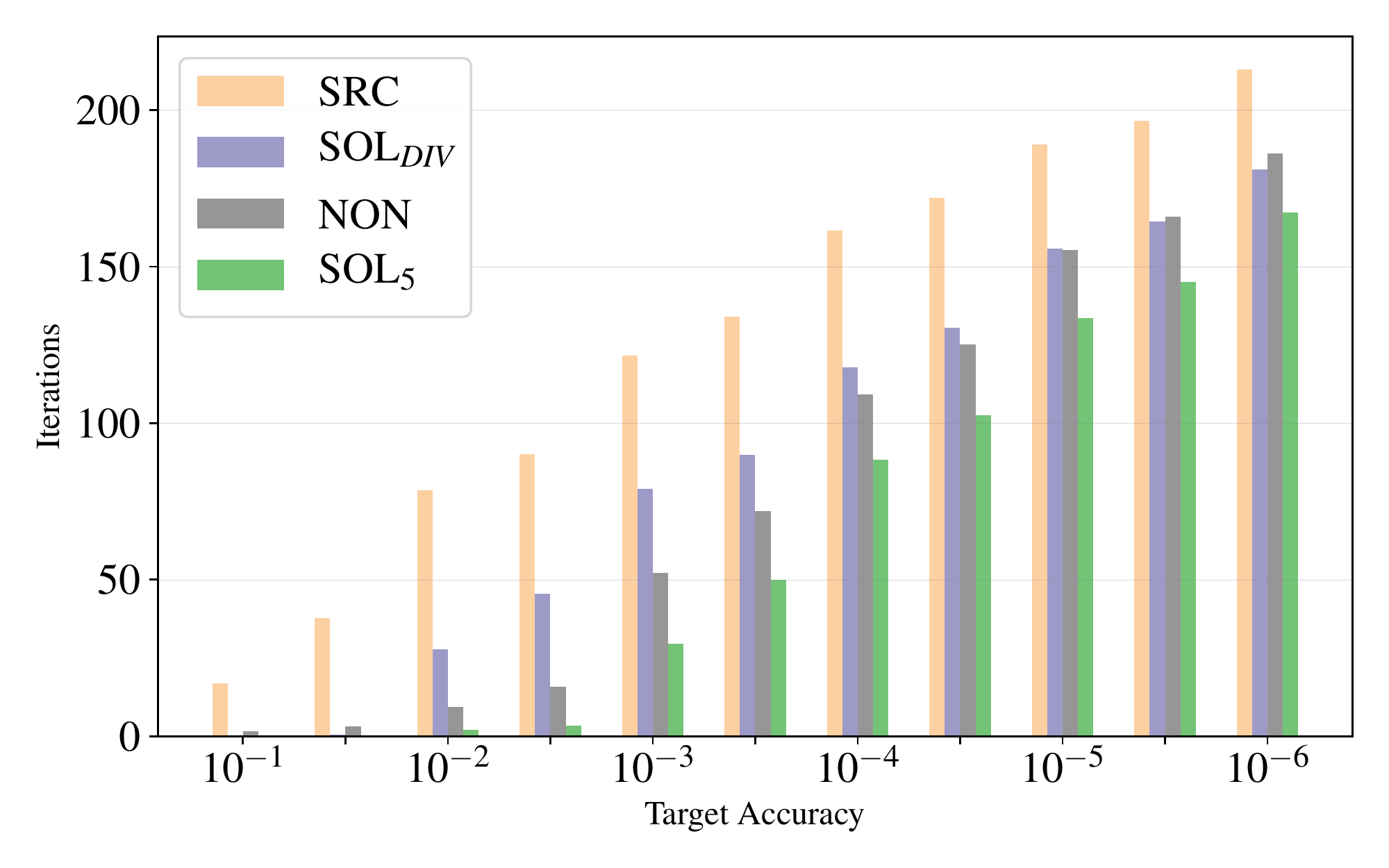}
    \put(-5, 55) {(a)}
  \end{overpic}
  \hspace{0.05\linewidth}
  \begin{overpic}[width=0.45\linewidth]{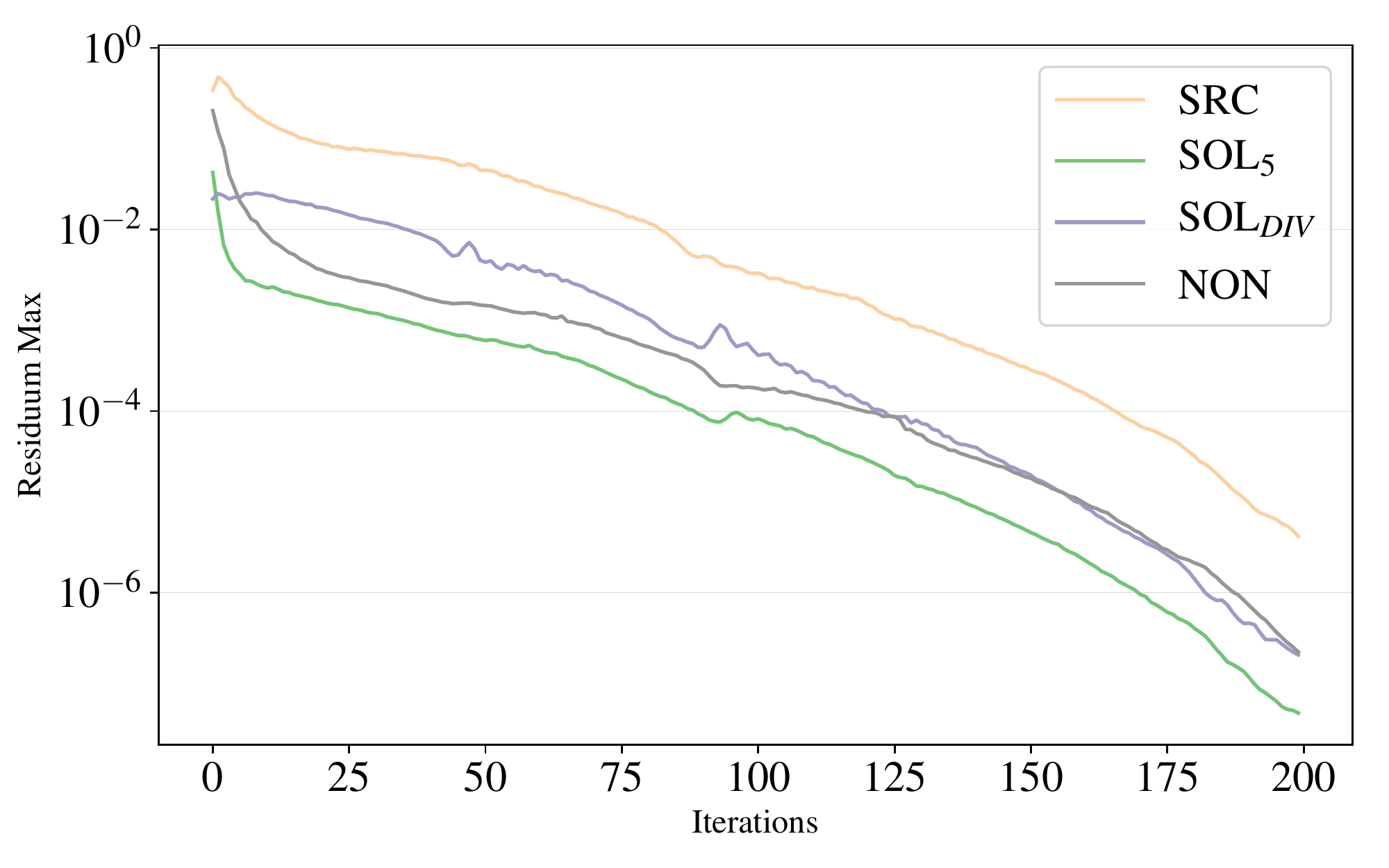}
    \put(-5, 55) {(b)}
  \end{overpic}
  \caption{(a) Iterations needed to reach target accuracy and (b) comparison of
    maximum residual error over iterations.}
  \label{fig:cgresults}
\end{figure}

We first compare how many CG iterations are required to reach a certain target accuracy given the inferred solutions by the three different types of models.
The results are shown in \mytabref{tab:cgresults} and visualized in \myfigref{fig:cgresults}.
Initially, \sol{\text{DIV}} reaches an accuracy of almost $10^{-2}$, closely followed by \sol{5}.
While the supervised NON version produces pressure predictions that seem quite close to the reference, its initial accuracy is only slightly better than the zero guess employed by SRC.
This is due to the error being measured locally per grid point, while the correctness of larger structures becomes more important after in interactions with the CG solvers.
Over the first five to ten CG iterations, the accuracy of \sol{5} improves very quickly, overtaking the other methods.
To reach an accuracy of $10^{-2}$, the CG solver requires an average of around two steps in conjunction with \sol{5}, nine steps with NON, 28 steps with \sol{\text{DIV}} and 78 steps 
starting from zero.
When running the CG solver for more iterations, the accuracy increases similarly for all methods, with \sol{5} retaining its advantage.

Comparing \sol{5} to \sol{\text{DIV}} shows the importance of training with the solver in the loop: the \sol{\text{DIV}} model does not 
receive any feedback regarding the behavior of the solver.
It predicts solutions that satisfy the loss -- measured per grid point -- but do not match the large-scale structures of the true solution.
Consequently, this task is left to the CG solver, which requires many iterations to work out the correct global solution.
The \sol{5} model, however, sees the corrections performed by the CG solver at training time and can learn to adjust its guess accordingly.

\begin{figure}[htb]
  \centering
  \begin{overpic}[width=0.56\linewidth]{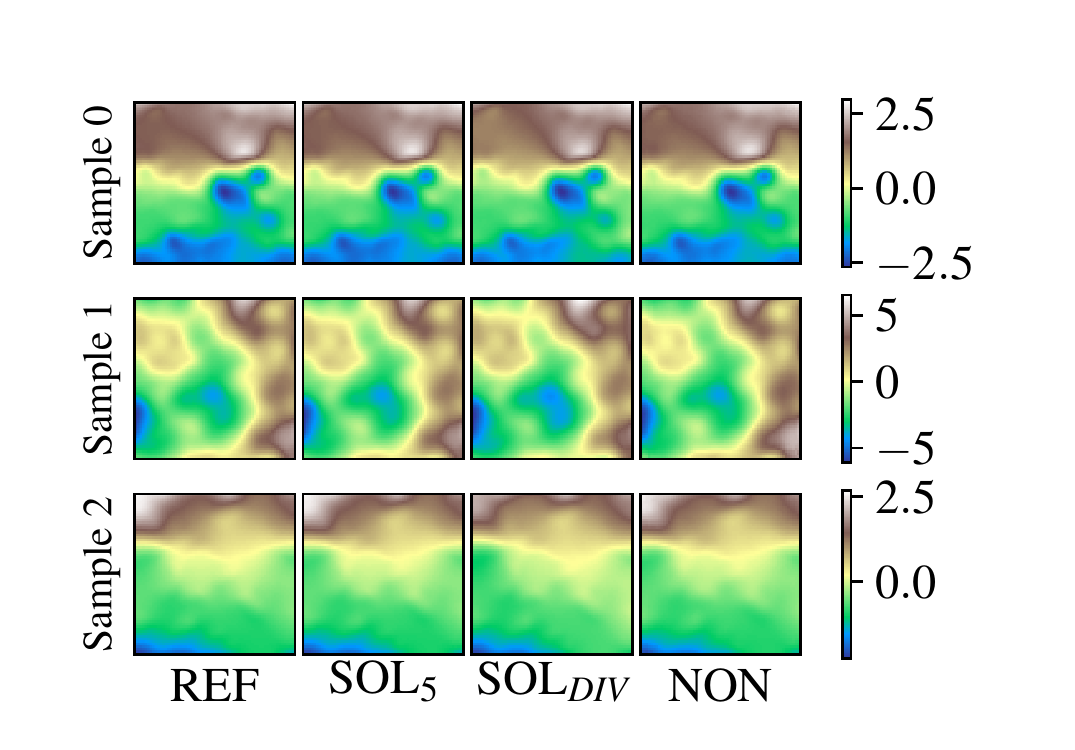}
    \put(-1, 55) {(a)}
  \end{overpic}
  \begin{overpic}[width=0.43\linewidth]{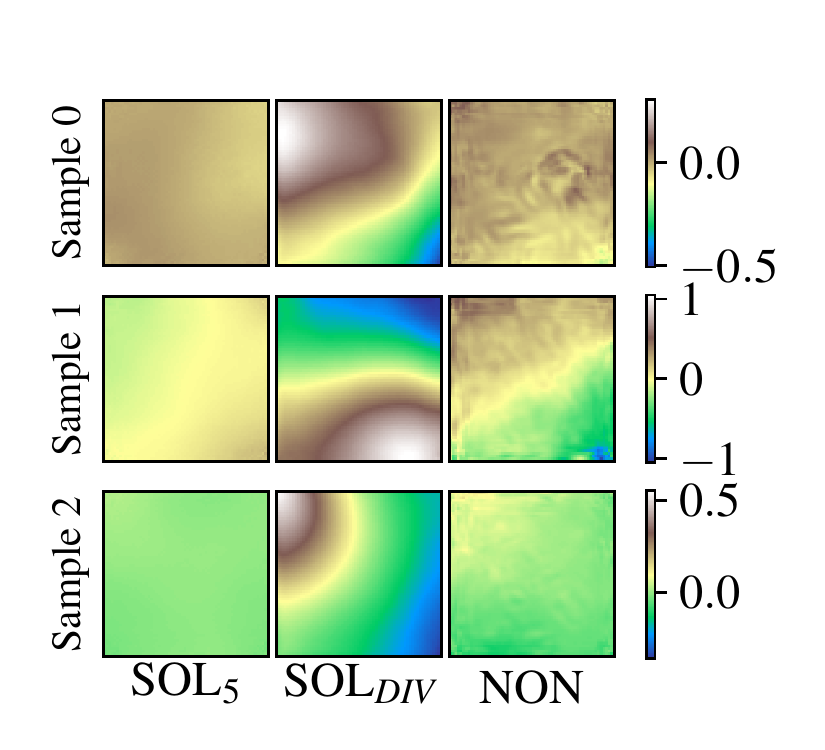}
    \put(-5, 73) {(b)}
  \end{overpic}
  \caption{(a) Sample outputs of the models and (b) difference of output from
    reference.}
  \label{fig:cg_img_results}
\end{figure}

\begin{figure}[htb]
  \centering
  \includegraphics[width=0.49\columnwidth]{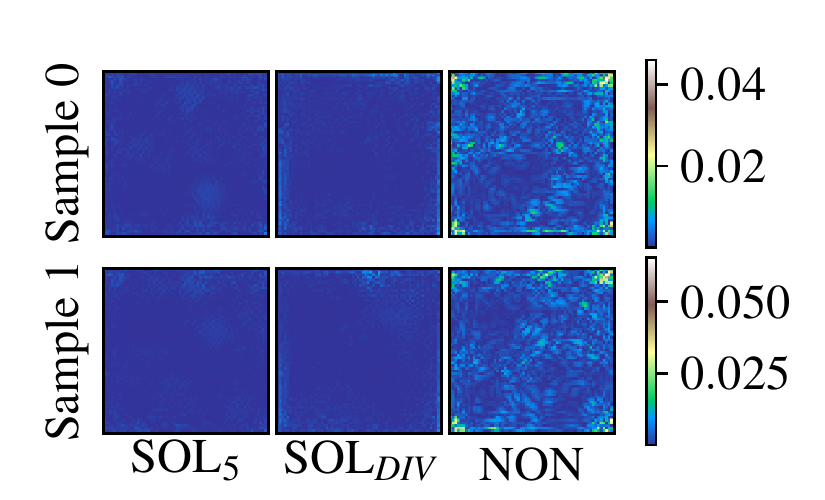}
  \includegraphics[width=0.49\columnwidth]{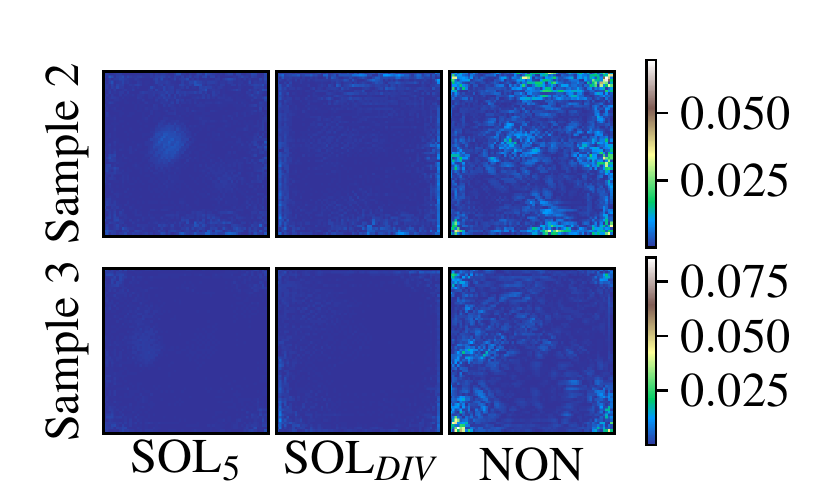}
  \caption{Residual error after one CG solver iteration.}
  \label{fig:cg_residual_img}
\end{figure}

When investigating the inferred pressure fields themselves (\myfigref{fig:cg_img_results}), we see that the guesses of the \sol{5} model come closest to the reference, followed by those of the NON variant. 
The \sol{\text{DIV}} differs more strongly,
and the residual divergence, shown in \myfigref{fig:cg_residual_img}, highlights that it 
has a noticeable error pattern near the outer border of the domain. This provides an explanation for the poor 
behavior of the \sol{\text{DIV}} model for the initial CG solver iterations:
while it minimizes the PDE-based loss in an absolute sense, it does not receive information about 
how different parts of the solution influence the future iterations of the solver. This ambiguity
is alleviated to some extent by the pre-computed reference solutions for NON, but especially 
the \sol{5} version receives this feedback in terms of gradient from the differentiable solver
and, in this way, can best adapt to the requirements for future iterations.

\begin{figure}[htb]
  \centering
  \includegraphics[width=0.49\columnwidth]{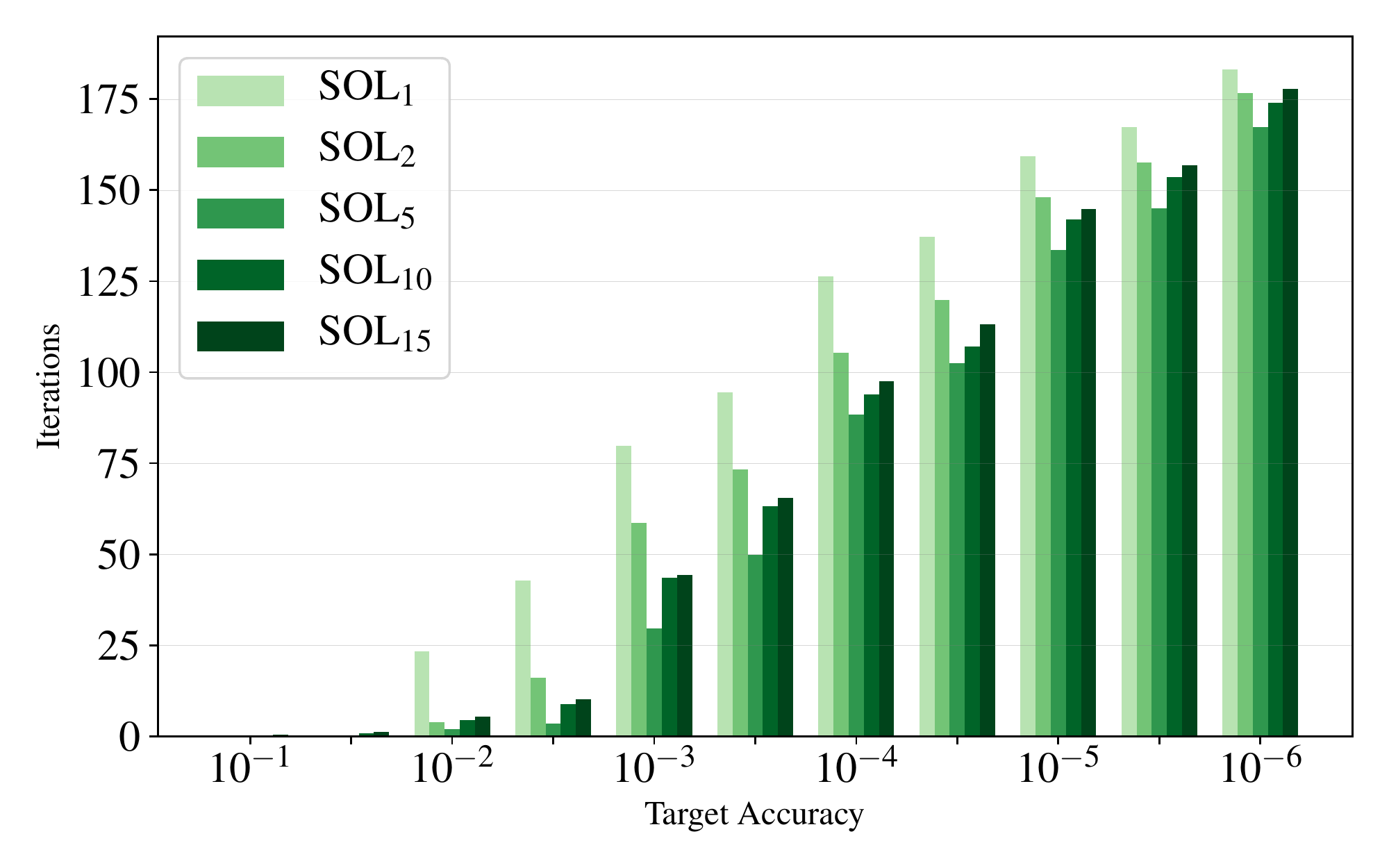}
  \includegraphics[width=0.49\columnwidth]{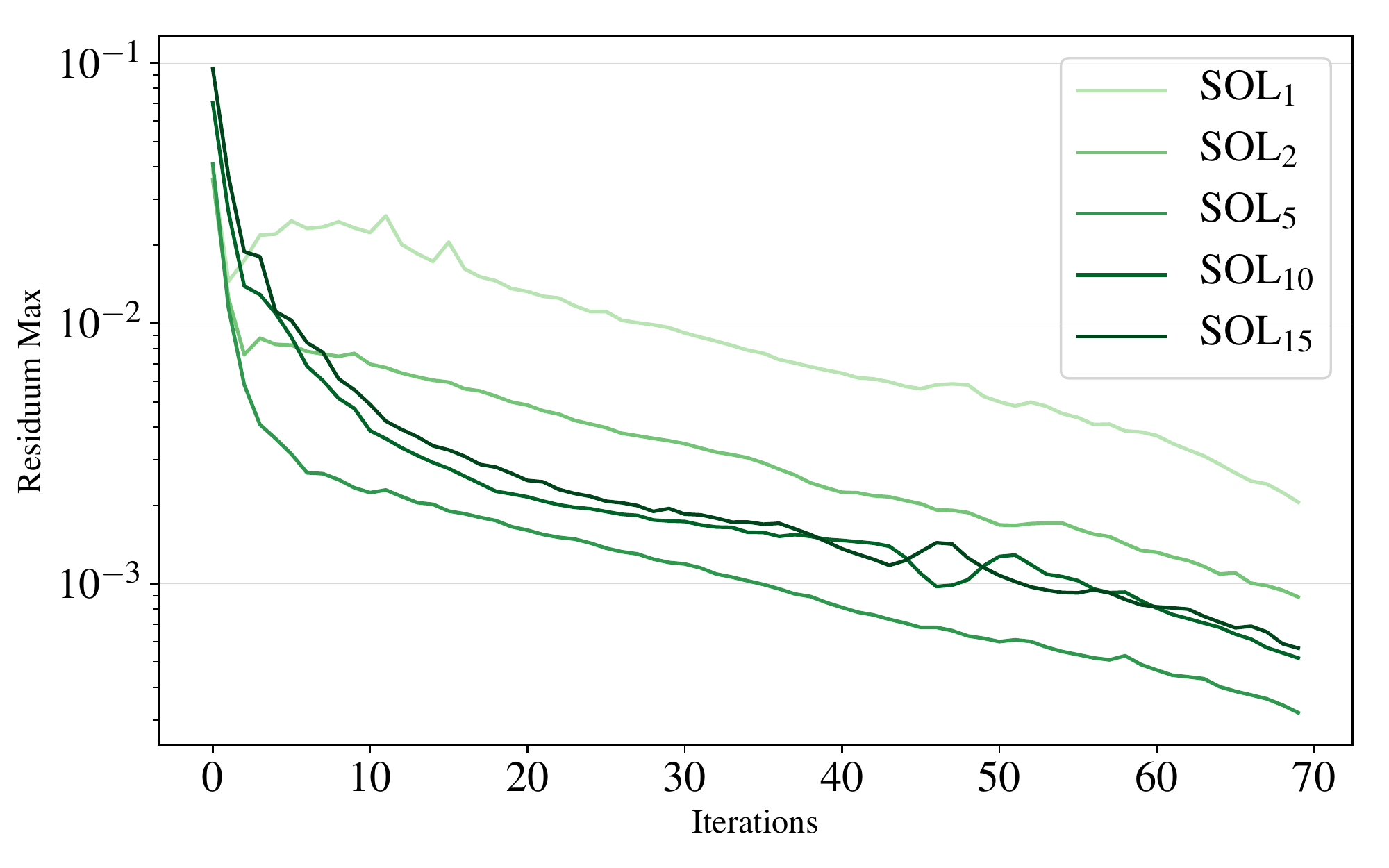}
  \caption{Comparison of 
    SOL models with different look-ahead steps.}
  \label{fig:cg_it_stepsizes}
\end{figure}

%
We also experimented with varying the number of look-ahead steps for \sol{n} models in the loss function of \myeqref{eq:cg_loss3}. 
This ablation study (\myfigref{fig:cg_it_stepsizes}) shows how too few iterations clearly deteriorate the performance, while more than 5 iterations 
lead to a slight increase in the required iterations. We assume that this behavior is potentially caused by evaluating the loss 
only for the final output of the $n$ iterations.

\paragraph{Discussion}
Our results highlight the advantages of training with the solver in the loop for fully 
implicit PDE solvers. Likewise, it shows that a physics-informed loss formulation alone 
yields only a partial view of the problem.
While a loss-based residual cannot adapt to iterative algorithms, the solver-in-the-loop 
models directly receive gradient-based feedback at training time.

The combination of an inferred initial guess with a traditional solver represents a particularly interesting
hybrid algorithm, as it gives convergence guarantees that a learned approach alone would not be able to provide.
Even if a trained model generates a sub-optimal solution, the solver can improve the solution 
until it matches the desired accuracy threshold. On the other hand, pre-training a model for a known problem domain can significantly reduce the required number of iterations and, consequently, reduce the workload in scenarios where PDEs from the same problem domain need to be solved repeatedly and in large numbers.
Here, the current hardware developments provide an additional promise: the advances in terms of highly specialized hardware for evaluating neural networks
can provide a substantial future speed-up even for a fixed, pre-trained model.


\begin{table}[htb]
  \caption{Evaluation of the CG solver performance for different models.}
  \label{tab:cgresults}
  \begin{center}
    \begin{tabular}{ccccccc}
      \toprule
      \textbf{Model}   & \multicolumn{6}{c}{\textbf{Iterations for Accuracy}, Mean (std. dev.)}     \\
      \cmidrule{2-7}
                       & $10^{-1}$ & $10^{-2}$ & $10^{-3}$ & $10^{-4}$ & $10^{-5}$ & $10^{-6}$ \\
      \midrule
      NON              & 1.67      & 9.33      & 52.16     & 109.12    & 155.37    & 186.12    \\
                       & (1.010)   & (5.428)   & (17.540)  & (15.875)  & (10.155)  & (5.719)   \\
      \sol{\text{DIV}} & 0.0       & 27.79     & 79.06     & 117.97    & 155.76    & 181.07    \\
                       & (0.0)     & (15.255)  & (10.042)  & (13.234)  & (9.403)   & (6.052)   \\
      \sol{5}          & 0.03      & 1.97      & 29.59     & 88.37     & 133.59    & 167.37    \\
                       & (0.171)   & (1.118)   & (14.832)  & (13.465)  & (11.605)  & (8.549)   \\
      \bottomrule
    \end{tabular}
  \end{center}
\end{table}

\clearpage

\subsection{Three-dimensional Unsteady Wake Flow}\label{app:expKarman3d}

As a final scenario, we target a three-dimensional 
fluid flow problem.
The third spatial dimension leads to a large increase in terms of 
degrees of freedom, especially in the finer reference manifold. Additionally,
the three axes of rotation lead to significantly more complicated flow 
structures. 

Overall, we target a setup that represents an extension of the 2D 
unsteady wake flow case of \myappref{app:expKarman2d}. Instead of a circular 
obstacle, the flow now faces a cylindrical obstacle in a 3D domain with 
extent of $1 \times 1 \times 2$. The cylinder 
with diameter $0.1$ is located at position $(1/2,1/2,0)^T$ and has an extent 
of 1 unit along the z-axis. We use the incompressible Navier-Stokes 
equations in three dimensions as underlying PDE:  
\begin{eqnarray}
  \label{eq:model-ns3d}
  \frac{\partial u_x}{\partial{t}} + \vu \cdot \nabla u_x =
  - \frac{1}{\rho}\nabla{p} + \nu \nabla\cdot \nabla u_x 
  \nonumber
  \\
  \frac{\partial u_y}{\partial{t}} + \vu \cdot \nabla u_y =
  - \frac{1}{\rho}\nabla{p} + \nu \nabla\cdot \nabla u_y 
  \nonumber
  \\
  \frac{\partial u_z}{\partial{t}} + \vu \cdot \nabla u_z =
  - \frac{1}{\rho}\nabla{p} + \nu \nabla\cdot \nabla u_z 
  \nonumber
  \\
  \text{subject to} \quad \nabla \cdot \vu = 0.
\end{eqnarray}

\begin{figure}[htb]
  \centering
  \begin{overpic}[width=1.0\linewidth]{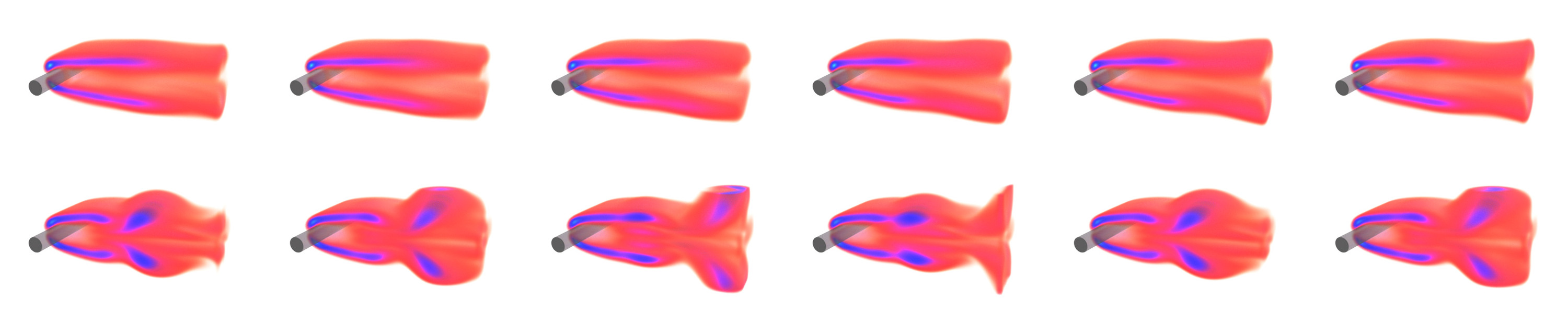}
    \put(100,17)      {(a)}
    \put(-1.6,0)    {\scriptsize\rotatebox{90}{\makebox[11.65\unitlength]{Reference}}}
    \put(-1.6,11.65){\scriptsize\rotatebox{90}{\makebox[11.65\unitlength]{SRC}}}
    \put(100.5,0){\includegraphics[height=11.65\unitlength]{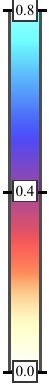}}
  \end{overpic}\vspace{10pt}
  \begin{overpic}[width=1.0\linewidth]{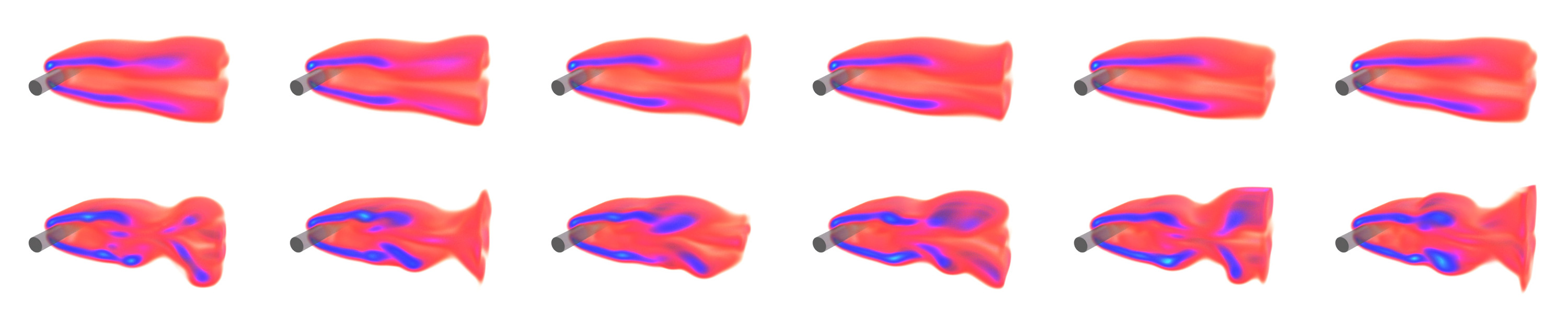}
    \put(100,17)      {(b)}
    \put(-1.6,0)    {\scriptsize\rotatebox{90}{\makebox[11.65\unitlength]{Reference}}}
    \put(-1.6,11.65){\scriptsize\rotatebox{90}{\makebox[11.65\unitlength]{SRC}}}
    \put(100.5,0){\includegraphics[height=11.65\unitlength]{figs/summary-legend-karman3d}}
  \end{overpic}
  \caption{Two example sequences with (a) Re=117.2 and (b) Re=273.4 of the
    three-dimensional wake flow from the training data set. 
    Each row shows 200 time steps for SRC (top) and
    reference versions (bottom) in terms of vorticity magnitude.
    }
  \label{fig:appx:karman3d-images-train}
\end{figure}

For reference simulations, the domain is discretized with 
$d_{r,x}\!=\!d_{r,y}\!=\!128$ and $d_{r,z}\!=256$ cells using a staggered layout for the velocity components. 
The source domain has a resolution of
$d_{s,x}\!=\!d_{s,y}\!=\!32$ and $d_{r,z}\!=64$ cells. Data sets from both domains contain 
phase space trajectories of 500 time steps.
For the training data, the viscosity coefficient $\nu$ is chosen
to yield Reynolds numbers 
Re$_{\text{train}} \in \{ 58.6, 78.1, 117.2, 156.3, 234.4, 312.5, 468.8, 625.0 \}$.
While the range of Reynolds numbers covers a slightly reduced range 
compared to the 2D case, there is still a factor of more than ten between 
largest and smallest ones, and the 3D nature of the flow introduces
a significant amount of complexity. The example visualizations of 
a training data set in \myfigref{fig:appx:karman3d-images-train} highlight the complexity 
of the flows.

For the test set, we use different Reynolds numbers, namely 
Re$_{\text{test}} \in \{$68.4, 97.7, 195.3, 136.7, 273.4, 390.6, 546.9$\}$.
The following test evaluations were computed for the seven Reynolds numbers in Re$_{\text{test}}$
over 300 time steps. Numeric values are given in \mytabref{tab:karman3d}.

\paragraph{Training Procedure} 

For the 3D case, we use a ResNet that largely follows the architecture of the 2D 
cases, but employs 3D convolutions instead. The ResNet contains six blocks 
with kernel sizes of 5$\times$5$\times$5 and 3$\times$3$\times$3 for the two convolutional layers per block.
The number of filters is increases to 48 in the center of the network, 
yielding 1002k trainable parameters (also see \myappref{app:models}). 
\new{As for the 2D case, the inputs for the 3D models contain a constant field indicating 
the targeted Reynolds number.}
All models were trained for 300k iterations using a
learning rate of $10^{-4}$ and a batch-size of four.
We then use three validation simulations with
Re$_{\text{val}} \in \{$61.0 , 305.2 , 470.0$\}$ to select the best performing model.

Due to the increased computational workload to train the 3D models,
we focus on a NON variant and a \sol{16} version, which 
uses the same differentiable Navier-Stokes solver for producing gradient 
information over the course of up to 16 unrolled simulation and inference steps 
for each iteration at training time. This version was trained 
with \sol{8} for 200k iterations and then for an additional 100k iterations 
as \sol{16}.

\paragraph{Results} 

The 3D flow represents a significant increase in terms 
of complexity for the deep learning models. Among others, we were not 
able to train a stable NON version despite numerous tests. While 
the models performed well for ca. 100 to 150 time steps, small scale oscillations
induced by the corrections accumulate and start to strongly distort the flow.
This is a good example of the undesirable shift of distributions for the inputs:
once the phase space trajectories produced by the hybrid method leave the 
distribution of the regular source states seen at training time, the 
model fails to infer reasonable corrections.

In contrast, the \sol{16} version retains its stability over the course 
of long simulations with several hundred steps. This is reflected in the MAE measurements 
of the velocity fields over the test cases: the regular source simulation induces 
an error of 0.167, which the NON version reduces to 0.143. The \sol{16}
reduces the error to 0.130 instead, which however only gives a partial view of 
the overall behavior of the different versions. 
The graphs over time shown in \myfigref{fig:appx:karman3d-freq:mae}
illustrate the diverging behavior of the NON version. While it does very well initially, 
even slightly surpassing \sol{16} around frame 100, the errors quickly grow 
afterwards, eventually leading to a performance that is worse than the source simulation.

The frequency graphs of the kinetic energy in \myfigref{fig:appx:karman3d-freq:ke},
measured for an array of $5^3$ sample points at the center of the domain,
instead show that the \sol{16}
simulations closely match the frequency distribution of the reference simulations. 
It succeeds in restoring the change of frequencies across the different temporal scales of the flow
significantly better than the SRC and NON models.
The source simulation instead underestimates larger frequencies and over-estimates 
smaller ones.

\begin{figure}[tb]
  \centering
  \begin{overpic}[width=1.0\linewidth]{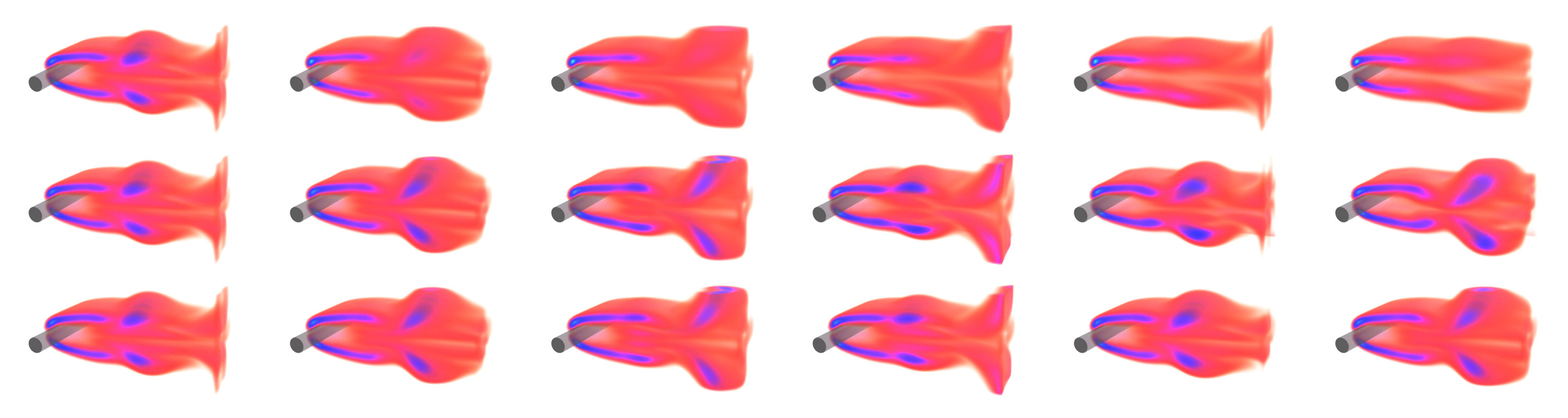}
    \put(100,22)   {(a)}
    \put(-1.6,0)   {\scriptsize\rotatebox{90}{\makebox[11.65\unitlength]{Ref.}}}
    \put(-1.6,8.0) {\scriptsize\rotatebox{90}{\makebox[11.65\unitlength]{\sol{16}}}}
    \put(-1.6,16.0){\scriptsize\rotatebox{90}{\makebox[11.65\unitlength]{SRC}}}
    \put(100.5,0)  {\includegraphics[height=11.65\unitlength]{figs/summary-legend-karman3d}}
  \end{overpic}\vspace{10pt}
  \begin{overpic}[width=1.00\linewidth]{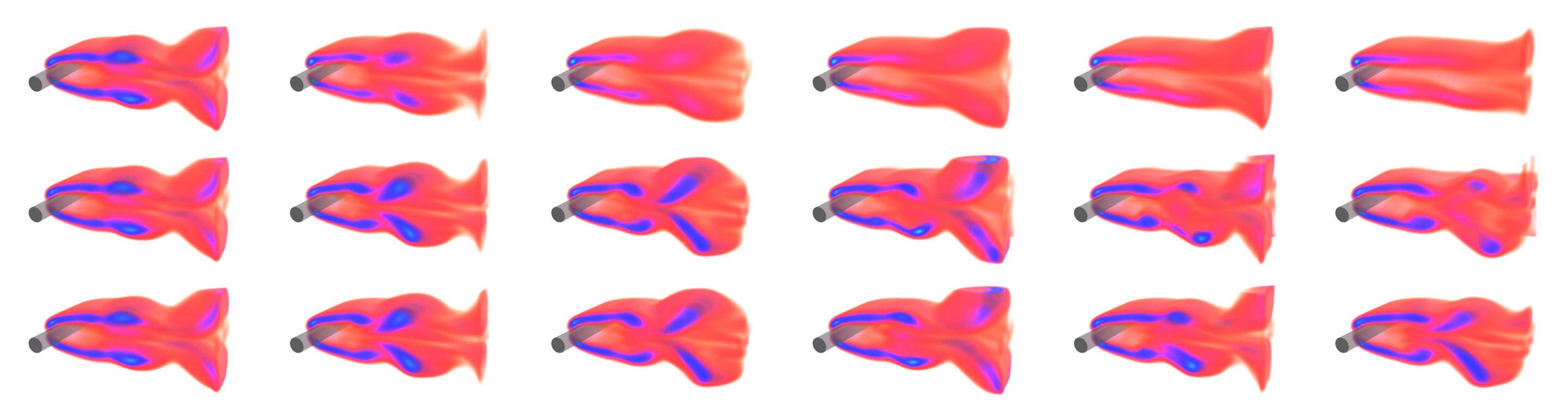}
    \put(100,22)   {(b)}
    \put(-1.6,0)   {\scriptsize\rotatebox{90}{\makebox[11.65\unitlength]{Ref.}}}
    \put(-1.6,8.0) {\scriptsize\rotatebox{90}{\makebox[11.65\unitlength]{\sol{16}}}}
    \put(-1.6,16.0){\scriptsize\rotatebox{90}{\makebox[11.65\unitlength]{SRC}}}
    \put(100.5,0)  {\includegraphics[height=11.65\unitlength]{figs/summary-legend-karman3d}}
  \end{overpic}\vspace{10pt}
  \begin{overpic}[width=1.00\linewidth]{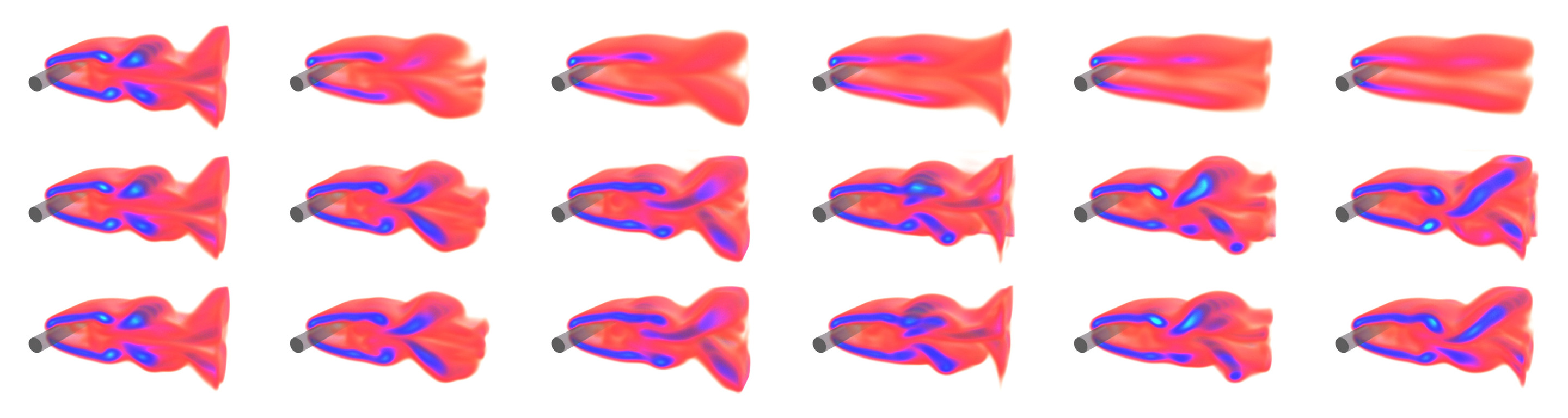}
    \put(100,22)   {(c)}
    \put(-1.6,0)   {\scriptsize\rotatebox{90}{\makebox[11.65\unitlength]{Ref.}}}
    \put(-1.6,8.0) {\scriptsize\rotatebox{90}{\makebox[11.65\unitlength]{\sol{16}}}}
    \put(-1.6,16.0){\scriptsize\rotatebox{90}{\makebox[11.65\unitlength]{SRC}}}
    \put(100.5,0)  {\includegraphics[height=11.65\unitlength]{figs/summary-legend-karman3d}}
  \end{overpic}
  \caption{Three test cases with (a) Re=68.4, (b) Re=136.7, and (c) Re=546.9.
    Each row shows time steps over the course of 200 time steps for SRC,
    \sol{16}, and the reference (top to bottom). The \sol{16} model interacting
    with the source solver successfully preserves the complex rotating motions
    behind the cylindrical obstacle (middle), which the regular source solver
    cannot resolve (top).}
  \label{fig:appx:karman3d-images-test}
\end{figure}

\myfigref{fig:appx:karman3d-images-test} visualizes the vorticity magnitude of 
several test cases with Reynolds numbers not seen during training.
The \sol{16} model manages to correct the vortex shedding behavior of the source simulation 
and closely matches the reference. As we visualize in the supplemental video,
the NON version starts to oscillate, injecting undesirable 
distortions into the velocity field.


\newcommand{\myFigHKthree}{0.37\columnwidth}
\begin{figure}[htb]
  \centering
  \subcaptionbox{Velocity error over time\label{fig:appx:karman3d-freq:mae}}{\includegraphics[height=\myFigHKthree,page=15]{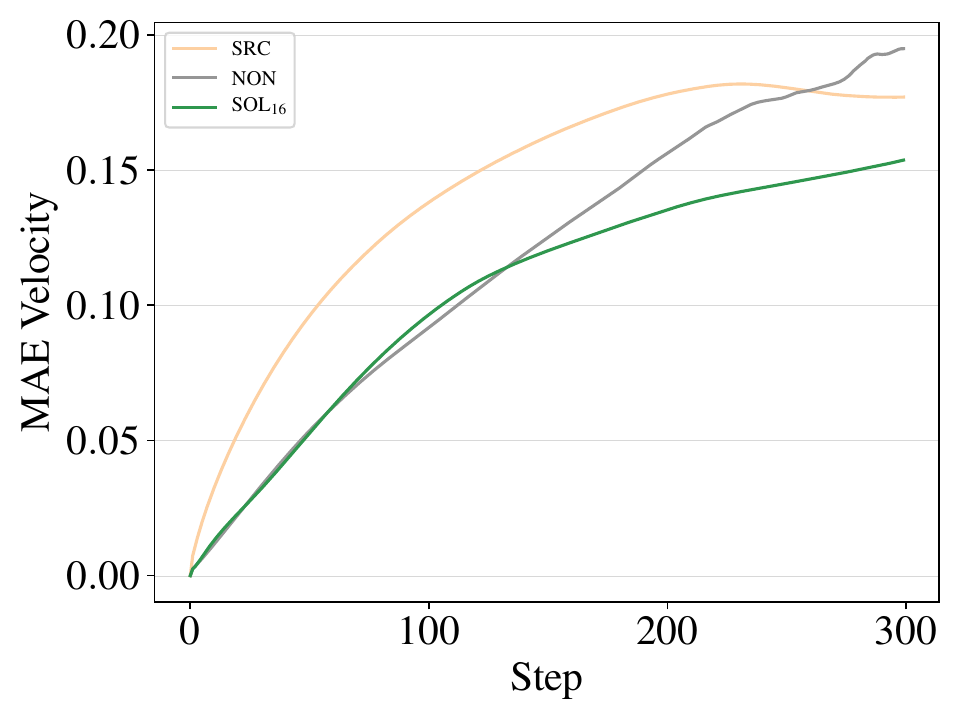}}
  \subcaptionbox{Frequency error\label{fig:appx:karman3d-freq:ke}}{\includegraphics[height=\myFigHKthree,page=16]{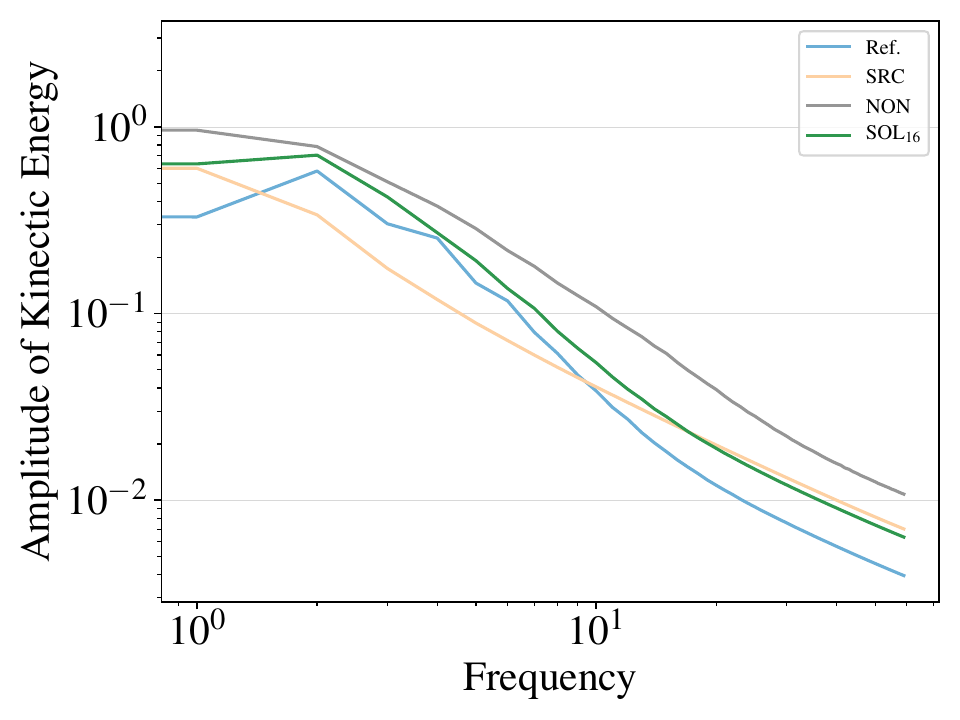}}
  \caption{Evolutions of velocity MAE and frequency errors over the course of 300 time steps 
    averaged for the seven test cases of the three-dimensional wake flow. (a) The NON versions perform well initially,
    but strongly diverges for later frames. (b) The \sol{16} shows a clearly improvement in terms of the frequency 
    distribution of the kinetic energies. The overall curve of \sol{16} closely follows the reference with an initial 
    offset over the reference, 
    which inherits from the source simulation.
    }
  \label{fig:appx:karman3d-freq}
\end{figure}

\begin{table}[htb]
  \caption{Quantitative evaluation of different models for the three-dimensional 
    wake flow scenario.}
  \label{tab:karman3d}
  \begin{center}
  \small
    \begin{tabular}{cccccc}
      \toprule
      \multicolumn{3}{c}{\textbf{MAE Velocity}, Mean (std. dev.)} & \multicolumn{3}{c}{\textbf{Freq. MAE Kinetic Energy}, Mean (std. dev.)}   \\
      \cmidrule(r){1-3} \cmidrule(l){4-6}
      SRC             & NON           & SOL$_{16}$    & SRC            & NON           & SOL$_{16}$    \\
      \cmidrule(r){1-3} \cmidrule(l){4-6}
      0.167 (0.035)   & 0.143 (0.070) & 0.130 (0.024) & 0.0614 (0.133) & 0.074 (0.209) & 0.058 (0.088) \\
      \bottomrule
    \end{tabular}
  \end{center}
\end{table}




\section{Performance}\label{app:perf}


We measure the computational performance of our models 
in comparison to a reference simulation 
on a workstation with an
Intel Xeon E5-1650 CPU with 12 virtual cores at 3.60GHz and
an NVIDIA GeForce GTX 1080 Ti GPU.
As reference solver, we employ a CPU-based simulator 
using OpenMP parallelization.
We compare this with our (relatively un-optimized)
differentiable physics framework, which evaluates the 
PDE and the trained model within \emph{TensorFlow} on the GPU.

For the buoyancy-driven flow simulation, the CPU-based reference simulation 
requires 5.79 seconds on average for 100 time steps. 
Instead, 
\new{evaluating the \sol{128} neural network model itself requires an accumulated 0.43 seconds. 
For comparison, computing 100 time steps of the source solver takes 0.476 seconds.}
\new{
In comparison to the inference for forward simulations with a pre-trained model, each iteration during training is significantly more expensive: for the \sol{8}, \sol{16}, and \sol{32} models of the 2D wake flow case, a training iteration took 0.6, 1.3, and 2.5 seconds on average, respectively. As this is a one-time, pre-processing cost, the gains in performance of the resulting hybrid solver can quickly offset the computational expense for training a model.}

The computational workload for PDE solvers typically rises super-linearly 
with the number of degrees of freedom. Hence, the gap is even more 
pronounced when considering the 3D wake flow case.
\new{Here, the reference simulation requires 913.2 seconds for 100 time steps, 
while the \sol{16} version requires 13.3 seconds on average. Thus, the 
source simulation with  learned corrections is more than 68 times 
faster than 
the reference simulation.}

Despite the substantial reduction in terms of runtime, we believe these performance 
results are preliminary, and far from the speed-up that could be achieved 
in optimal settings with a learning-augmented PDE solver. An inherent advantage of combining 
an approximate PDE solver with a deep-learning-based corrector ANN 
stems from the fact that a relatively simple solver suffices as a basis. Hence, 
while existing reference solvers in scientific computing fields might come 
with vast existing code-bases, the source solver could encompass only a small subset 
of the full solver and introduce the residual dynamics via a learned component.
This would also reduce the work to provide gradients for the source solver, 
which many existing simulation frameworks do not readily offer.
Due to its reduced scope, the source solver would also be significantly easier to optimize. 
Additionally, the learned corrector component would trivially benefit from all 
future hardware advances for efficient evaluations of neural networks.
Hence, we believe that, in practice, a much more substantial speed-up will be achievable
than the ones we have measured for the two- and three-dimensional simulation
scenarios of this work.



\section{Neural Network Architectures}\label{app:models}

\begin{figure}[htb]
\centering
$\vcenter{\hbox{\includegraphics[width=0.49\columnwidth]{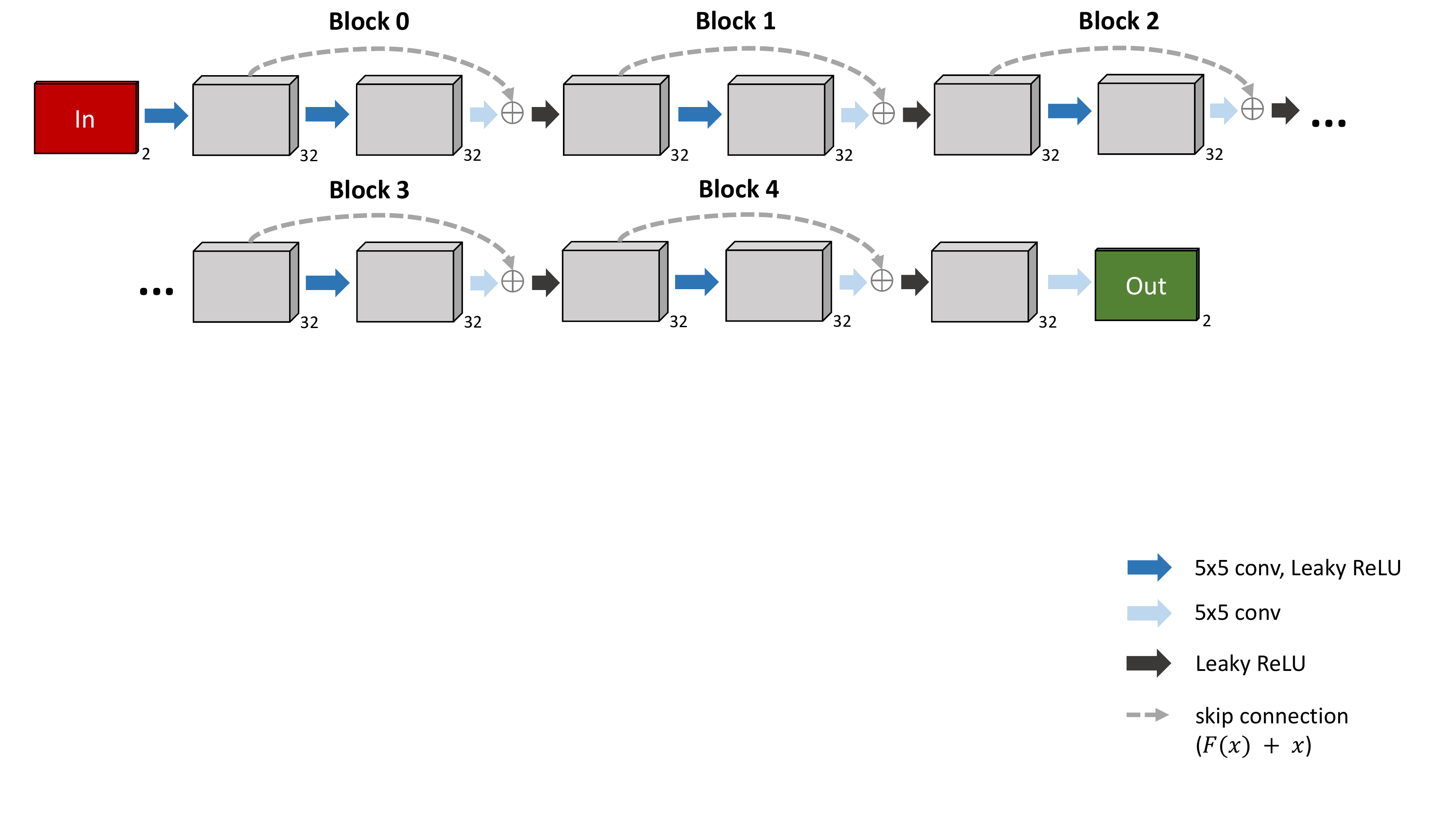}}}$
$\vcenter{\hbox{\includegraphics[width=0.49\columnwidth]{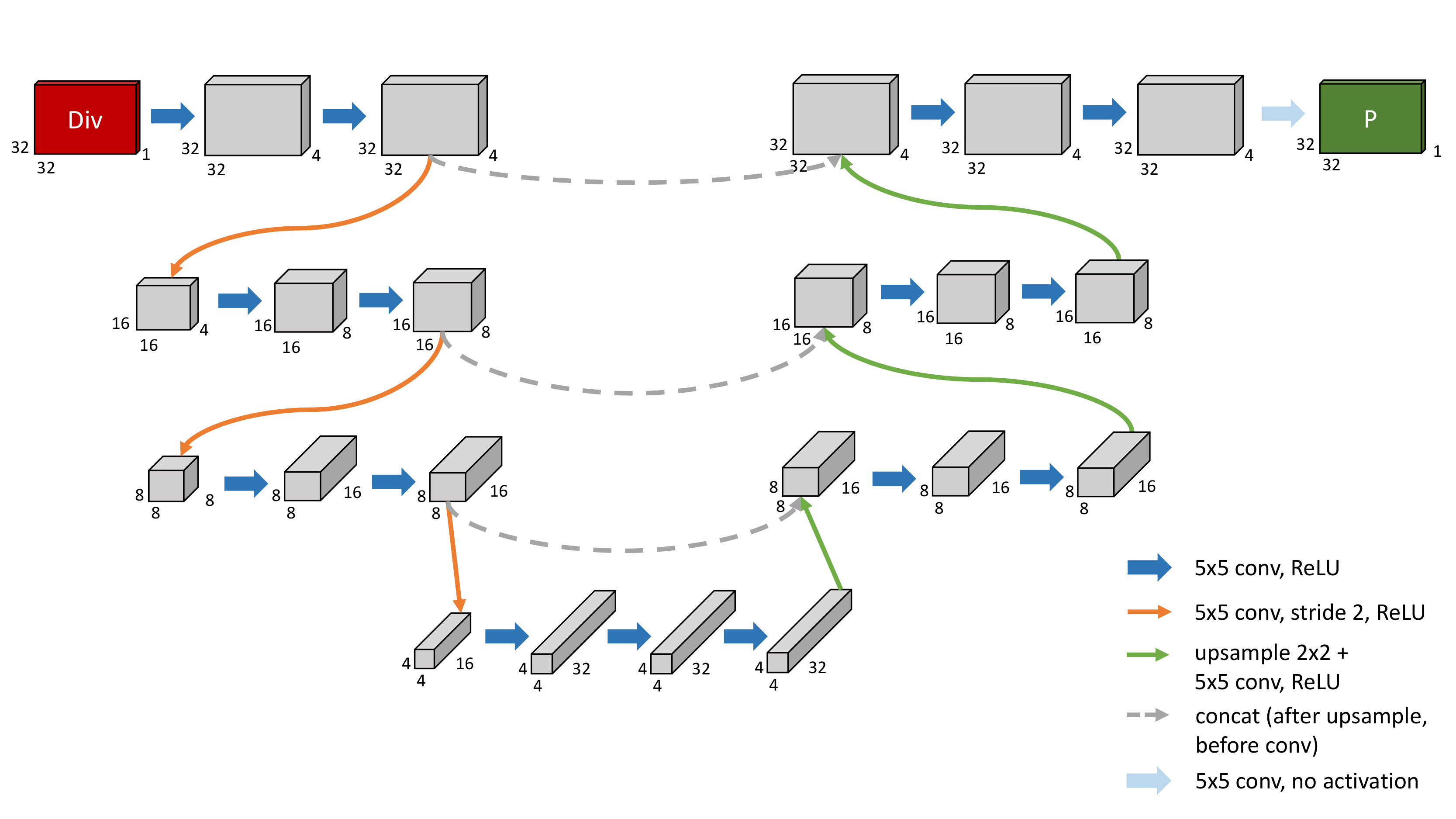}}}$
\caption{A visual summary of the two main architectures of the neural networks used for
  Sect.~\ref{app:expKarman2d} to \ref{app:expBurgers} (left), 
  and Sect.~\ref{app:expCgSolver} (right).
}
\label{fig:nn_architecture}
\end{figure}

Below, we give additional details of the network architectures
used for the five different scenarios. We intentionally 
slightly vary the architecture to demonstrate that our solver-in-the-loop 
approach does not rely on a single, specific architecture.
We employ ResNets for the large majority of the PDE interaction models
as the correction task resembles a translation from phase space input 
quantities to a field of localized corrections.
The CG solver scenario, on the other hand, requires a more global view,
which motivates our choice of a U-net architecture.
The overall structure with kernel sizes and feature maps 
of both types of networks is illustrated in \myfigref{fig:nn_architecture}.
We additionally list hyperparameters for each architecture 
in \mytabref{tab:nn_architecture}.


%
\begin{table}[htb]
  \caption{Hyperparameters of neural network architectures.}
  \label{tab:nn_architecture}
  \begin{center}
    \small
    \begin{tabular}{lccccr}
      \toprule
      Experiment                                & Arch.      & Layers & Features     & Conv. Kernels & Train. Weights \\
      \midrule
      2D Wake Flow \ref{app:expKarman2d}        & Res-Net    & 12     & 32           & $5^2$         & 260,354        \\
      2D Wake Flow \ref{app:expKarman2d}, Small & Sequential & 3      & 32, 64       & $5^2$         & 56,898         \\
      Buoyancy \ref{app:expBuoy}, M$_{XS}$      & Res-Net    & 6      & 4            & $5^2,3^2$     & 1,310          \\
      Buoyancy \ref{app:expBuoy}, M$_{S}$       & Res-Net    & 8      & 8            & $5^2,3^2$     & 5,114          \\
      Buoyancy \ref{app:expBuoy}, Regular       & Res-Net    & 10     & 16           & $5^2,3^2$     & 35,954         \\
      Buoyancy \ref{app:expBuoy}, M$_{L}$       & Res-Net    & 14     & 16, 32       & $5^2,3^2$     & 100,114        \\
      Buoyancy \ref{app:expBuoy}, M$_{XL}$      & Res-Net    & 14     & 32, 64       & $5^2,3^2$     & 400,930        \\
      Forced Adv.-diff. \ref{app:expBurgers}    & Res-Net    & 12     & 32           & $5^2$         & 261,154        \\
      CG Solver \myappref{app:expCgSolver}      & U-Net      & 22     & 4, 8, 16, 32 & $5^2$         & 127,265        \\
      3D Wake Flow \ref{app:expKarman3d}        & Res-Net    & 14     & 24,48        & $5^3,3^3$     & 1,002,411      \\
      \bottomrule
    \end{tabular}
  \end{center}
\end{table}


\end{document}